\documentclass[12pt]{spieman}  
\usepackage{tocloft}
\usepackage{amsmath,amsfonts,amssymb}
\usepackage{mathabx}
\usepackage{deluxetable}
\usepackage{graphicx}
\usepackage{setspace}
\usepackage{float}
\usepackage{hhline}
\usepackage{lineno}
\usepackage{pdflscape}

\newcommand{\rom}[1]{\uppercase\expandafter{\romannumeral #1\relax}}
\newcommand{\PSUAA}{Department of Astronomy \& Astrophysics, 525 Davey Laboratory, The Pennsylvania State University, University Park, PA, 16802, USA}
\newcommand{\PSUCEHW}{Center for Exoplanets and Habitable Worlds, 525 Davey Laboratory, The Pennsylvania State University, University Park, PA, 16802, USA}
\newcommand{\PSUStats}{Center for Astrostatistics, 525 Davey Laboratory, The Pennsylvania State University, University Park, PA, 16802, USA}
\newcommand{\PSUICDS}{Institute for Computational and Data Sciences, The Pennsylvania State University, University Park, PA 16802, USA}

\def\gappeq{\mathrel{ \rlap{\raise.5ex\hbox{$>$}}
                      {\lower.5ex\hbox{$\sim$}}  } }

\title{Paths to Robust Exoplanet Science Yield Margin for the Habitable Worlds Observatory}

\author[a]{Christopher C. Stark}
\author[b]{Bertrand Mennesson}
\author[c]{Steve Bryson}
\author[d,e,f,g]{Eric B. Ford}
\author[h]{Tyler D. Robinson}
\author[c]{Ruslan Belikov}
\author[a]{Matthew R. Bolcar}
\author[a]{Lee D. Feinberg}
\author[i]{Olivier Guyon}
\author[a,j]{Natasha Latouf}
\author[a]{Avi M. Mandell}
\author[a]{Bernard J. Rauscher}
\author[c]{Dan Sirbu}
\author[a]{Noah W. Tuchow}

\affil[a]{NASA Goddard Space Flight Center, Greenbelt, MD 20771, USA}
\affil[b]{Jet Propulsion Laboratory, California Institute of Technology, 4800 Oak Grove Drive, Pasadena, CA 91109, USA}
\affil[c]{NASA Ames Research Center, Moffett Field, CA 94035, USA}
\affil[d]{\PSUAA}
\affil[e]{\PSUCEHW}
\affil[f]{\PSUICDS}
\affil[g]{\PSUStats}
\affil[h]{Lunar \& Planetary Laboratory, University of Arizona, Tucson, AZ 85721 USA}
\affil[i]{Steward Observatory, University of Arizona, Tucson, AZ 87521, USA}
\affil[j]{George Mason University, 4400 University Dr, Fairfax, VA 22030, USA}

\cftpagenumbersoff{figure}
\cftpagenumbersoff{table} 
\begin{document} 
\maketitle

\begin{abstract}

The Habitable Worlds Observatory (HWO) will seek to detect and characterize potentially Earth-like planets around other stars. To ensure that the mission achieves the Astro2020 Decadal's recommended goal of 25 exoEarth candidates (EECs), we must take into account the probabilistic nature of exoplanet detections and provide ``science margin" to budget for astrophysical uncertainties with a reasonable level of confidence. In this study, we explore the probabilistic distributions of yields to be expected from a blind exoEarth survey conducted by such a mission. We identify and estimate the impact of all major known sources of astrophysical uncertainty on the exoEarth candidate yield. As expected, $\eta_{\Earth}$ uncertainties dominate the uncertainty in EEC yield, but we show that sampling uncertainties inherent to a blind survey are another important source of uncertainty that should be budgeted for during mission design. We adopt the Large UV/Optical/IR Surveyor Design B (LUVOIR-B) as a baseline and modify the telescope diameter to estimate the science margin provided by a larger telescope. We then depart from the LUVOIR-B baseline design and identify six possible design changes that, when compiled, provide large gains in exoEarth candidate yield and more than an order of magnitude reduction in exposure times for the highest priority targets. We conclude that a combination of telescope diameter increase and design improvements could provide robust exoplanet science margins for HWO. 

\end{abstract}


\begin{spacing}{2}

\section{Introduction}
\label{intro}

A primary driving science case for the Habitable Worlds Observatory (HWO) is the high-contrast imaging of potentially Earth-like planets, or exoEarth candidates (EECs). The Astro2020 Decadal Survey recommended a quantitative science goal of detecting and characterizing 25 EECs\cite{astro2020} with a $\sim$6 m inscribed diameter (ID) telescope, roughly based on the expectation value of blind survey EEC yields from Ref.~\citenum{stark2019}. However, assuming few EECs have been detected prior to launch\cite{morgan2021a}, HWO's blind survey detection rates will be probabilistic, with many factors affecting our chances of success. As such, the EEC yield for HWO cannot be known exactly in advance and is more accurately represented as a distribution. 

There are multiple astrophysical uncertainties that will ultimately lead to uncertainties in HWO's direct imaging exoplanet yield for a blind survey. Some of these are unavoidable while others could conceivably be reduced with future observations. Previous studies have examined some of these sources of uncertainty, but have mostly treated them in isolation from one another. Ref. \citenum{savransky2016} was the first to include exoplanet sampling uncertainties, but did not account for some other uncertainties like the exozodi distribution, while Ref. \citenum{stark2014} looked only at the impact of median exoplanet albedo and exozodi independently. Other studies that have combined multiple sources of uncertainty have been incomplete and adopted rudimentary methods. Refs. \citenum{luvoirfinalreport} and \citenum{habexfinalreport} attempted to simultaneously incorporate many sources of astrophysical uncertainty into the yield calculations, but ignored exoplanet albedo, simplified the treatment of exozodi uncertainty, and did not have a well-informed distribution of possible exoplanet occurrence rates. Here we perform a more complete study of the impact of astrophysical uncertainties on EEC yields. 

The Astro2020 Decadal Survey asserted that a sample size of 25 EECs ``provides robustness against the uncertainties in the occurrence rate of Earth-sized worlds and against the vagaries associated with the particular systems near Earth."\cite{astro2020} Here we quantitatively assess this statement by estimating the EEC yield distribution for HWO and using this distribution to estimate our probability of achieving a given EEC yield.  By adopting design choices that shift this distribution to higher yields, we show how building ``science margin" into the mission design can ensure HWO has a higher chance of achieving its goals. This same science margin, if designed properly, can also help budget against performance degradation.

We use the most recent version of the Altruistic Yield Optimizer (AYO), detailed in Ref. \citenum{stark2023}, to estimate EEC yield distributions for a blind exoEarth survey with a coronagraph-based mission in family with HWO. In Section \ref{section:methods} we briefly review the AYO methods and present our baseline mission assumptions. We then discuss the sources of astrophysical uncertainty one by one in Section \ref{section:astrounc}, show how each impacts the yield distribution, and present a final estimated yield distribution incorporating known sources of astrophysical uncertainties. In Section \ref{section:paths} we identify paths to improving EEC yields through tangible changes to our baseline mission, some of which are relatively straightforward and some of which require significant technological development. Finally, we discuss how the concept of ``science margin" can also help budget for performance degradation in the mission, or (relatedly) provide margin against cost growth by allowing relaxation of parameters that drive the mission cost.

The design of HWO will be informed by many metrics. Here we focus on a single well-defined metric: the detection and characterization of EECs. The Astro2020 Decadal Report did not define ``characterization," so we make the same assumption as the LUVOIR and HabEx reports and budget for the spectral characterization time required to search each EEC for water vapor. As shown by previous studies\cite{kopparapu2018, stark2019, howe2024}, a survey designed to detect and characterize EECs will, by its nature, detect many additional exoplanets. These additional exoplanets will be very diverse, spanning a broad range of phase space, and their yield will be an important metric for HWO design. However, the relative merits of diverse exoplanet yields is a topic beyond the scope of this paper.

\section{Methods and baseline assumptions\label{section:methods}}

We use AYO\cite{stark2014,stark2015,stark2023} to calculate exoplanet yields. Briefly, AYO distributes a large number ($\sim10^5$) of synthetic EECs around each star for thousands of nearby stars, sampling the range of possible orbits, phases, and planet radii consistent with the adopted definition of an EEC, calculates their exposure times given a model of the observatory/instrument and background sources, and then numerically determines the completeness $C$ as a function of exposure time $t$ (and importantly, its derivative $dC/dt$). AYO then uses an advanced version of the equal-slope method\cite{hunyadi2007} to determine the optimal value of $dC/dt$ for all observations, simultaneously optimizing the selected targets, the number of visits to each star, and the exposure and delay times for each visit. Briefly, the equal-slope method requires that $dC/dt$ is equal for all observations, ensuring that they are equally productive per unit time, as expected for optimally distributed exposure time\cite{stark2014, stark2015}.

We use the HWO Preliminary Input Catalog (HPIC) as our input target list\cite{tuchow2024}. This target list contains $\sim13$k stars within 50 pc complete to a TESS magnitude of 12, formed from the union of the TESS Input Catalog\cite{stassun2019} and the Gaia DR3 catalog\cite{gaiadr3}. The HPIC includes a wide range of stellar properties, including photometry, distance, effective temperature, spectral type, luminosity, radius, and mass. The HPIC also contains basic information on binarity from the Washington Double Star Catalog and the Gaia Catalog of Nearby Stars (GCNS)\cite{gaia2021}, which we use to calculate a stray light background for each star in a manner identical to Ref. \citenum{stark2019}. We do not consider multi-star wavefront control techniques like that being investigated for Roman CGI\cite{sirbu2018}. The HPIC has similar fidelity to the HWO Mission Stars List developed by the NASA Exoplanet Exploration Program (ExEP) office\cite{mamajek2024}, but provides the much larger sample of stars needed for accurate trade studies\cite{tuchow2024} (the latter provides just 160 stars, while we investigate scenarios that can survey as many as $\sim$500 stars).
 
Table \ref{table:astroassumptions} lists all high level astrophysical assumptions we make. Unless otherwise stated, we adopt the same EEC definition as in the HabEx and LUVOIR final reports\cite{habexfinalreport,luvoirfinalreport}. Specifically, the $10^5$ EECs distributed around each star are placed on circular orbits within the conservative HZ spanning $0.95$--$1.67$ AU for a solar twin, as described in Refs.~\citenum{kopparapu2013} and \citenum{kopparapu2014}. These planets have wavelength-independent geometric albedos of 0.2 (we address this assumption later in Section \ref{sec:exoplanet_albedo}), have maximum radii of 1.4 $R_{\Earth}$, and minimum radii given by $0.8 (a/{\rm EEID})^{-0.5}$ $R_{\Earth}$, where $a$ is semi-major axis and EEID is the Earth-equivalent insolation distance. We adopt a baseline EEC occurrence rate $\eta_{\Earth}=0.24$, consistent with the estimated occurrence rates for FGK stars integrated over our EEC boundaries\cite{sag13_report,bryson2021}, and maintain constant $\eta_{\Earth}$ independent of spectral type (we address uncertainty in $\eta_{\Earth}$ in Section \ref{section:eta_earth}).

We adopt the same zodi and exozodi definitions as the LUVOIR and HabEx studies, but reduce the median exozodi level from 4.5 zodis to 3.0 zodis, in line with the latest results from the Large Binocular Telescope Inteferometer (LBTI) HOSTS survey\cite{ertel2020}. The LUVOIR and HabEx studies adopted a distribution of exozodi values and assigned individual values to each star. Here we assign the same 3 zodis to every star for our baseline calculations. We will show the impact of an exozodi distribution on yields in subsequent sections. The adopted exozodi model has the same color as the host star, no azimuthal dependence, no planet-induced structures, and has a flux that falls with the inverse square of the circumstellar distance, as described in Ref.~\citenum{stark2014}.

\begin{deluxetable}{ccl}
\tablewidth{0pt}
\footnotesize
\tablecaption{Baseline Astrophysical Parameters\label{table:astroassumptions}}
\tablehead{
\colhead{Parameter} & \colhead{Value} & \colhead{Description} \\
}
\startdata
$\eta_{\Earth}$ & $0.24$\tablenotemark{a} & Fraction of Sun-like stars with an exoEarth candidate \\
$R_{\rm p}$ & $[0.6\tablenotemark{b}\;,1.4]$ $R_{\Earth}$ & ExoEarth candidate radius range \\
$a$ & $[0.95,1.67]$ AU & ExoEarth candidate semi-major axis range\tablenotemark{c} \\
$e$ & $0$ & Eccentricity (circular orbits) \\
$\cos{i}$ & $[-1,1]$ & Cosine of inclination (uniform distribution) \\
$\Omega$ & $[0,2\pi)$ & Argument of pericenter (uniform distribution) \\
$M$ & $[0,2\pi)$ & Mean anomaly (uniform distribution) \\
$\Phi$ & Lambertian & Phase function \\ 
$A_G$ & $0.2$ & Geometric albedo of exoEarth candidate at 550 and 1000 nm \\
$z$ & 23 mag arcsec$^{-2}$ & Average V band surface brightness of zodiacal light\tablenotemark{d} \\
$z'$ & 22 mag arcsec$^{-2}$  & V band surface brightness of 1 zodi of exozodiacal dust\tablenotemark{e} \\
$n$ & $3.0$ & Median exozodi level \\
\enddata
\vspace{-0.1in}
\tablenotetext{a}{Corresponds roughly to $\Gamma_{\Earth}\sim0.4$ for the adopted EEC definition.}
\tablenotetext{b}{At the HZ outer edge. Minimum planet radius given by $0.8(a/{\rm EEID})^{-0.5}$ $R_{\Earth}$}
\tablenotetext{c}{For a solar twin.  The habitable zone is scaled by $\sqrt{L_{\star}/L_{\Sun}}$.}
\tablenotetext{d}{Varies with ecliptic latitude.}
\tablenotetext{e}{For Solar twin. Varies with spectral type and planet-star separation---see Appendix C in Ref.~\citenum{stark2014}.}
\end{deluxetable}

\begin{figure}[H]
\centering
\includegraphics[width=6in]{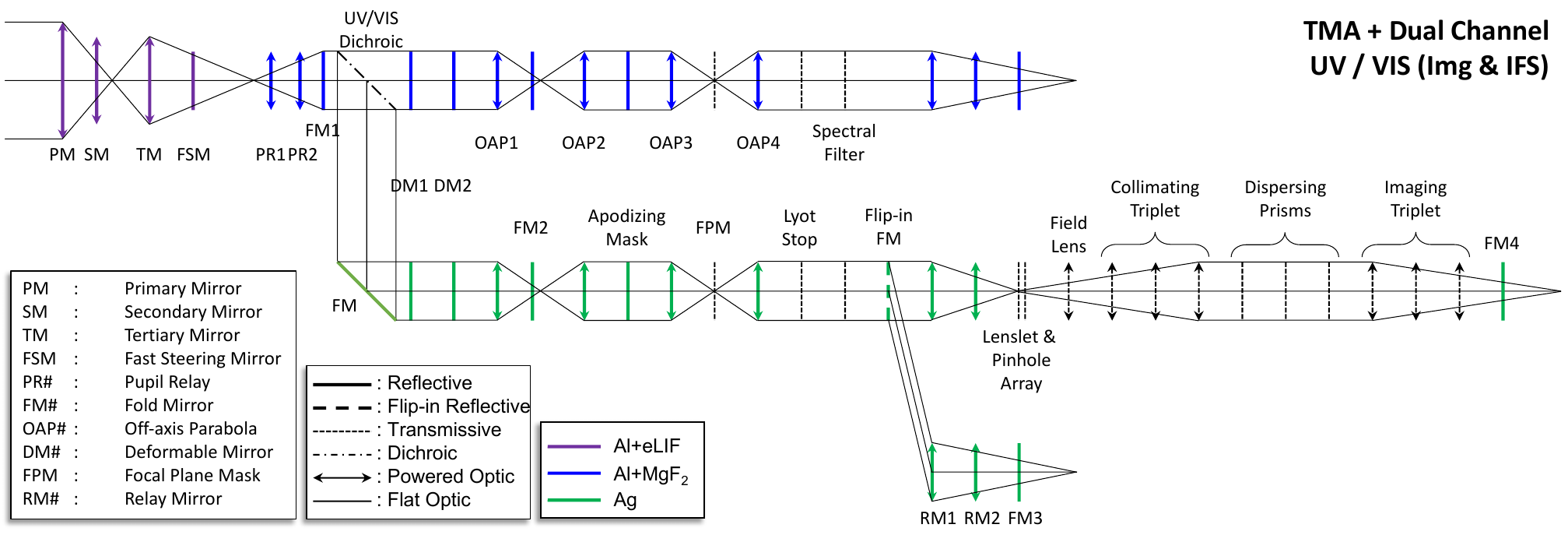}
\caption{Optical layout for our LUVOIR-B baseline mission parameters. We do not explicitly show dual parallel polarization channels for each wavelength channel, which we assume for the baseline coronagraph design.\label{fig:optical_layout-LUVOIR-B}}
\end{figure}

For our baseline mission, we start with the same assumptions as LUVOIR-B, with the ID scaled down to 6 m (from 6.7 m). LUVOIR-B adopted an off-axis segmented primary mirror with three coronagraph channels operating in the UV, VIS, and NIR. Although all three coronagraph channels were parallelized, separated by dichroics, the LUVOIR study assumed two coronagraph channels could operate in parallel at a time. Given that the NIR channel would have a larger inner working angle (IWA), the LUVOIR study chose to operate the UV and VIS channels in parallel for detection. As such, the UV channel was designed to extend to a maximum wavelength of $\sim$500 nm, where there are more stellar photons. Figure \ref{fig:optical_layout-LUVOIR-B} illustrates the end-to-end optical layout of our baseline mission. We assume dual polarization channels that can operate in parallel for both the UV and VIS wavelength channels, which we do not explicitly show in the illustration for the sake of clarity. The throughputs and reflectivities of all optics were calculated as functions of wavelength. The VIS channel was assumed to operate from 500-1000 nm to cover the water band short of 1000 nm. Because our yield analyses in this paper will not address wavelengths longer than 1000 nm, we largely ignore the NIR channel and leave discussion of it to future work.

\begin{figure}[H]
\centering
\includegraphics[width=6in]{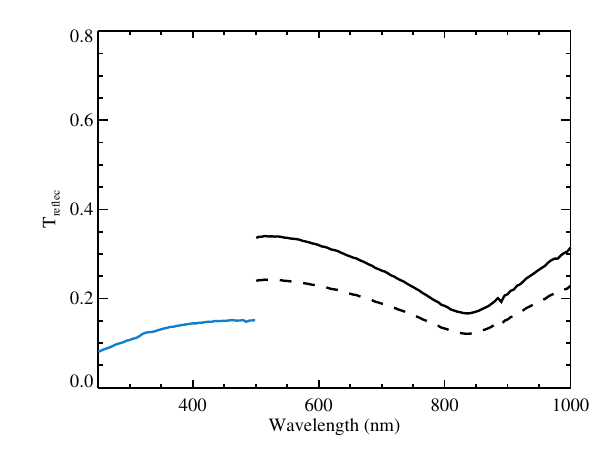}
\caption{Wavelength dependent optical throughput for our LUVOIR-B baseline mission. The optical throughput of the UV and VIS imagers are shown as solid blue and black lines, respectively, while the IFS is shown as a black dashed line. \label{fig:throughput-TMA-UVVIS}}
\end{figure}

Figure \ref{fig:throughput-TMA-UVVIS} shows the optical throughput for our baseline mission. The blue and black solid lines show the optical throughput for the UV and VIS imagers, respectively. The dashed line shows the throughput of the VIS integral field spectrograph (IFS) used for spectral characterizations. Our baseline quantum efficiency (QE) response curve is $0.9$, independent of wavelength. When calculating detection exposure times, we adopt the throughput and QE evaluated at the central wavelength of the bandpass, an approximation that is reasonable for slowly varying responses like that shown in Figure \ref{fig:throughput-TMA-UVVIS}. For spectral characterization exposure times, we adopt the throughput and QE at the long-wavelength edge of the bandpass, under the conservative expectation that characterizations will predominantly occur at wavelengths where the coronagraph's IWA plays an important role. One exception to this approach is in our treatment of the Skipper CCD in Section \ref{section:Cass_2VIS_WFSPSF_Skipper}, which exhibits a fast-varying QE response curve and warrants a bandpass-averaged approach. A more realistic handling of bandpass-varying response curves would require an understanding of how such variances affect spectral retrievals, something that has yet to be studied within the community.

\begin{figure}[H]
\centering
\includegraphics[width=6in]{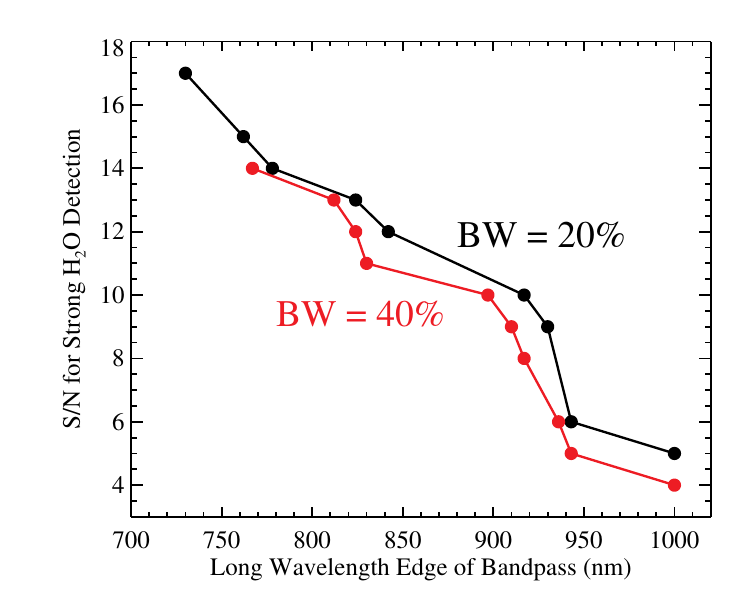}
\caption{Continuum SNR required for a strong detection of water vapor on an Earth-like exoplanet as a function of the long-wavelength edge of the bandpass for 20\% (black) and 40\% (red) bandwidth\cite{latouf2024}. A broader bandwidth covers more water lines, allowing for detection at a lower continuum SNR. We adopt 20\% bandwidth for our baseline mission and examine doubling the number of visible wavelength coronagraph channels in Section \ref{section:Cass_2VIS}.\label{fig:snr_vs_lambda}}
\end{figure}

Building off of the work described in Ref. \citenum{stark2023}, we include bandpass optimization in our analyses. AYO optimizes bandpass selection on a star-by-star basis by calculating multiple possible exposure times for different bandpasses, then choosing the option that provides the maximum value of $C/t$. For our baseline mission assumptions, we allow for spectral characterization bandpass optimization. We budget for the detection of water vapor on all detected EECs at spectral resolution $R=140$ (notably larger than the $R=70$ asummed in the LUVOIR final report and motivated by the work of Ref. \citenum{latouf2023}), allowing AYO to optimize the bandpass for each star using the $S/N$ and wavelength options shown in Fig.~8 of Ref.~\citenum{stark2023}, which we have reproduced as a black line in Figure \ref{fig:snr_vs_lambda}. Given these options, AYO typically chooses the $S/N = 5$ at 1000 nm option. 

We do not allow detection bandpass optimization for our baseline mission assumptions and require that all detections be performed at a wavelength of 500 nm. This limitation is motivated by the fact that the LUVOIR-B mission concept split the UV and VIS coronagraph channels at 500 nm to enable efficient parallel usage of both channels for exoplanet detections. Detection bandpasses must obviously stay within their respective channels and the current bandpass optimization method does not currently allow for independent limits on each coronagraph channel. Ultimately this restriction on detection bandpass will have negligible impact on the baseline mission yield, as the vast majority of targets prefer detections at 500 nm anyway\cite{stark2023}.  We note that we relax this assumption later in Section \ref{sec:change_design}, at which point we alter the coronagraph design from the baseline assumptions. For all broadband exoplanet detections, we require $S/N > 7$, a conservative estimate that provides a low mission-long probability of false positives\cite{bijan2020} under the assumption of Gaussian noise (deviation from Gaussian noise would lead to even longer exposure times).

We adopt the Deformable Mirror-assisted Vortex Coronagraph (DMVC) used in the LUVOIR-B study. This coronagraph provides a 360$^{\circ}$ dark zone and is estimated to achieve an azimuthally-averaged contrast $\sim10^{-10}$ in a single polarization, an IWA of $\sim$3.5 $\lambda$/D, a core throughput of $\sim45$\% at large separations, and a bandwidth of 20\% (further details can be found in Ref.~\citenum{stark2019}). As previously noted, we implicitly assume that the coronagraph design has parallel polarization channels. We note that the core throughput of the off-axis PSF is a smooth function of separation, such that exoplanets can be detected interior to the formal IWA. 

We make the same detector assumptions as the LUVOIR study, namely a red-enhanced electron multiplying charge capture device (EMCCD) based on future improvements to the Roman Coronagraph's EMCCD. While the QE of the Roman Coronagraph EMCCD is just a few percent at 1000 nm\footnote{\url{https://roman.ipac.caltech.edu/sims/Param_db.html}}, the LUVOIR study adopted an optimistic wavelength-independent QE of 0.9. However, we note that Ref.~\citenum{stark2023} explicitly showed that if such a QE were not possible, we could search for water at shorter wavelengths at the expense of some EEC yield. We also adopted a clock induced charge (CIC) of $1.3\times10^{-3}$ $e^-$ pix$^{-1}$ frame$^{-1}$, roughly an order of magnitude better than what has been demonstrated by the Roman Coronagraph EMCCD. Clock induced charge is a noise term that becomes apparent when operating an EMCCD in Geiger mode (also called photon-counting mode).

The EMCCD was paired with an IFS in the VIS channel to obtain exoplanet and debris disk spectra. An IFS requires additional optics, illustrated in Fig. \ref{fig:optical_layout-LUVOIR-B}, that reduce throughput and disperse the exoplanet's light over a large number of pixels, effectively amplifying the impact of the EMCCD detector noise. We carry forward the LUVOIR-B IFS assumptions, adopting a 30\% reduction in throughput due to IFS optics and 96 pixels per PSF core at 1 $\mu$m (16 lenslets at 1 $\mu$m assuming Nyquist sampling at 500 nm, 2 $\times$ 3 pixels per dispersed lenslet). Table \ref{table:missionassumptions} summarizes the high-level assumptions fed to AYO for our baseline mission parameters.

\begin{deluxetable}{ccl}
\tablewidth{0pt}
\footnotesize
\tablecaption{Coronagraph-based Mission Parameters\label{table:missionassumptions}}
\tablehead{
\colhead{Parameter} & \colhead{Value} & \colhead{Description} \\
}
\startdata
& & \bf{General Parameters} \\
$\Sigma \tau$ & $2$ yrs & Total exoplanet science time of the mission \\
$\tau_{\rm slew}$  & $1$ hr & Static overhead for slew and settling time \\
$\tau_{\rm WFC}$ & $2.7$ hrs\tablenotemark{a} & Static overhead to dig dark hole \\
$\tau'_{\rm WFC}$ & $1.1$ & Multiplicative overhead to touch up dark hole \\
$X$ & $0.7$ & Photometric aperture radius in $\lambda/D_{\rm LS}$\tablenotemark{b} \\
$\Omega$ & $\pi(X\lambda/D_{\rm LS})^2$ radians & Solid angle subtended by photometric aperture\tablenotemark{b} \\
$\zeta_{\rm floor}$ & $10^{-10}$ & Raw contrast floor \\
$\Delta$mag$_{\rm floor}$ & $26.5$ & Noise floor (faintest detectable point source at S/N$_{\rm d}$) \\
$T_{\rm contam}$ & $0.95$ & Effective throughput due to contamination \\

\hline
& & \bf{Detection Parameters} \\
$\lambda_{\rm d,1}$ & 450 nm\tablenotemark{c} & Central wavelength for detection in SW coronagraph \\
$\lambda_{\rm d,2}$ & 550 nm\tablenotemark{c} & Central wavelength for detection in LW coronagraph \\
S/N$_{\rm d}$ & $7$ & S/N required for detection (summed over both coronagraphs) \\
$T_{\rm optical,1}$ & $0.15$\tablenotemark{c} & End-to-end reflectivity/transmissivity at $\lambda_{\rm d,1}$ \\
$T_{\rm optical,2}$ & $0.34$\tablenotemark{c} & End-to-end reflectivity/transmissivity at $\lambda_{\rm d,2}$ \\
$\tau_{\rm d,limit}$ & 2 mos & Detection time limit including overheads \\

\hline
& & \bf{Characterization Parameters} \\
$\lambda_{\rm c}$ & 1000 nm\tablenotemark{c} & Wavelength for characterization in LW coronagraph IFS \\
S/N$_{\rm c}$ & $5$\tablenotemark{c} & Signal to noise per spectral bin evaluated in continuum \\
R & 140 & Spectral resolving power \\
$T_{\rm optical,IFS}$ & $0.23$\tablenotemark{c} & End-to-end reflectivity/transmissivity at $\lambda_{\rm c}$ \\
$\tau_{\rm c,limit}$ & 2 mos & Characterization time limit including overheads \\

\hline
& & \bf{Detector Parameters} \\
$n_{\rm pix,d}$ & 4\tablenotemark{c} & \# of pixels in photometric aperture of each imager at $\lambda_{\rm d,\#}$ \\
$n_{\rm pix,c}$ & 96\tablenotemark{c} & \# of pixels per spectral bin in LW coronagraph IFS at $\lambda_{\rm c}$ \\
$\xi$ & $3\times10^{-5}$ $e^-$ pix$^{-1}$ s$^{-1}$ & Dark current\\
RN & 0 $e^-$ pix$^{-1}$ read$^{-1}$ & Read noise\\
$\tau_{\rm read}$& N/A & Time between reads\\
CIC & $1.3\times10^{-3}$ $e^-$ pix$^{-1}$ frame$^{-1}$ & Clock induced charge\\
$T_{\rm QE}$ & $0.9$ & Raw QE of the detector at all wavelengths \\
$T_{\rm dQE}$ & $0.75$ & Effective throughput due to bad pixel/cosmic ray mitigation \\

\hline
\enddata
\vspace{-0.1in}
\tablenotetext{a}{See Eq.~17 from Ref.~\citenum{stark2019}}
\tablenotetext{b}{$D_{\rm LS}$ is the diameter of Lyot stop projected onto the primary mirror}
\tablenotetext{c}{Example provided at most likely bandpass; AYO optimizes bandpass and adjusts values accordingly.}

\end{deluxetable}

The HabEx and LUVOIR studies mandated six visits to every target to account for orbit determination of the EECs, assuming this results in $\sim$3 detections. Recent work has shown that two detections of a planet in reflected light may be adequate to constrain the orbit when including photometry\cite{bruna2023}. As such, we drop the six visit mandate for this study. For a coronagraph-based mission, the impact of such a mandate on yields is small\cite{stark2016}, so we expect our results to be approximately valid even when including a six visit mandate.

\section{Astrophysical sources of yield uncertainty\label{section:astrounc}}

Estimating yield uncertainties for a future HWO mission requires an understanding of how a given source of uncertainty will impact observations. Some uncertainties can be retired as the mission survey is conducted, which we dub as ``actionable." Others likely can't be measured early enough in the survey to fully react, which we dub as ``static." For example, Ref.~\citenum{stark2015} showed that as long as we can measure the exozodi background of each star after the first visit, the achievable yield approaches that of having perfect prior exozodi knowledge---exozodi is therefore an actionable source of uncertainty. However, EEC albedo is likely to be more of a static source of uncertainty---with a target sample size of 25 EECs, we will not have much of an understanding of the albedo distribution until the majority of the survey has been conducted.

We address actionable and static sources of uncertainty differently in our yield calculations. To estimate actionable sources of uncertainty, we run a large number of independent yield calculations to estimate the yield distribution. For each calculation, we draw from a distribution of values describing the source of uncertainty and pass that information to AYO. AYO then optimizes the observations based on the information provided to it and returns an EEC yield. This process effectively assumes perfect prior knowledge of the parameter, an assumption that is approximately valid for actionable sources of uncertainty\cite{stark2015}. For static sources of uncertainty, we cannot pass any information about the astrophysical property to AYO. Instead, we use AYO to optimize observations under our baseline astrophysical assumptions, and then vary the parameter after the observation plan has been ``set in stone." 

Below we inspect each prominent source of astrophysical uncertainty one at a time, model it as an actionable or static source of uncertainty, and discuss its impact on the yield. We compile these uncertainties as we go, building up to a final combined yield uncertainty.

\subsection{Exoplanet sampling uncertainty\label{section:sampling}}

The most fundamental source of uncertainty for a blind survey is due to exoplanet sampling, or the uncertainty that results from ``luck of the draw." Occurrence rates describe the mean rate of planets per star. Even if we knew the occurrence rates perfectly, we are not guaranteed to find the exact expectation value of planets in our population of target stars\cite{savransky2016}. Additionally, the chances of a planet occurring around one star versus another can affect yields. 

Exoplanet sampling uncertainty is a fundamental limit of a blind survey---the only way to fully retire it is to have perfect prior knowledge of every EEC. Precursor extreme-precision radial velocity (EPRV) surveys could identify which stars host EECs and help reduce exoplanet sampling uncertainty while improving the efficiency of HWO\cite{morgan2021a}. Simulations of EPRV surveys suggest exoEarths could be detectable around nearly $\sim100$ high priority stars for a range of future dedicated EPRV telescope architectures\cite{crass2021}. However, these simulations represent best case scenarios and would take at least a decade after EPRV instrument commissioning. While such precursor information will be useful when conducting HWO's EEC survey, allowing it to achieve faster EEC yields\cite{morgan2021a}, it will come too late to affect the early stages of HWO design when the scale of the mission and key telescope/instrument trades will ultimately dictate the range of accessible targets (addressed in Section \ref{section:paths}).

We note that a Poisson draw treatment for exoplanet sampling uncertainty is not strictly correct, as it would assume the presence of one planet in a given system does not affect the presence of another planet, which Newton would roundly reject. On the one hand, the presence of a planet should rule out nearby planets that would be gravitationally unstable, such that a Poisson draw would tend to concentrate planets around fewer stars and thus be a conservative choice. On the other hand, exoplanets may ``flock together," such that the presence of a planet implies a higher likelihood of another planet, making a Poisson draw an optimistic choice. Ideally we would distribute planets consistent with known multiplicity rates and check for orbital stability, but empirical multiplicity rates in the HZs of FGK stars are unknown. We therefore proceed with Poisson draws and note the possibility that we will underestimate the exoplanet sampling uncertainty. We note that compared to the simple numerical alternative of a Monte Carlo success-based draw, in which a maximum of one planet per star is assigned, our method is conservative.

To estimate the exoplanet sampling uncertainty for an HWO blind survey, we first ran a single AYO calculation using our baseline mission and astrophysical parameters. This resulted in an expected yield of 22.5 EECs, in agreement with Ref.~\citenum{stark2019}. As part of the calculation, we saved many properties of the $10^5$ EECs injected around each star: their orbital elements, fluxes, positions, exposure times, and critically, the visit during which they were detected. With this information in hand, we then performed a Poisson draw on each individual star using an occurrence rate equal to $\eta_{\Earth}$. We then randomly selected the appropriate number of planets from the star's population of $10^5$ EECs. Randomly selected planets with a valid visit record were counted as detected, whereas those without a valid visit record were treated as undetected. We repeated this Poisson draw 10k times, building up a distribution of yields.

\begin{figure}[H]
\centering
\includegraphics[width=6in]{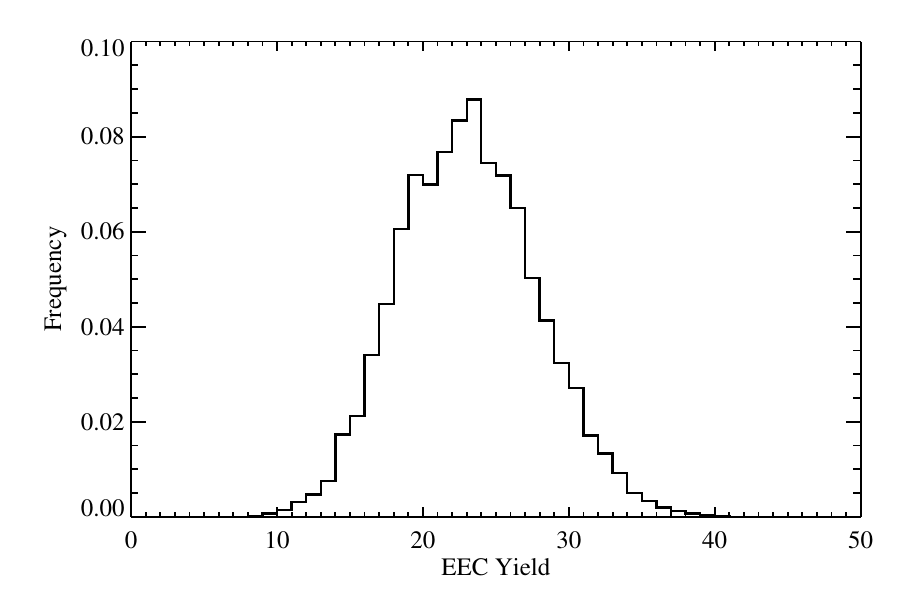}
\caption{EEC yield distribution of our baseline mission considering only ``luck of the draw" exoplanet sampling uncertainty. Without prior knowledge of which stars host EECs, there is no way to reduce this fundamental uncertainty. This distribution assumes Poisson draws ignoring exoplanet multiplicity and should be regarded as a lower limit on the amount of uncertainty.\label{fig:exoplanet_sampling_noise}}
\end{figure}

Figure \ref{fig:exoplanet_sampling_noise} shows the impact of sampling noise for our baseline mission. Given our assumption of Poisson draws that ignore multiplicity, our estimate for exoplanet sampling uncertainty should be considered a lower limit. With a mean-normalized standard deviation of $0.21$, the spread in yields is substantial. \emph{HWO will need to contend with relatively large uncertainties inherent to a blind exoplanet survey.}

\subsection{Exoplanet albedo uncertainty\label{sec:exoplanet_albedo}}

The yield distribution above assumes all EECs have a uniform geometric albedo $A_G=0.2$, equivalent to that of an Earth-twin\cite{robinson2011}. In reality, EEC albedos will vary. Unfortunately, we have no way of knowing the distribution of EEC albedos in advance. Given an expected sample size of $\sim$25 EECs, it's likely that we won't understand the distribution of albedos until well into the survey. We therefore treat albedo uncertainty as a static uncertainty. 

To constrain the effect of a static albedo uncertainty, we implement a more detailed exoplanet sampling treatment than previously described, allowing for randomly assigned albedos among the drawn EECs. To do so, we first use AYO to optimize observations under the assumption that all EECs have $A_G=0.2$. We then perform a Poisson draw on each star and randomly select the appropriate number of random EEC orbits and phases, just like in Section \ref{section:sampling}. Next, we assign each randomly drawn planet an albedo that differs from the AYO-assumed $A_G=0.2$. With this difference in albedo and the saved planet fluxes from AYO, we can determine the adjusted flux of every randomly selected EEC under the assumption of Lambertian phase functions. For each visit to the star, we advance the randomly drawn planet along its orbit based on the orbital properties saved by AYO, determine its visit-updated separation and albedo-adjusted flux, and determine if it would have been detected during the visit. 

To determine if the randomly drawn planet would have been detected during a visit, we take advantage of the fact that AYO resolves every orbit into 100 evenly spaced mean anomalies. Each AYO observation effectively detects planets along a segment of an orbit, such that any planets detected along the orbit segment occupy a finite range of fluxes. The faintest detected planet flux along the orbit segment (usually corresponding to crescent phase) is limited by the exposure time, while the brightest planet flux along the segment (usually corresponding to gibbous phase) is constrained by the IWA of the coronagraph. Therefore, we can approximately determine whether a randomly drawn EEC with \emph{differing} albedo is detectable during a given visit by requiring 1) its albedo-adjusted flux to be greater than the minimum exoplanet flux detected during that visit by AYO along the same orbit, and 2) its stellar separation to be greater than the minimum separation of any EEC on the same orbit detected by AYO during that visit. To verify that this new approximate sampling treatment produced adequate results, we first randomly drew planets, all with $A_G=0.2$, and compared results with the simple procedure described in Section \ref{section:sampling}. After 10k random draws, the mean and standard deviation of the distributions were statistically identical, validating our method.

We note one significant limitation of our method for estimating the impact of albedo uncertainties. AYO ``designs" the survey under the assumption that $A_G=0.2$. Detection times are used to determine whether a planet of differing albedo would have been detected, but characterization times are not considered. Characterization times are budgeted for by AYO under the assumption that the planets have a single $A_G=0.2$. Obviously darker planets would require longer characterization times while brighter planets would require less. However, because the observation plan is already ``set in stone" by AYO when doing the albedo draw, we cannot adjust the characterization times after the fact. To first order, we don't expect this issue to be significant driver of yields as long as the albedo distribution is roughly symmetric about $A_G=0.2$ (which will be true for our final preferred albedo distribution). We do expect this limitation to overestimate the yield of faint planets in systems that are already near the 2-month characterization limit---planets with $A_G<0.2$ in these systems could require characterization times in excess of the limit and should not count toward yield. However, planets with $A_G<0.2$ and characterization times close to the 2-month limit will represent a minority of the planets contributing to the yield (see right panel of Figure \ref{fig:albedo_unc_dist}). We leave refining this method to future work and note that we may be underestimating the yield degradation due to the exoplanet albedo distribution.

We consider two extreme scenarios to constrain the impact of exoplanet albedo: relatively dark, completely cloud-free water worlds, and relatively bright, completely cloud-covered water worlds. Both models were 100\% ocean covered. The dark extreme imagines a cloudless ocean world whose full-phase brightness would be relatively small, owing to the low reflectivity of deep ocean water seen in backscatter. At the opposite (but still habitable) extreme, a completely cloud-covered ocean is relatively bright at full phase and has a phase function that is distinct from that generated by ocean glint. These extremes bound the expected reflectance behaviors for ocean worlds which, more realistically, would present cloud-covered and cloud-free scenes across the disk. Phase-dependent reflectance models were computed using an existing 3-D tool for producing disk-integrated synthetic observations of a pixelated planetary disk\cite{tinetti2005,tinetti2006,robinson2011}. We assume Earth-like liquid water clouds and ocean wind speeds (which are a necessary input to the ocean specular reflectance model\cite{cox1954}).

The phase functions of these models are relatively close to Lambertian up to phase angles of $\sim$100$^{\circ}$. Given that the majority of detections occur near phase angles of 90$^{\circ}$ (i.e., quadrature), we choose to treat these models as Lambertian spheres and calculate albedos that reproduce the proper reflectance at quadrature. This results in $A_G=0.08$ for the cloud-free model and $A_G=0.56$ for the cloud-covered model.

\begin{figure}[H]
\centering
\includegraphics[width=6in]{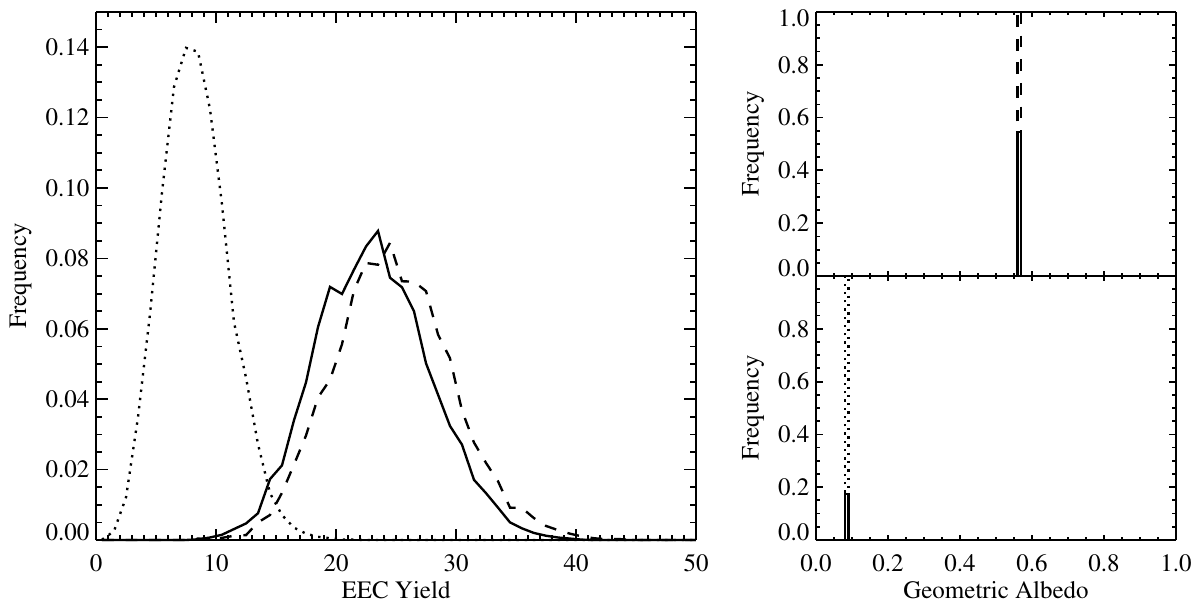}
\caption{\emph{Left:} EEC yield distribution of our baseline mission with exoplanet sampling and albedo uncertainties for the extreme scenarios in which all EECs turn out to be water-covered cloudless planets (dotted line) or cloud-covered water worlds (dashed line), compared to Earth-twins (solid line). \emph{Right:} Geometric albedo distributions of injected planets (dotted and dashed) compared to detected planets (solid). The detection rate for dark water worlds is $<$20\%, explaining the shift in the yield distribution to lower values.  \label{fig:extreme_albedo_unc}}
\end{figure}

\begin{figure}[H]
\centering
\includegraphics[width=6in]{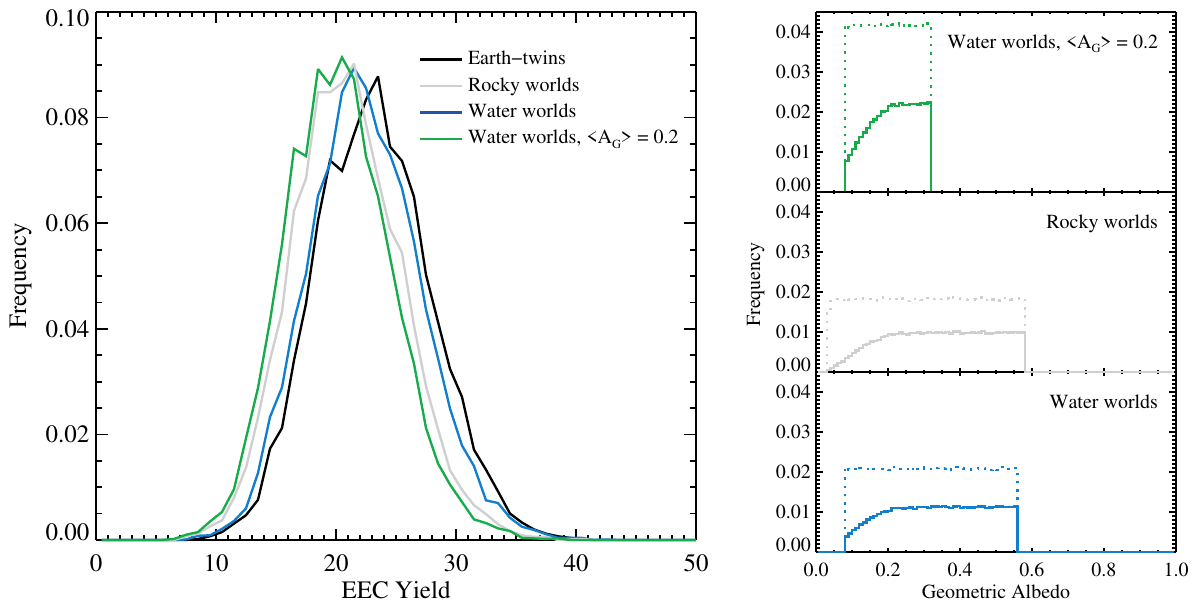}
\caption{\emph{Left:} EEC yield distribution of our baseline mission with exoplanet sampling and albedo uncertainties assuming a distribution of albedo values. Yields are shown for a uniform distribution of water worlds (blue), a uniform distribution of rocky worlds (gray), a uniform distribution of water worlds with $\langle A_G \rangle=0.2$ (green), and Earth-twins for comparison (black). \emph{Right:} Geometric albedo distributions of injected planets (dotted) compared to detected planets (solid) for each of the three albedo distributions assumed. Observations are ``tuned" to $A_G=0.2$, resulting in lower detection rates for $A_G<0.2$. We adopt the green curves as our fiducial albedo distribution.  \label{fig:albedo_unc_dist}}
\end{figure}

The left panel of Figure \ref{fig:extreme_albedo_unc} shows the yield distributions that result when observations are tuned to $A_G=0.2$. The solid line shows a benchmark: the results when 100\% of EECs are drawn with $A_G=0.2$. The dotted and dashed lines show the results when 100\% of EECs are drawn with $A_G=0.08$ and $A_G=0.56$, respectively. While the width of the distribution does decrease in the case of $A_G=0.08$, to first order, the dominant effect of changing the exoplanet albedo is that the peak of the distribution shifts. Given the limitations of our method discussed above, we note that we are likely overestimating the yield in the case $A_G=0.08$. The right panel shows the albedo distribution of injected planets (dotted and dashed lines), as well as the albedo distribution of detected planets (solid lines). These distributions are effectively Dirac functions, broadened by our choice of histogram binning. The distributions show that while the detection rate for bright cloud-covered worlds is $\sim$55\%, the detection rate for dark water worlds is $<$ 20\%. This highlights the impact of an observational bias: planets brighter than expected don't substantially help yields, but planets fainter than expected can go undetected. This sort of observational bias will creep up again in subsequent sections as we consider additional sources of astrophysical uncertainty.

The dotted and dashed lines shown in Figure \ref{fig:extreme_albedo_unc} are extremes that set rough constraints on the impact of albedo uncertainty on exoplanet yield. In reality, we expect EEC albedos to occupy a distribution of values, but we don't yet know the nature of that distribution. In spite of our ignorance, we look into the effects of adopting three different uniform distributions. First, we adopt our full range of water worlds with $0.08 < A_G < 0.56$. Second, we adopt an even broader range of $0.03 < A_G < 0.58$, roughly representing the range of reflectances one might expect near quadrature for rocky worlds as dark as Mercury and as bright as Venus. Both of these uniform distributions have $\langle A_G \rangle > 0.2$, so we adopt a third distribution with $0.08 < A_G < 0.32$, which has a mean geometric albedo equal to that of an Earth-twin. The left panel in Figure \ref{fig:albedo_unc_dist} shows the results of these uniform distributions, with the blue, gray, and green curves corresponding to our full range of water worlds, full range of rocky worlds, and narrow range of water worlds, respectively. The $A_G=0.2$ benchmark is shown in black. The blue curve coincidently mirrors that of the benchmark, while the gray and green curves are shifted to lower yield values. 

The dotted lines in the panels on the right show the distribution of injected planets as a function of albedo for each of our three assumed distributions. Solid lines show the distribution of detected planets. The detection rate of planets with $A_G > 0.2$ is relatively flat, but decreases linearly with albedo for $A_G < 0.2$. This explains the shift in the yield distribution: many planets at the faint end of the albedo distribution will go undetected. In all scenarios, a minority of the yield is comprised of planets with $A_G<0.2$, and only a fraction of those would exceed the 2-month exposure time limit, suggesting that the limitations of our albedo draw method would have a relatively small affect on the estimated yields.

We experimented with ``tuning" the observations to different exoplanet albedo. First we ran AYO with $A_G=0.15$ for all EECs (tuning observations to fainter-than-Earth-twin planets), then drew the same three distributions. EEC yields were lower in all three cases, as AYO devoted more time to searching for fainter planets and ended up observing fewer stars during the two year time budget. We note that the limitations of our albedo draw method should have an even smaller impact in this case, as our observations are tuned closer to the faint edge of the albedo distribution. Next we ran AYO with $A_G=0.25$ for all EECS (tuning observations to brighter-than-Earth-twin planets). EEC yields were slightly larger for the black, blue, and gray curves, as AYO opted to ``pick off" the brighter portions of the distribution, while the green curve with $\langle A_G \rangle=0.2$ remained approximately the same. However, we have less confidence in these result as the limitations of our albedo draw method should become more pronounced as observations are tuned to brighter planets, which should tend to overestimate yields to a greater degree. It's possible that if we are willing to assume that there are EECs with albedo greater than that of Earth, we could gain some yield at the expense of finding fewer planets fainter than the Earth. However, this assumption seems both poorly founded and risky. We conclude that observation optimization cannot significantly improve the yields of planets fainter than the Earth---we would need to improve the mission performance parameters to accomplish this. 

None of the yield distributions shown in Figure \ref{fig:albedo_unc_dist} are correct. Completely cloud free water worlds are probably unlikely, as are completely cloud-covered water worlds. This suggests that our uniform albedo distributions are pessimistic. However, some EECs may in fact be as dark as Mercury, or as bright as Venus. The actual albedo distribution may even be multi-modal. Given a need to budget for albedo uncertainty at some level, we choose to move forward with the most pessimistic uniform albedo distribution ($0.08 < A_G < 0.32$), which produces the green yield distribution shown in Figure \ref{fig:albedo_unc_dist}. The green yield distribution in Figure \ref{fig:albedo_unc_dist} has a mean value of 19.8 EECS; budgeting for albedo uncertainty decreases the expected yields by $\sim$12\%. Notably the standard deviation of this green distribution is 22\% of the mean; the albedo uncertainty did not significantly increase the fractional width of the yield distribution, i.e., exoplanet sampling dominates the uncertainty in yield. We note that the general resilience of EEC detections against albedo uncertainties does not imply that the characterization time for the EECs is also resilient against uncertainties in target spectra. Characterization time\,---\,in terms of, e.g., exposure time required to detect key atmospheric species\,---\,will depend strongly on the details of atmospheric composition, cloud distributions, and surface reflectivity.

\subsection{Exozodi sampling uncertainty\label{section:exozodi_sampling}}

So far we have combined exoplanet sampling uncertainty with albedo uncertainty, but we have assumed all stars are assigned the same amount of exozodiacal dust. In reality each star will have a different brightness of exozodiacal dust. While we may know some individual exozodi levels in advance of the HWO survey, we will not know all of them. However, we can learn the rest of the individual exozodi levels ``on the fly" and adapt to them as the survey progresses. Ref. \citenum{stark2015} showed that the EEC yield when adapting to exozodi levels after the first observation is nearly equal to the yield if they were all known in advance. Real-time adaptation to exozodi levels should lead to even higher yields. Exozodi sampling uncertainty is therefore an actionable uncertainty.

To estimate the impact of exozodi sampling uncertainty on yield distributions, we adopt the best fit exozodi distribution from the LBTI HOSTS survey, which has a median exozodi level of three zodis and is multi-modal, with several peaks at higher zodi levels\cite{ertel2020}. From this distribution, we randomly draw exozodi levels, assign them to individual stars, provide that information to AYO, and then calculate an optimized yield. We repeat this process 500 times. We include the exoplanet sampling and albedo uncertainties previously discussed by performing 1000 draws of random EECs with $0.08 < A_G < 0.32$ for each of the 500 exozodi draws. The LBTI HOSTS survey detected dust around four potential HWO targets: $297\pm56$ zodis around Eps Eri, $148\pm28$ zodis around Tet Boo, $588\pm121$ zodis around 72 Her, and $235\pm45$ zodis around 110 Her; for yield calculations, we assigned these stars their LBTI-measured nominal exozodi levels\cite{ertel2020}.

\begin{figure}[H]
\centering
\includegraphics[width=6in]{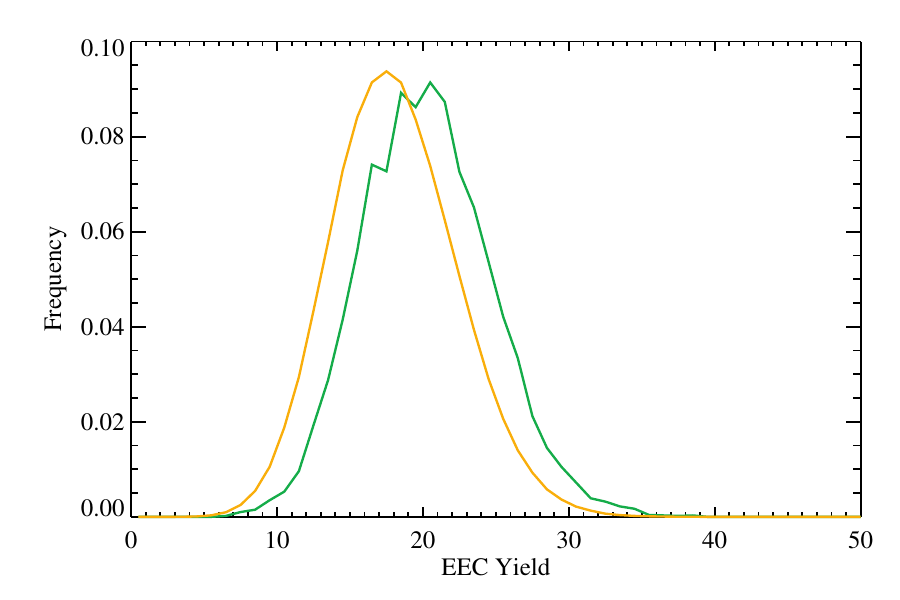}
\caption{EEC yield distribution of our baseline mission with exoplanet sampling, albedo, and exozodi sampling uncertainties (orange), compared to the green yield distribution calculated in Section \ref{sec:exoplanet_albedo}. Drawing exozodi values from a distribution (as opposed to assigning all stars the same median value) shifts the yield distribution to lower values, as some high priority targets are assigned higher exozodi values. \label{fig:exozodi_sampling}}
\end{figure}

Figure \ref{fig:exozodi_sampling} shows the new EEC yield distribution including exozodi sampling uncertainty in orange, along with the previous green distribution from Figure \ref{fig:albedo_unc_dist}. The yield distribution again maintains roughly the same width but shifts to the left, with a mean of 17.6 EECs. This is due to another observational bias analogous to the albedo distribution discussed in Section \ref{sec:exoplanet_albedo}. If a high-priority target is assigned a zodi value less than three zodis, there is little yield to be gained, as detection exposure times were already short. However, randomly assigning a larger exozodi value to a high-priority star can significantly extend exposure times. Even if the yield code replaces these high-zodi stars with other low-zodi stars, the limited pool of targets means the code must replace a previously productive target with a lower-productivity target, driving the yield distribution systematically to lower values. Roughly one third of this shift can be explained by the four specific stars discussed above being assigned high zodi values---these otherwise high-priority stars are effectively scrubbed from the target list, reducing the expected EEC yield by one. 

We note that the mean-normalized standard deviation of the orange curve in Figure \ref{fig:exozodi_sampling} is $0.24$, not too dissimilar from the $0.22$ mean-normalized standard deviation of the green curve. This shows that exozodi sampling uncertainty is a minor contribution to the total uncertainty budget, which at this point remains dominated by exoplanet sampling uncertainty.

\subsection{Exozodi distribution uncertainty\label{section:exozodi_distribution}}

Not only is the individual exozodi level of each star unknown, but our understanding of the exozodi level \emph{distribution} is uncertain. While the LBTI HOSTS survey fit the data to derive a single maximum likelihood distribution\cite{ertel2020}, many other distributions are also consistent with the data, albeit at lower likelihoods. Here we add exozodi distribution uncertainty to the planet sampling, albedo, and exozodi sampling uncertainties already estimated.

To estimate the impact of exozodi distribution uncertainty, we must first form a set of possible exozodi distributions and calculate their likelihoods. We start by considering the maximum likelihood methods used in the LBTI HOSTS analysis, which do not formally rely on Bayesian priors (discussed later). To do this, we generate a series of 300k exozodi values from the maximum likelihood LBTI HOSTS exozodi distribution\cite{ertel2020} following the iterative approach described in Section 4.6.3 of Ref. \citenum{mennesson2014}. We then generate 30k ``perturbed" distributions that differ from the best fit distributions. We note that in practice, the method for perturbing the maximum likelihood distribution is not well defined, but we have examined multiple methods, all producing similar results. We then calculate the likelihood $L$ of having observed the data from each of those distributions using Equations 15 \& 16 from Ref. \citenum{mennesson2014}, and compare with the maximum likelihood $L_{\rm max}$. For a given likelihood ratio, there are many possible perturbed distributions, and the perturbed distributions we generated do not follow a normal distribution. We therefore enforce $X= \left(-2\ln{\left(L/L_{\rm max}\right)}\right)^{1/2}$ to follow a unit normal distribution by defining 31 bins ranging from $-3 \le X \le 3$ and randomly draw the correct number of distributions with the proper likelihood value. The end result is a ``set" of $\sim$20k distributions that follow the LBTI HOSTS maximum likelihood approach, with the statistics for the set following a normal distribution based on likelihood.

\begin{figure}[H]
\centering
\includegraphics[width=6in]{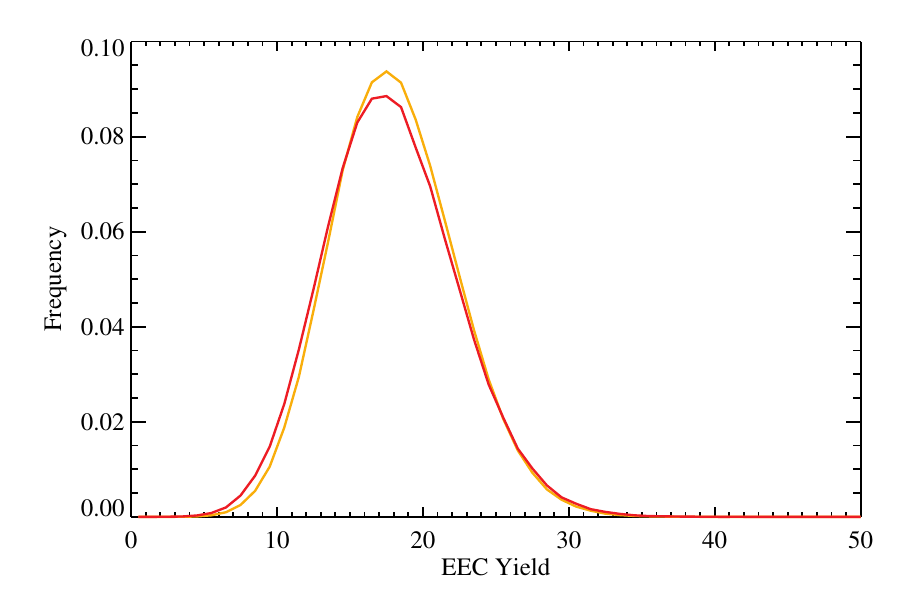}
\caption{EEC yield distribution of our baseline mission with exoplanet sampling, albedo, exozodi sampling, and exozodi distribution uncertainties (red), compared to the orange yield distribution calculated in Section \ref{section:exozodi_sampling}. Drawing exozodi values from all distributions consistent with the LBTI HOSTS data negligibly impacts yield uncertainty. \label{fig:exozodi_distribution}}
\end{figure}

With these distributions in hand, we perform 498 distinct yield calculations. For each yield calculation, we draw a random exozodi distribution and then assign each star a random exozodi level drawn from that distribution. We continue to include exoplanet albedo and exoplanet sampling uncertainties following the methods previous described. Figure \ref{fig:exozodi_distribution} shows the results of including the uncertainty in the exozodi distribution as a red line, compared to our previous yield distribution without it in orange. While the red curve with exozodi distribution uncertainty is slightly shifted to lower values and slightly broader (mean-normalized standard deviation of $0.26$ compared to $0.24$), there is remarkably little difference in the two yield distributions. This suggests that following the LBTI HOSTS formalism and assuming the LBTI HOSTS results are not systematically biased, any remaining uncertainty in the exozodiacal dust distribution has a smaller impact than the inherent exoplanet sampling uncertainty. 

While it has been previously shown that EEC yield is a weak function of median exozodi level\cite{stark2014}, this is the first estimate suggesting that the uncertainty in the distribution has a negligible impact on an HWO blind survey. As such, we briefly consider alternative approaches to deriving a set of exozodi distributions consistent with the LBTI data set. To do so, we adopt a Bayesian approach to fitting the LBTI data set. We consider two possible functional forms: a log-normal distribution and a non-parametric 40$^{\rm th}$ degree Bernstein polynomial basis distribution. For the non-parametric approach, we adopt two different priors described by a Dirichlet distribution with parameter $\alpha = [0.1, 0.25]$, where $\alpha=0.1$ weights the result toward a smooth fit and $\alpha=0.25$ allows for a higher degree of modality in the distribution (i.e., multi-peaked). 

\begin{figure}[H]
\centering
\includegraphics[width=6.5in]{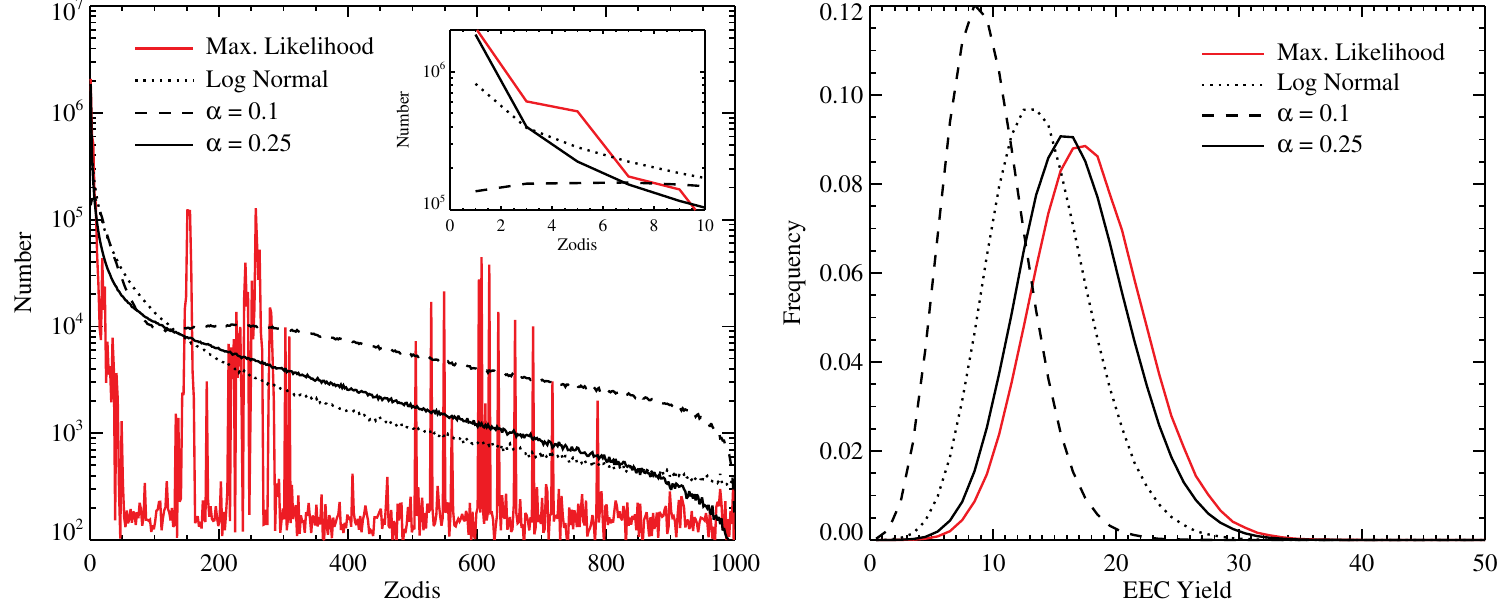}
\caption{\emph{Left:} 10k randomly drawn exozodi levels from 500 randomly drawn distributions when fitting the LBTI HOSTS data with maximum likelihood (solid red), log-normal (dotted black), non-parametric with $\alpha=0.1$ (dashed black), and non-parametric with $\alpha=0.25$ (solid black) approaches.  \emph{Right:} Yield distributions including exozodi distribution uncertainty for each of these approaches; the red curve is identical to the red curve in Figure \ref{fig:exozodi_distribution}. Adopting a multi-modal non-parametric fit ($\alpha=0.25$) produces similar yields to the LBTI HOSTS maximum likelihood approach\cite{ertel2020} because the fraction of distributions $\lesssim 3$ zodis is similar. \label{fig:exozodi_distribution_comparison}}
\end{figure}

The left panel of Figure \ref{fig:exozodi_distribution_comparison} shows the result of 10k exozodi draws from 500 randomly drawn distributions for all four of our approaches. The red curve shows the LBTI HOSTS maximum likelihood approach, while the black curves show the log-normal and non-parametric approaches. The right panel of Figure \ref{fig:exozodi_distribution_comparison} shows the yield distribution for each of these approaches, calculated in a similar fashion as previously described. The log-normal approach, which does not accommodate multi-modality, and the $\alpha=0.1$ prior approach, which allows only for modest multi-modality, predict higher exozodi levels on average and significantly shift the yield curve to lower numbers. The $\alpha=0.25$ prior approach, which does accommodate some degree of multi-modality, produces results that are very similar to the LBTI HOSTS maximum likelihood approach. We note that this is in spite of the median exozodi level for the $\alpha=0.25$ prior approach being twice that of the maximum likelihood approach---the reason for this is that while the approaches have different medians, both have similar fractions of the distribution $\lesssim 3$ zodis, as shown by the inset panel in Figure \ref{fig:exozodi_distribution_comparison}.

We caution that the choice of priors appears to significantly affect the implied exozodi distribution and ultimately the yield distribution, and the yield distribution is most sensitive to the fraction of the exozodi distribution at very low exozodi levels. Determining which approach is best is difficult without more/better data and is therefore beyond the scope of this paper. However, given that the LBTI data set appears to clearly be multi-modal\cite{ertel2020} and that the $\alpha=0.25$ and maximum likelihood approaches largely agree, we proceed with the LBTI HOSTS maximum likelihood approach to estimate exozodi distribution uncertainty, i.e. the solid red curve shown in Figure \ref{fig:exozodi_distribution}. 

We note that all of the yield distribution curves shown in Figures \ref{fig:exozodi_distribution} and \ref{fig:exozodi_distribution_comparison} come with several major caveats. For example, our calculations explicitly assume that we can subtract exozodi to the Poisson noise limit without impacting the planet's signal. While this has been shown to be true for inclined, smooth exozodi less dense than a few tens of zodis\cite{kammerer2022}, it may be difficult for smooth edge-on disks\cite{kammerer2022} as well as edge-on disks with structure\cite{defrere2012,currie2023}. Additionally, the presence of hot dust near the star\cite{absil2013,ertel2014} or cold pseudo-zodi at small projected distances in edge-on disks\cite{pseudozodi} may cause contrast degradation and make PSF subtraction difficult. We leave these issues for future investigations.

Of all the sources of astrophysical uncertainty we have considered thus far, the exoplanet sampling uncertainty inherent to a blind survey appears to be the dominant term. Our estimate of the impact of this source of uncertainty, which should be regarded as a lower limit, resulted in a mean-normalized standard deviation of $0.21$, whereas all other terms combined only increased the mean-normalized standard deviation to $0.26$. Assuming uncertainties add in quadrature, this suggests all other terms combined result in a mean-normalized standard deviation of $0.15$. The dominant effect of these other sources of uncertainty has been to reduce the expectation value of the yield by nearly 25\%, from 22.5 EECs to 17.3 EECs due to observational biases. Assuming we do not find the majority of EECS or measure the majority of target star's exozodi levels in advance of the HWO mission, most of these uncertainties cannot be mitigated prior to the mission and thus must be budgeted for with yield margin. Next we consider the one remaining source of astrophysical uncertainty, which will dominate over all other terms, but also has the potential to be somewhat mitigated prior to launch.

\subsection{Uncertainty in $\eta_{\Earth}$\label{section:eta_earth}}

The final source of astrophysical uncertainty we consider is the occurrence rate of EECs, $\eta_{\Earth}$. While $\eta_{\Earth}$ uncertainty may seem like a static source of uncertainty, for a fixed mission lifetime we must budget spectral characterization time for each detected EEC. E.g., in the event we detect more planets than expected, we must devote time to characterize them, reducing the time we can spend searching for EECs. We therefore treat $\eta_{\Earth}$ as an actionable source of uncertainty.

To incorporate the uncertainty in $\eta_{\Earth}$, which we refer to as $\sigma_{\eta_{\Earth}}$, we use the $\eta_{\Earth}$ values from Ref.~\citenum{bryson2021}. Ref.~\citenum{bryson2021} calculated occurrence rates using the Kepler DR25 exoplanet data catalog\cite{thompson2018} supplemented by Gaia-based stellar and exoplanet properties\cite{berger2020a, berger2020b} and corrected for catalog completeness and reliability.  Ref.~\citenum{bryson2021} used a power-law population model with a rate that depended on exoplanet radius, exoplanet insolation flux, and host star effective temperature, with the power law parameters inferred using a Poisson likelihood. To compute $\eta_{\Earth}$, we integrate the power law model using the posterior power law parameter values from their analysis of planets in the conservative habitable zone of quiet, isolated FGK main sequence dwarfs. Our domain of integration, defining our rocky habitable zone population, is:
\begin{itemize}
  \item The exoplanet radius range $0.8 R_{\Earth} \sqrt{{\mathrm EEID}/a} < R < 1.4 R_{\Earth} $, where $R$ is radius, $a$ is the semi-major axis, and EEID is the Earth-Equivalent Insolation Distance (the distance at which the planet would have the same insolation as Earth), chosen to be consistent with the radius range adopted for yield calculations herein.
  \item The exoplanet insolation flux range defined for each stellar effective temperature by the conservative habitable zone \cite{kopparapu2013}.
  \item The stellar effective temperature range $3900 \mathrm{K} < T_\mathrm{eff} < 7300 \mathrm{K}$.
\end{itemize}

Ref.~\citenum{bryson2021} accounted for the lack of information about DR25 catalog completeness beyond 500-day orbital periods by computing power law posteriors for two bounding cases: case 1 assumed completeness was zero beyond 500 days, and case 2 assumed the completeness beyond 500 days was equal to the completeness at 500 days.  We computed the $\eta_{\Earth}$ distribution for each of these cases, and created our final $\eta_{\Earth}$ distribution by uniformly randomly drawing from both cases. 

Performing this computation for each element of our posterior, we get an $\eta_{\Earth}$ with mean and 86\% confidence interval $\eta_{\Earth} = 0.26^{+0.29}_{-0.14}$.  The large uncertainties are primarily due to the very small number of detections of habitable zone planets orbiting FGK stars whose planet radius is within our desired range.

\begin{figure}[H]
\centering
\includegraphics[width=6in]{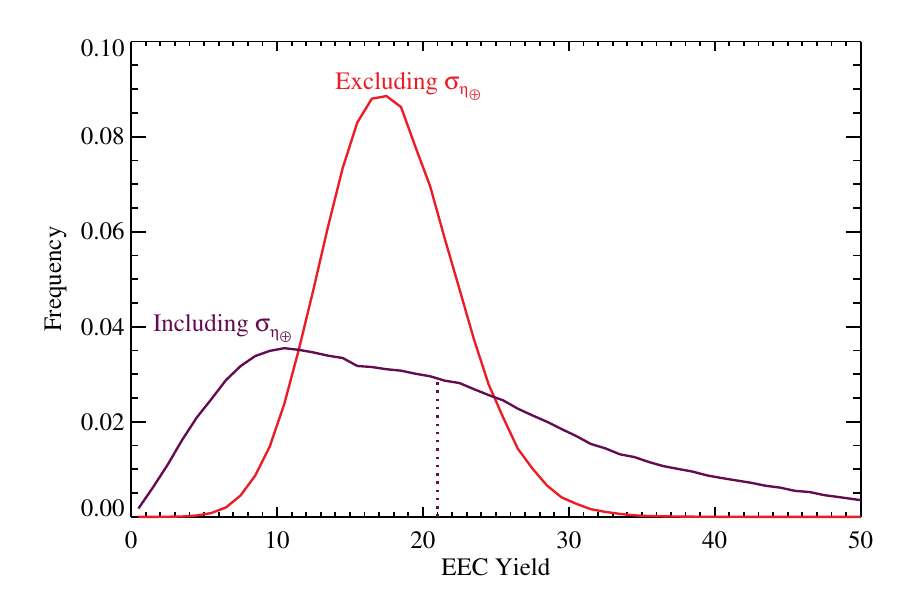}
\caption{EEC yield distribution of our baseline mission with all known sources of astrophysical uncertainty: exoplanet sampling, albedo, exozodi sampling, exozodi distribution, and $\eta_{\Earth}$ uncertainties (purple). The red yield distribution is the same as was calculated in Section \ref{section:exozodi_distribution}, which excludes uncertainty in $\eta_{\Earth}$. The dotted purple line marks the mean of the purple distribution. Uncertainties in $\eta_{\Earth}$ substantially broaden the EEC yield distribution. \label{fig:eta_earth_uncertainty}}
\end{figure}

To include $\sigma_{\eta_{\Earth}}$ in our yield calculations, we repeat the 498 calculations performed in Section \ref{section:exozodi_distribution}, but draw a unique value of $\eta_{\Earth}$ for each one using the methods described above. Figure \ref{fig:eta_earth_uncertainty} shows the results in purple compared to our previous results from Section \ref{fig:eta_earth_uncertainty} shown in red. While the median of the purple distribution is similar to that of the red distribution, the mean (shown as a vertical purple dotted line) is substantially higher due to a tail of large $\eta_{\Earth}$ values; including uncertainty in $\eta_{\Earth}$ leads to higher mean yields. Our uncertainty in $\eta_{\Earth}$ has a major impact on the breadth of the expected yield distribution. In short, even for missions with an expectation value close to two dozen, the uncertainties in $\eta_{\Earth}$ are large enough to produce non-negligible chances of single digit EEC yields.

We note that the mean of the red distribution shown in Figure \ref{fig:eta_earth_uncertainty} is $\sim$20\% lower than the yield predicted for a 6 m ID telescope by Ref.~\citenum{stark2019}. Although we made several changes to assumed inputs, notably $R=140$ instead of $R=70$ to detect H$_2$O and an updated LBTI best-fit exozodi distribution, these changes mostly offset. The majority of the decrease in expected yields is due to the geometric albedo distribution (which decreases yield by $\sim$15\%) and exozodi distribution uncertainty (which decreases yield by $\sim$4\%), which were not included in Ref.~\citenum{stark2019}.

In summary, assuming plausible distributions in exoplanet albedo, uncertainty in the HWO EEC yield appears to be dominated by two sources of astrophysical uncertainty. The first is simply exoplanet sampling, which is inherent to a blind survey and may only be partially overcome by precursor detection of the EECs. To be maximally useful, this must happen prior to design of the mission. The second and most significant source of astrophysical uncertainty is $\eta_{\Earth}$. We note that a goal of 25 EECs ignoring uncertainties in $\eta_{\Earth}$ is therefore not equivalent to a goal of 100 cumulative HZs. A goal of 100 cumulative HZs implicitly ignores \emph{both} dominant sources of uncertainty, $\sigma_{\eta_{\Earth}}$ and exoplanet sampling uncertainties. 

The Astro2020 Decadal Survey asserted that a sample size of 25 EECs ``provides robustness against the uncertainties in the occurrence rate of Earth-sized worlds and against the vagaries associated with the particular systems near Earth."\cite{astro2020} Here, robustness can be defined as the probability of achieving a given yield goal. With some minimum yield goal defined, we can use the distributions shown in Figure \ref{fig:eta_earth_uncertainty} to calculate this probability. As the Astro2020 Decadal Report did not define the minimum acceptable yield goal, the meaning of spectral characterization, nor the vagaries of particular systems, we must define them here. We therefore adopt two minimum yield goals throughout the rest of this study: the detection of 25 EECs and subsequent search for water vapor including all sources of astrophysical uncertainty, and the detection of 25 EECs and subsequent search for water vapor including all sources of astrophysical uncertainty except $\eta_{\Earth}$ uncertainties. These two goals correspond to the two yield distribution curves shown in Figure \ref{fig:eta_earth_uncertainty}.

We define the probability of detecting and searching 25 EECs for water vapor, $P_{25}$, as the fraction of the yield distribution that exceeds 25 EECs. We calculate this quantity for each of the distributions shown in Figure \ref{fig:eta_earth_uncertainty}. For our baseline mission parameters with a 6 m inscribed diameter, we find $P_{25}$ is just 6\% when ignoring $\eta_{\Earth}$ uncertainties and 32\% when including $\eta_{\Earth}$ uncertainties. In the following section we investigate modifications to the LUVOIR-B design that could improve scientific performance.

\section{Paths to budget for yield uncertainty\label{section:paths}}

Here we explore paths to shift the final two distributions shown in Figure \ref{fig:eta_earth_uncertainty} to larger values and increase the confidence in achieving the goal of 25 EECs. We will explore multiple possible improvements to our baseline mission assumptions. For each, we will describe the impact of the change on the fundamental mission parameters, EEC yield, and data quality. 

\begin{figure}[H]
\centering
\includegraphics[width=6in]{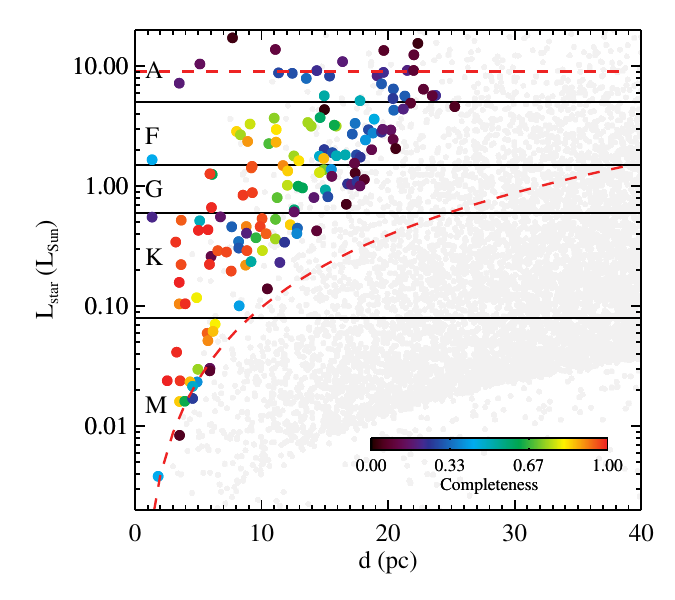}
\caption{HZ completeness of selected targets for one representative run of the LUVOIR-B baseline scenario. Star-to-star variation in completeness is due to random assignment of exozodi levels. Red dashed lines roughly mark the boundaries of the observable targets, while gray dots indicate the range of the input target list. \label{fig:target_plots_baseline}}
\end{figure}

To understand how to improve yields, we must first understand the astrophysical performance of our baseline mission. Figure \ref{fig:target_plots_baseline} shows the targets selected for observation by AYO for a single representative simulation of our baseline LUVOIR-B scenario. Targets are color coded by HZ completeness, while the full input target list is shown in gray. Black horizontal lines mark the luminosity boundaries for different stellar types. The red dashed lines roughly mark the boundaries of accessible targets. EEC contrast becomes more challenging for early type stars, reducing HZ completeness for targets at higher luminosity. The upper horizontal line marks where a 1.4 $R_{\Earth}$ planet at the EEID has a contrast equal to the systematic noise floor, a rough visual guide marking a limit imposed on target accessibility by the noise floor.

The curved red line indicates the luminosity at which the outer edge of the HZ is located at 1.5 $\lambda/D$ when $\lambda$=1000 nm. Because the angular scale of the HZ decreases with distance and for late type stars observing these targets requires operation at smaller working angles where coronagraph throughput is lower and contrast is degraded. The lower curved dashed line therefore represents another visual guide marking limits imposed by operating at small working angles. Of course operating near 1.5 $\lambda/D$ means the exoplanet is only marginally resolved and would likely blend together with other planets in the scene\cite{saxena2022}. We expect that spatial resolution limitations will ultimately place a strict limit on working angle that we do not enforce here. Yield estimates would benefit from future studies firmly establishing this working angle limit.

Based on the red dashed boundaries and the selected targets shown in Figure \ref{fig:target_plots_baseline}, it is clear that the baseline mission has many accessible targets that go unobserved, or are under-observed. If additional mission time were available, or exposure times were shortened, the additional observations would increase completeness toward the upper right corner of the plot. Throughout this section we will therefore highlight the importance of achieving shorter exposure times. 

In addition to improved yields, there are other important motivators for decreasing exposure times. The AYO calculations here adopt the same exposure time limit as the HabEx and LUVOIR studies: 2 months. While exposure times are budgeted for properly in the yield code, exposure times this long are problematic. First, exposure times lasting several weeks can make additional spectral characterization, beyond just the H$_2$O detection that we include, unlikely, and a single 20\% bandpass to search for water vapor may not constitute adequate ``characterization" of EECs for HWO. Further, exposure times on the order of a month will be complicated by the motion of the planet, which may disappear behind the IWA or move into crescent phase. Finally, long exposure times present real-world scheduling constraints that may be difficult to overcome and optimize. The shorter we can make exposure times, the easier the observations become and the more spectral information we can obtain on each EEC. 

We note that extending the mission lifetime is fundamentally different from shortening exposure times. The former provides more time to observe any target that currently meets the 2-month limit---this would increase the yield, but would not extend the range of accessible targets. The latter changes the exposure time of every star, extending the range of accessible targets such that more targets are compliant with the 2-month limit. The targets selected for observation (colored dots) in the upper right corner of Figure \ref{fig:target_plots_baseline} have exposure times approaching the 2-month limit. An extended mission lifetime would increase the completeness of targets already selected for observation (colored dots), while a reduction in exposure times would allow more of the unobserved targets (gray dots to the upper-right) to be observed.

\subsection{Build a bigger telescope\label{sec:bigger_telescope}}

\begin{figure}[H]
\centering
\includegraphics[width=6in]{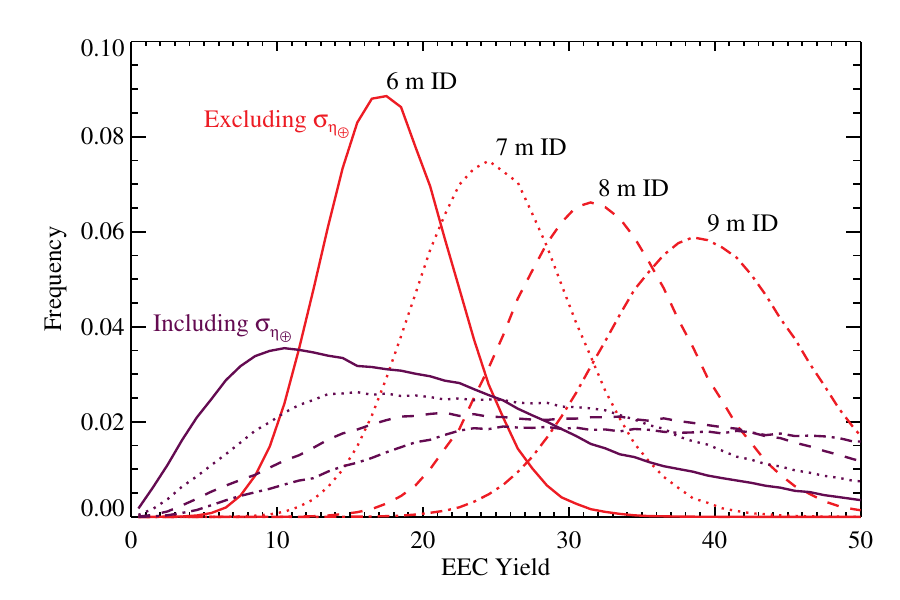}
\caption{EEC yield distributions for four different telescope diameters when adopting our baseline mission parameters including $\eta_{\Earth}$ uncertainty (purple) and excluding it (red). Solid, dotted, dashed, and dot-dashed lines correspond to IDs of 6, 7, 8, and 9 m, respectively.  \label{fig:yield_dist_vs_D}}
\end{figure}

As shown by Ref.~\citenum{stark2014}, EEC yield is most sensitive to telescope diameter. Therefore, we study how the distributions shown in Figure \ref{fig:eta_earth_uncertainty} vary with telescope diameter. To do so, we repeat the calculations from Sections \ref{section:exozodi_distribution} and \ref{section:eta_earth} for inscribed diameters, IDs, ranging from 6 m to 9 m. Figure \ref{fig:yield_dist_vs_D} shows the resulting EEC yield distributions, with solid, dashed, dotted, and dash-dotted lines corresponding to 6, 7, 8 and 9 m IDs, respectively. Purple distributions correspond to those with all known astrophysical uncertainties included and red lines correspond to those without $\sigma_{\eta_{\Earth}}$ included. Excluding $\sigma_{\eta_{\Earth}}$, a 9 m ID telescope increases yield by a factor of 2.2, in agreement with power law relationships established by previous works\cite{stark2014,stark2015}.

\begin{figure}[H]
\centering
\includegraphics[width=6in]{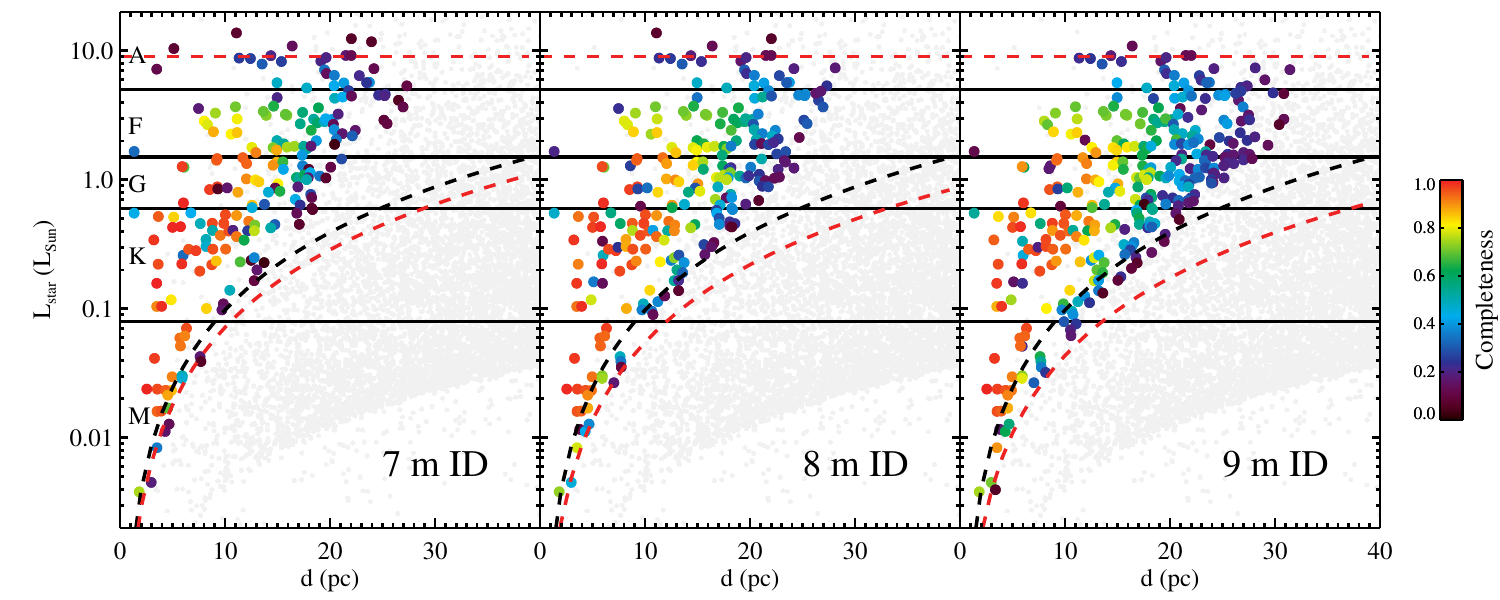}
\caption{Targets selected for observation for one simulation of the 7, 8, and 9 m ID telescope scenarios, color-coded by total completeness, assuming the same single representative exozodi draw as in Fig.~\ref{fig:target_plots_baseline}. The red dashed lines indicate the boundaries of the accessible targets---the horizontal line roughly marks the assumed astrophysical noise floor while the curved line roughly indicates the working angle limit. The black dashed line indicates the working angle limit of a 6 m ID telescope for reference. Larger telescopes expand the range of accessible targets.\label{fig:target_plots_vs_D}}
\end{figure}

Figure \ref{fig:target_plots_vs_D} shows the targets selected for observation as a function of telescope diameter. These plots assume the same single representative exozodi draw as in Fig.~\ref{fig:target_plots_baseline}. The black dashed line roughly marks the working angle limit of a 6 m ID telescope. As the telescope diameter increases, the working angle limit shifts downward, as shown by the curved red line. Larger telescope diameters allow targets with smaller angular HZs to be selected, whether they are at larger distances or lower luminosities. Notably, none of the scenarios in Fig.~\ref{fig:target_plots_vs_D} ``use up" all of the potentially accessible targets. Regardless of aperture size, the LUVOIR-B baseline parameters therefore lead to missions that are limited by exposure times. Only by by reducing exposure times can a mission take advantage of the full extent of the target list, a concept we explore in Section \ref{sec:change_design}.

\begin{figure}[H]
\centering
\includegraphics[width=6in]{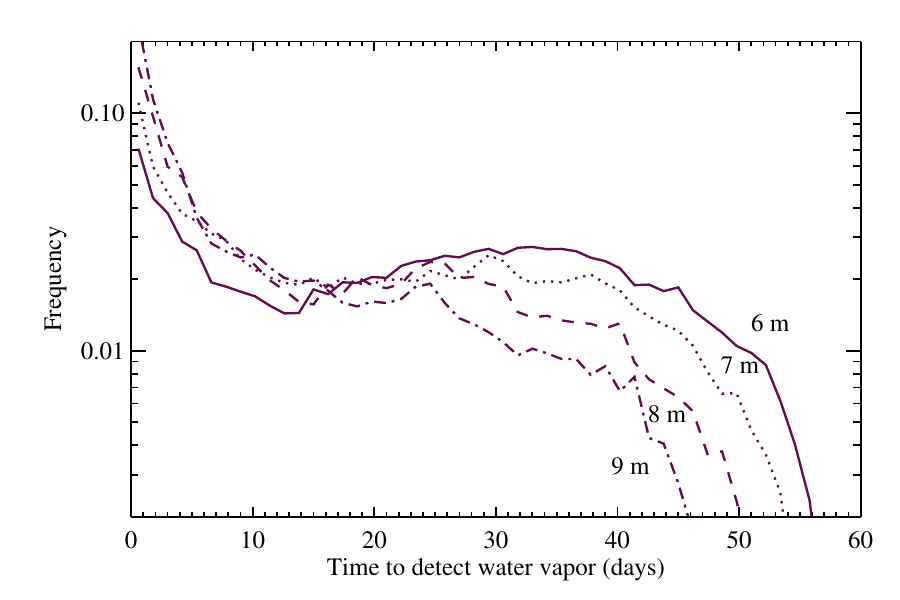}
\caption{Possible spectral characterization time distributions of the first 18 EECS for our baseline mission parameters with a 6, 7, 8, and 9 m ID telescope shown in solid, dotted, dashed, and dash-dotted lines, respectively. A 9 m ID telescope reduces spectral characterization times by a factor of 6.\label{fig:tspecdist_vs_D}}
\end{figure}

Figure \ref{fig:tspecdist_vs_D} shows the distribution of possible spectral characterization times to detect water vapor absorption for all 498 yield calculations performed. We only show the scenario in which $\eta_{\Earth}$ uncertainty is excluded. Of course, as we shorten exposure times by going to larger telescopes, more observations can be included which will necessarily be around more challenging targets and extend the distributions, making it difficult to see the exposure time impacts for the highest priority stars. Therefore, to make the distributions shown in Figure \ref{fig:tspecdist_vs_D}, we consider only the first 18 EECs of any simulation. 

Not surprisingly, the spectral characterization time distribution of the telescope with a 9 m ID is more sharply peaked toward shorter exposure times. In the non-background-limited regime, exposure times should scale as $D^{-2}$, while in background-limited regime they should scale as $D^{-4}$. If all else were equal, we'd therefore expect exposure times for our highest priority targets to be reduced by a factor of 2.3--5.1 when increasing telescope diameter from 6 m to 9 m. However, all else being equal, larger telescopes also provide higher coronagraphic throughput at a fixed angular separation for the DMVC, as well as an expanded target list to choose from, further reducing exposure times. We find the mean exposure time decreases by a factor of 6.2 going from 6 to 9 m ID; \emph{the mean spectral characterization time of the first 18 EECs is 22 days for a 6 m ID telescope, but can be shortened to just 3.5 days for a 9 m ID telescope.}

\begin{figure}[H]
\centering
\includegraphics[width=6in]{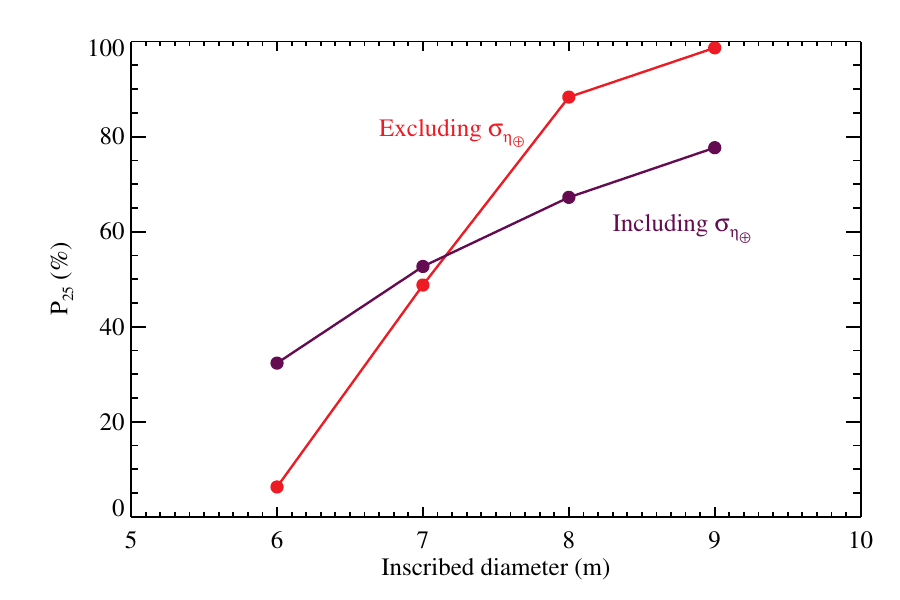}
\caption{Fraction of yield distribution $>$25 EECs, $P_{25}$,  when budgeting for detection and water vapor characterization, as a a function of telescope diameter for our baseline mission parameters. The purple line includes $\eta_{\Earth}$ uncertainty and the red line excludes it. \label{fig:conf_vs_D}}
\end{figure}

We calculate $P_{25}$ for each of the distributions shown in Figure \ref{fig:yield_dist_vs_D} and plot it as a function of telescope diameter in Figure \ref{fig:conf_vs_D}. For our baseline mission parameters, an $\sim$8.2 m ID telescope can achieve $P_{25}>90$\% when ignoring uncertainties in $\eta_{\Earth}$, but even a 9 m ID telescope cannot achieve $P_{25}\gtrsim80$\%  when budgeting for $\eta_{\Earth}$ uncertainties. Unless we consider diameters significantly larger than 6 m, telescope diameter alone cannot build in robust science margin for uncertainty in $\eta_{\Earth}$ under our baseline mission parameters. In the following section we investigate modifications to the LUVOIR-B design that could improve scientific performance without increasing the primary mirror diameter.

\subsection{Improve the mission design\label{sec:change_design}}

There are a number of changes to the LUVOIR-B baseline design that could improve the quantity and quality of data. Some of these are relatively straightforward changes, while others require development of new technologies. Here we highlight six possible improvements, noting that many others likely exist. We examine each design change one at a time, starting from the simplest and adding them up as we go. All of these changes will focus on reductions in exposure time. While previous works have shown that improvements to parameters controlling exposure time have only modest impact on the EEC yield\cite{stark2014,stark2015}, \emph{these impacts will ultimately compile, resulting in significant changes to expected EEC yield}. The end result demonstrates that some design changes are truly synergistic.

\subsubsection{Scenario A: Minimize aluminum reflections\label{section:Cass_VIS}}

The LUVOIR-B design adopted a three mirror anastigmat (TMA) optical telescope assembly (OTA) with a fourth fast steering mirror. Three additional pre-coronagraph optics were necessary prior to the UV-VIS channel dichroic. All seven of these mirrors were aluminum coated. Typical reflectivities for protected aluminum coated mirrors are $\sim$90\%, 87\%, and 92\% at 500, 760, and 1000 nm, respectively. Silver coated mirrors have reflectivities of $\sim$98\%, 97\%, and 96\% at the same three wavelengths. Just five aluminum coated mirrors will reduce throughput at 760 nm, a key wavelength for detection of the molecular oxygen (a biosignature gas), \emph{by almost a factor of two compared to silver}. As every exoplanet photon is precious, we should strive to reduce the number of aluminum coated mirrors. We note that Ref.~\citenum{stark2019} estimated the impact of some of these changes already; here we break these choices down in detail.

\begin{figure}[H]
\centering
\includegraphics[width=6in]{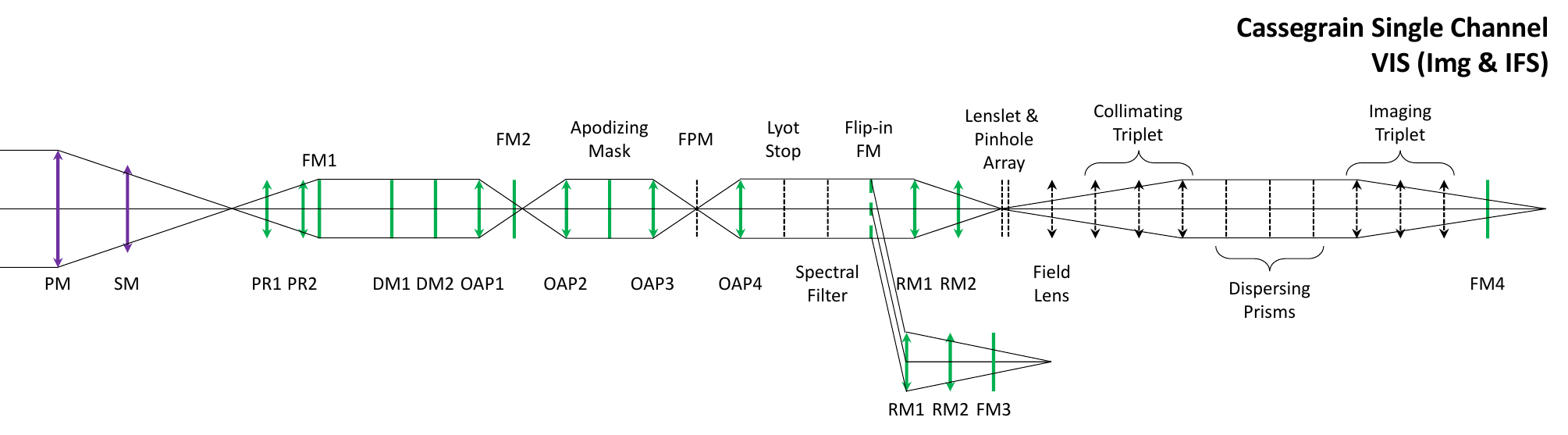}
\caption{Optical layout for a Cassegrain design with single VIS channel. We do not explicitly show dual parallel polarization channels, which we assume for the baseline coronagraph design. See Figure \ref{fig:optical_layout-LUVOIR-B} for legend. \label{fig:optical_layout-Cass-VIS}}
\end{figure}

Our first potential design change is to adopt a Cassegrain telescope with only two aluminum coated telescope mirrors. These two mirrors preserve UV science for any other instruments in the observatory, including a UV coronagraph. While it may still be possible to operate a UV coronagraph in parallel with a VIS coronagraph under this assumption, here we make the conservative assumption that the UV coronagraph cannot be parallelized, simply to illustrate a point:  in terms of EEC yield, the increased throughput will more than make up for the lack of a parallelized UV channel. Figure \ref{fig:optical_layout-Cass-VIS} shows the optical layout adopted for this design change and includes both IFS and broadband imaging modes for the single VIS channel. A possible UV channel is not shown. We assume dual polarization operation, but do not show the potential split needed for separate polarization channels; a detailed study of the need for separate polarization channels is beyond the scope of this paper.

Without a static dichroic present to split the parallel UV and VIS wavelengths, we are now able to implement the detection bandpass optimization feature within AYO. As shown by Ref.~\citenum{stark2023}, this usually results in V band detections for the majority of stars, but can select longer wavelengths for nearby and late type stars. We carry this feature forward through the rest of the analyses in this study.

\begin{figure}[H]
\centering
\includegraphics[width=6in]{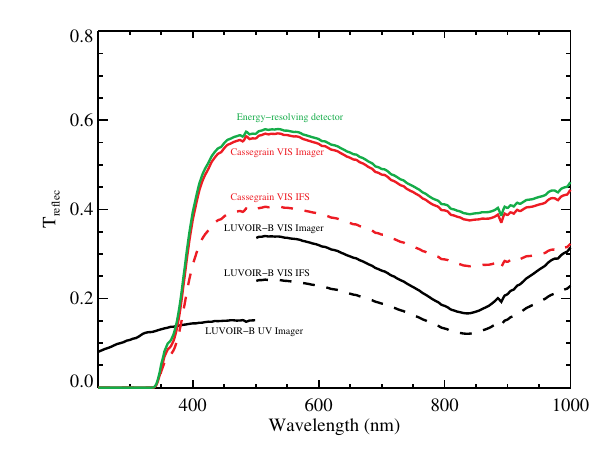}
\caption{Wavelength dependent optical throughput for a Cassegrain design with single VIS channel imager and IFS (red), an energy-resolving detector (green), and the LUVOIR-B baseline (black). Minimizing aluminum reflections can increase the throughput by 50\% at 1000 nm and nearly double it at $\sim$700 nm. Opting for an energy resolving detector can increase the throughput by an additional 40\% compared to an IFS.\label{fig:throughput-overlaid}}
\end{figure}

Figure \ref{fig:throughput-overlaid} shows the end-to-end optical throughput (not including the coronagraph's core throughput) of this Cassegrain VIS channel (red) compared to the LUVOIR-B baseline (black).  At 500 nm, where most detections will be performed, the optical throughput is now $\sim$1.7$\times$ that of our baseline design's VIS coronagraph, and is 1.2$\times$ the effective throughput of the baseline design's combined parallel UV and VIS coronagraphs.

\begin{figure}[H]
\centering
\includegraphics[width=6in]{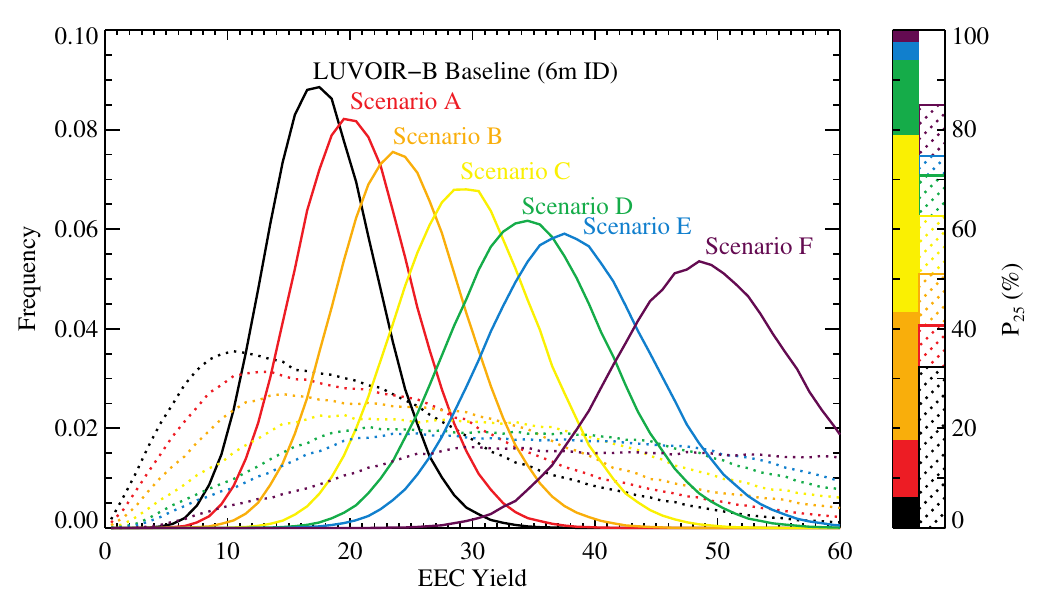}
\caption{\emph{Left:} Yield distributions for all design change scenarios considered, with dotted and solid lines including and excluding $\eta_{\Earth}$ uncertainty, respectively. Table \ref{table:design_change_summary} summarizes the design changes included in each scenario. \emph{Right:} Fraction of yield distribution $>$25 EECs, $P_{25}$,  when budgeting for detection and water vapor characterization. Dotted and solid bar plots correspond to including and excluding $\eta_{\Earth}$ uncertainty, respectively.\label{fig:yield_vs_design_change}}
\end{figure}

Figure \ref{fig:yield_vs_design_change} shows the impact of this change on EEC yield. The red curve shows the estimated yield distribution for Scenario A compared to the LUVOIR-B baseline yield distribution in black. Dashed and solid lines correspond to including and excluding $\eta_{\Earth}$ uncertainties, respectively. In spite of assuming only a single coronagraph channel, by minimizing the number of aluminum reflections the EEC yield increases by 15\%. The bar plots on the right in Figure \ref{fig:yield_vs_design_change} show $P_{25}$ increases as well, to 0.18 and 0.41 when excluding and including $\sigma_{\eta_{\Earth}}$, respectively.

\begin{figure}[H]
\centering
\includegraphics[width=6in]{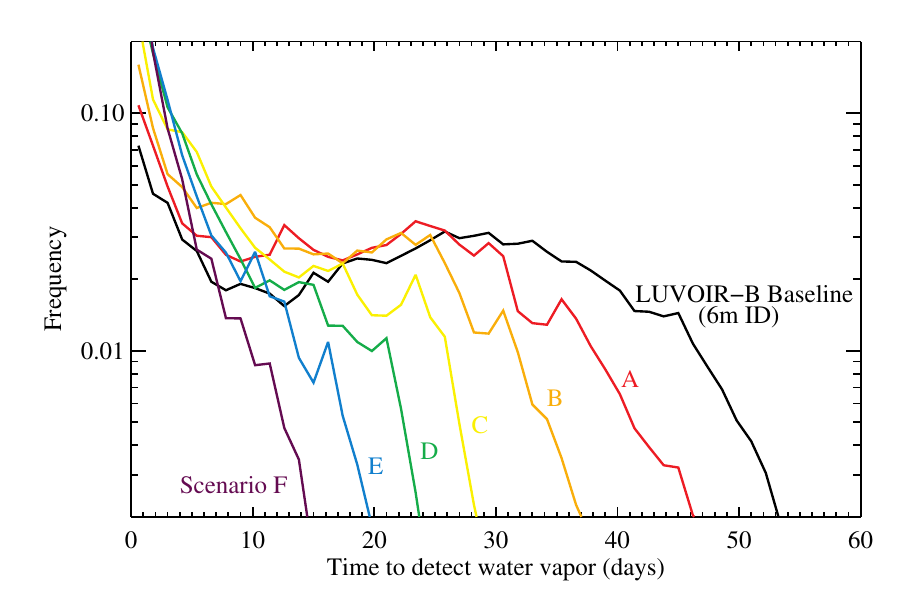}
\caption{Spectral characterization time distributions for all design change scenarios considered when excluding $\eta_{\Earth}$ uncertainty. Table \ref{table:design_change_summary} summarizes the design changes included in each scenario.\label{fig:tdist_vs_design_change}}
\end{figure}

This change also improves exposure times in the visible channel.  Figure \ref{fig:tdist_vs_design_change} shows the spectral characterization time distribution for Scenario A (red) compared to the LUVOIR-B baseline (black). The throughput at $\sim$700 nm is nearly twice that of our baseline design and is 50\% greater at 1000 nm (cf. Figure \ref{fig:throughput-overlaid}). Ultimately this translates into mean characterization exposure times 1.4$\times$ shorter.

\begin{figure}[H]
\centering
\includegraphics[width=6in]{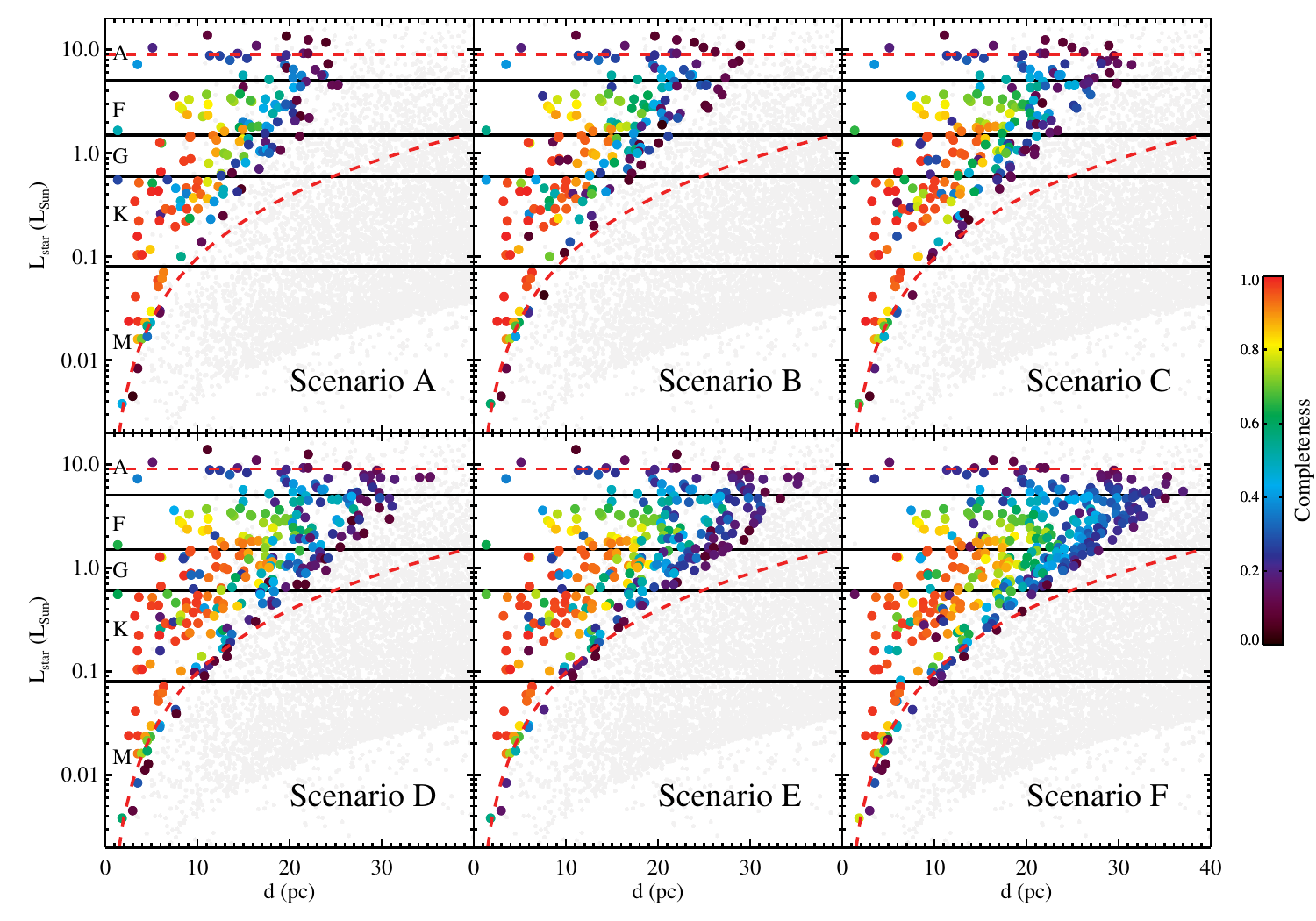}
\caption{The targets selected for each scenario color-coded by total completeness. The scenarios are cumulative, in that we compile multiple design improvements. The dashed lines indicate the boundaries of the accessible targets---the horizontal line roughly marks the assumed astrophysical noise floor while the curved line roughly indicates the working angle limit.\label{fig:target_list}}
\end{figure}

The top-left panel of Figure \ref{fig:target_list} plots each target selected for observation in Scenario A, color-coded by the total completeness achieved on each target. The horizontal dashed line indicates the stellar luminosity at which a $1.4$ R$_{\Earth}$ planet at quadrature has a $\Delta$mag equal to the assumed astrophysical noise floor, $\Delta$mag$_{\rm floor}$. The curved dashed line indicates the luminosity at which the outer edge of the HZ (1.67 AU) is located at $1.5$ $\lambda/D$ for $\lambda=1000$ nm. These two dashed lines roughly indicate the boundaries of the accessible target list. The stars at the top-right corner of this ``wedge" of targets require longer exposure times. Therefore, expanding the yield without reducing the IWA requires shorter detection times.

\begin{figure}[H]
\centering
\includegraphics[width=6in]{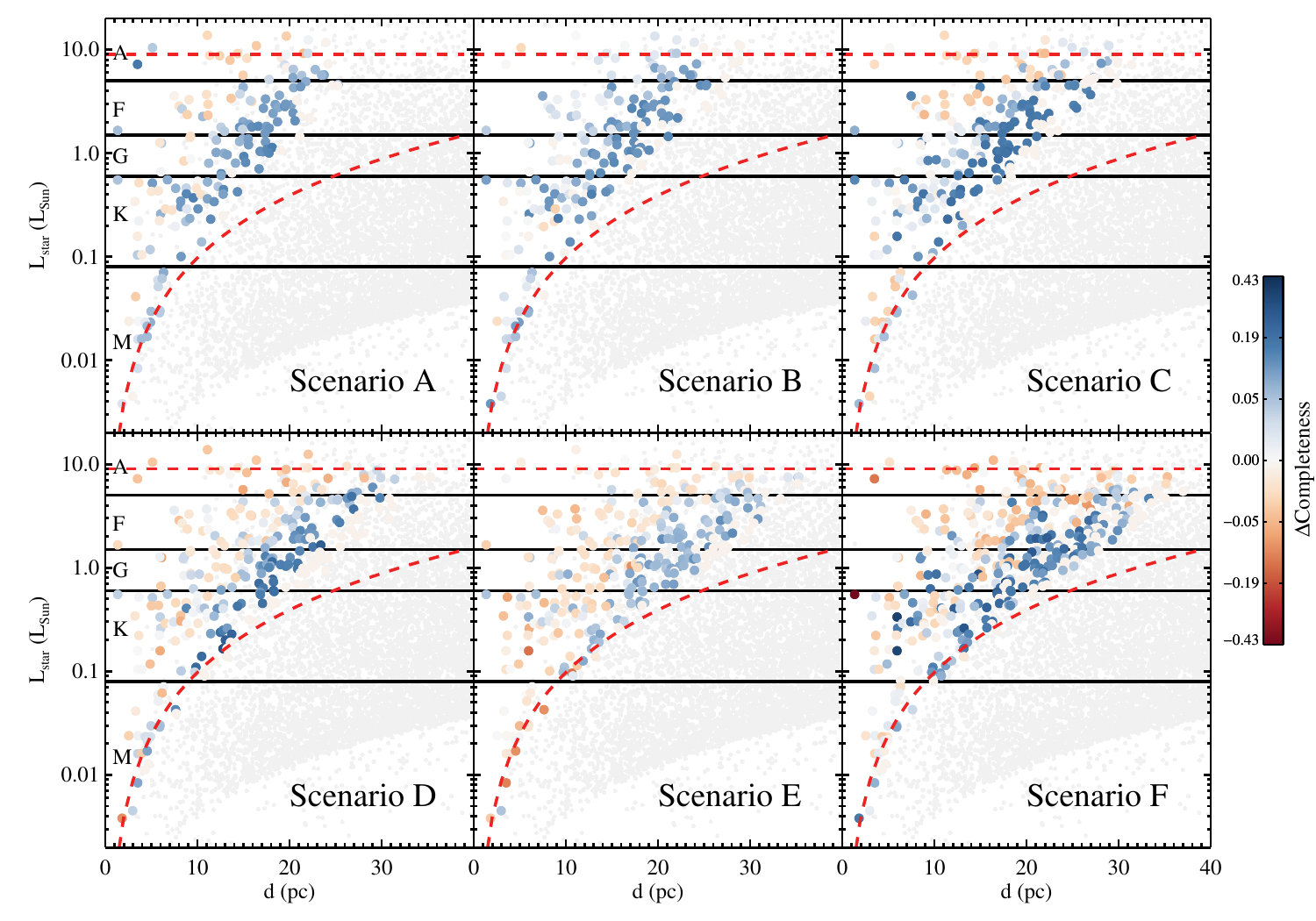}
\caption{The incremental change in completeness between successive scenarios. Each design improvement allows the mission to access more distant stars.\label{fig:targets_deltac}}
\end{figure}

The top-left panel of Figure \ref{fig:targets_deltac} shows the incremental change in completeness of each star for Scenario A compared to the baseline LUVOIR-B scenario. Blue indicates an increase in completeness while red indicates a decline in completeness. The scale has been stretched to emphasize the sign of the change, as shown by the color bar on the right. As shown in Figure \ref{fig:targets_deltac}, as the mission becomes more capable, the optimal redistribution of exposure time toward what were previously more challenging stars slightly reduces the completeness of nearby stars. While it may seem counter-intuitive to reduce exposure times of nearby stars in favor of more distant stars, this effect is the result of the less capable missions having "over-invested" time in nearby stars due to exposure time limitations.

\subsubsection{Scenario B: Operate two visible coronagraphs in parallel\label{section:Cass_2VIS}}

The LUVOIR-B study adopted a single visible wavelength coronagraph channel covering 500-1000 nm. In Scenario B we consider adding a second, parallel VIS coronagraph channel to Scenario A, as shown in Figure \ref{fig:optical_layout-LUVOIR-B}. This design change will significantly improve the yield of EECs that we can search for water vapor and will also substantially improve spectral data quality.

\begin{figure}[H]
\centering
\includegraphics[width=6in]{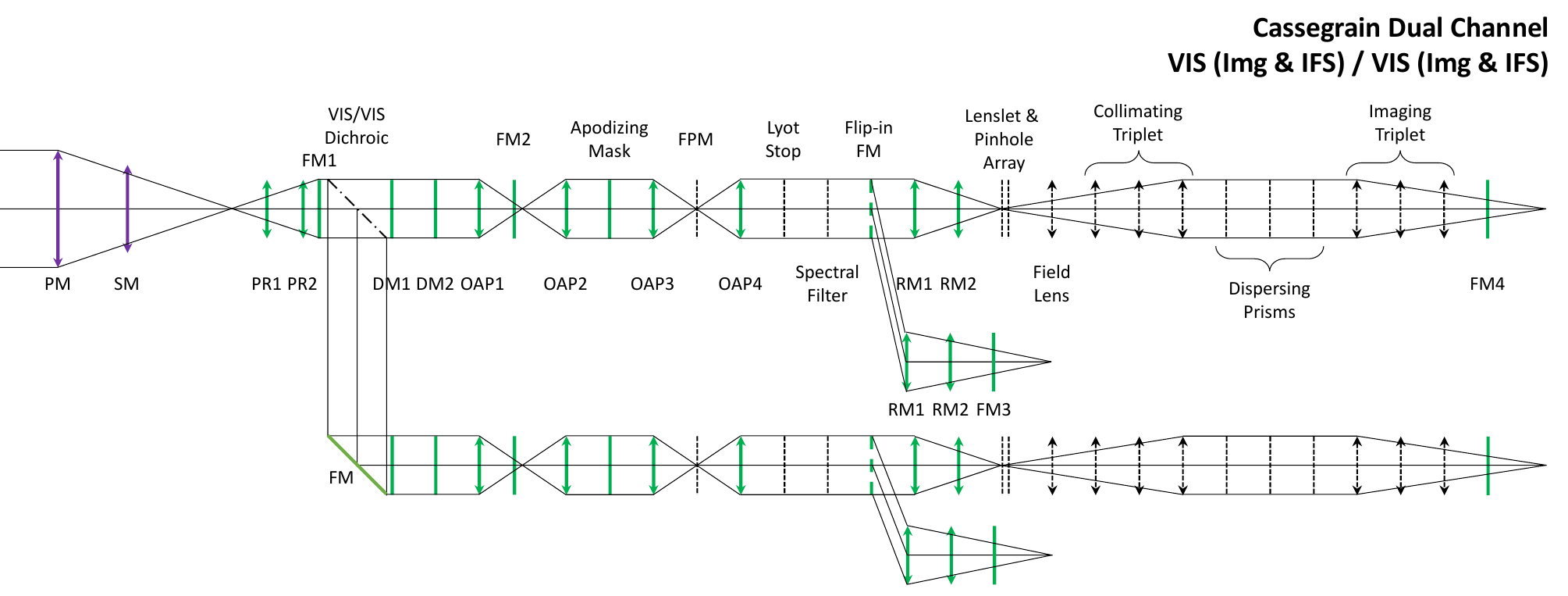}
\caption{Optical layout for a Cassegrain design with two parallel VIS channels. We do not explicitly show dual parallel polarization channels for each wavelength channel, which we assume for the baseline coronagraph design. See Figure \ref{fig:optical_layout-LUVOIR-B} for legend. \label{fig:optical_layout-Cass-2VIS}}
\end{figure}

There are a number of reasons why two parallel VIS channels would be an improvement. First, detection efficiency would increase. If the two coronagraph channels could both observe near 500 nm, where detections are efficient, then the bandpass would effectively double, decreasing detection exposure times by a factor of two (ignoring overheads) and providing color information in every detection. 

Second, the broader bandpass could improve the instantaneous spectral coverage for characterization. There are a number of key absorption features in the $\sim$750-1000 nm range, including molecular oxygen at 760 nm and water at 950 nm. These features may not be simultaneously observable with a single coronagraph channel without using advanced WFSC methods that sacrifice field of view for bandwidth. This means that under the assumption of a single VIS channel, the spectrum from $\sim$750-1000 nm may have to be pieced together using observations at different epochs. This will be complicated by the fact that the planet changes phases during its orbit; stitching together the spectrum of an orbiting planet may be a challenging task. By covering the majority of the desired spectrum in a single observation, this challenge is eased.

Accessing more water absorption lines simultaneously via a broader bandwidth can substantially reduce the time needed to detect water. Ref. \citenum{latouf2024} showed that by doubling the coronagraphic bandpass to 40\%, the SNR needed to detect water near near 1 $\mu$m is reduced from 5 to 4, which should equate to a $\sim$40\% reduction in exposure time.

Finally, two visible coronagraphs would provide redundancy of HWO's primary science instrument, a critical requirement of a Class A mission. This redundancy could also provide the capability to observe two polarizations simultaneously in the event that a coronagraph channel required a single polarization to achieve the necessary contrast.

To operate two VIS coronagraph channels efficiently for EEC detections, we would desire both coronagraphs to operate near 500 nm. On the other hand, efficient spectral characterizations means we'd want to operate the coronagraphs in the $\sim$750-1000 nm range. One way to enable both is to spit the channels with a \emph{selectable} dichroic, allowing us to split channels near $\sim$500 nm during detections, or $\sim$850 nm during characterizations.

To estimate the impact of a dual VIS design, we perform the same calculations as described in Section \ref{section:Cass_VIS}, but double the bandwidth of the detection coronagraph and the number of detector pixels under the assumption of a selectable dichroic. As such, we assume that channels must observe in adjacent bandpasses. To account for the reduced SNR required to detect water with 40\% bandwidth, we adopt the red curve shown in Figure \ref{fig:snr_vs_lambda} for spectral characterizations. The orange curve in Figure \ref{fig:yield_vs_design_change} shows the distribution of yields for Scenario B. This change has a substantial impact on the EEC yield, increasing it by 19\%.

Doubling the number of visible coronagraph channels has a significant impact on yield because it effectively doubles the photon collection rate for photometric detections and reduces spectral characterization time by 40\% by lowering the SNR required to detect H$_2$O, as shown by the orange curve in Figure \ref{fig:tdist_vs_design_change}.  We note that H$_{2}$O may be one of few atmospheric species that can take advantage of the broader bandwidth (another notably being CH$_{4}$). In comparison, O$_2$ has a single sharp feature near 760 nm that would not benefit from broader bandwidth. 

Dual visible coronagraph channels have several other scientific benefits that are note reflected in the yield or exposure time numbers. Specifically, dual photometric detections provide rudimentary color information that can help distinguish between planets from epoch to epoch\cite{krissansentotton2016}, which may be important as planets shift in position relative to the host star. Second, dual visible coronagraphs can provide an enhanced ability to simultaneously detect multiple atmospheric species in a single observations. Ref. \citenum{latouf2024} showed that two 20\% bandpasses can simultaneously detect H$_2$O, O$_2$, and $O_3$ at $R=140$ and SNR$=11$ assuming present atmospheric levels for an Earth twin, something a single 20\% bandpass simply cannot do in a single observation.

\subsubsection{Scenario C: Adopt model-based PSF subtraction\label{section:Cass_2VIS_WFSPSF}}

The LUVOIR study baselined coronagraphs with raw contrast of $10^{-10}$. However, HWO will need to detect planets with contrasts more challenging than $10^{-10}$ at SNR $>$ 10\cite{luvoirfinalreport}. This means that the speckle noise floor must, at least, be better than $10^{-11}$. Designing a coronagraph with raw contrast better than $10^{-11}$ would be very challenging, as restricting the raw contrast can limit the rest of the design phase space, potentially resulting in low throughput, a narrow bandpass, and/or greater sensitivities to wavefront aberrations\cite{stlaurent2018}. Therefore, we will need to perform PSF subtraction to reduce speckles to better than $10^{-11}$. There are many potential PSF subtraction methods, each with benefits and challenges. Here we address several key methods and discuss how adopting a model-based PSF subtraction method could substantially improve the science return. 

The Roman Coronagraph will baseline reference differential imaging (RDI). Under the RDI approach, the instrument observes a science target and a reference target back-to-back (or potentially interleaved). The reference target is ideally an exact match to the science target, but bright and isolated, with no astrophysical scene around it. However, at the $10^{-10}$ contrast level, we expect most stars to have some level of astrophysical contamination around them. Further, even though the bandpass for HWO may be relatively narrow, it will operate at wavelengths where the color of stars can be significantly different. Some coronagraphs are also sensitive to stellar diameter, meaning our reference star and science star would also have to match in terms of angular diameter. For these reasons, we suggest that RDI may be challenging for HWO.

One alternative is angular differential imaging (ADI), in which the science target is observed twice at two different roll angles. The speckles rotate with the telescope while the astrophysical scene remains fixed in the sky. As a result, we can co-align each exposure in the instrument frame to subtract the speckles, producing positive and negative copies of the astrophysical scene. This would provide a much better match in terms of the reference star and may help empirically subtract exozodiacal dust\cite{kammerer2022}, but comes at a cost. First, the differential roll angle to displace a PSF near the IWA would be $\sim$40$^{\circ}$ for HWO, placing strict constraints on wavefront stability as a function of roll angle. Second, the empirical ADI subtraction multiplies the count rate of all noise components by a factor of two. The LUVOIR and HabEx studies adopted ADI as the baseline PSF subtraction method and most yield calculations to date have included this factor of two on background count rates (c.f., Equation 11 in Ref.~\citenum{brown2005} and Equation 5 in Ref.~\citenum{stark2019}).

Model-based PSF subtraction operates differently. By combining and correlating high-cadence wavefront telemetry with the bright unobscured starlight, we may be able to reconstruct the coronagraphic PSF at any point in time during the science exposure\cite{guyon2021}. This could partially relax some telescope stability requirements, as we do not need to maintain PSF stability/repeatability at the $10^{-11}$ level---only to a level that provides the desired raw contrast. Model-based PSF subtraction could also reduce the systematic speckle noise floor\cite{guyon2021}, potentially to the Poisson noise limit, or to a level governed by the incoherence of the stellar leakage. Here we ignore this potential benefit of model-based PSF subtraction and maintain the same noise floor described by $\Delta$mag$_{\rm floor} = 26.5$, as prior studies have shown little gain in EEC yield when improving the noise floor\cite{stark2014,stark2015}. Most pertinent to this study, because no empirical background subtraction is required, \emph{model-based PSF subtraction removes the factor of two on all background count rates, effectively cutting all exposure times in half}. 

We define Scenario C as Scenario B with model-based PSF subtraction added. For this scenario, we repeat the calculations performed in Section \ref{section:Cass_2VIS}, but remove the factor of two in front of all background count rates in our exposure time calculator. The yellow curve in Figure \ref{fig:yield_vs_design_change} shows the results for Scenario C. Model-based PSF subtraction is estimated to improve yields by 23\%, from 23.8 EECs for Scenario B to 29.3 EECs for Scenario C. $P_{25}$ increases accordingly, to 0.79 and 0.63 when excluding and including $\sigma_{\eta_{\Earth}}$, respectively. As shown in Figure \ref{fig:tdist_vs_design_change}, the characterization exposure times are reduced by a factor of 1.6. These significant improvements are the result of reducing \emph{all} noise count rates by a factor of two, including detector noise.  

HWO's PSF subtraction method may ultimately end up being a combination of multiple approaches. This could range from using a library of empirical PSFs\cite{soummer2011} to spectral differential imaging\cite{fergus2014}. Regardless of the technique, reducing the background noise associated with some empirical methods would be a fruitful endeavor.

\subsubsection{Scenario D: Improve detector performance\label{section:Cass_2VIS_WFSPSF_Skipper}}

The LUVOIR and HabEx studies baselined an EMCCD as the visible wavelength coronagraph detector. The adopted parameters for this EMCCD were optimistic, based on assumed future improvements to the Roman Coronagraph EMCCD. We have carried these assumptions through to this study, as shown in Table \ref{table:missionassumptions}. Some of those assumptions have proven true. Roman's EMCCD dark current values are on par with the LUVOIR-B baseline assumption of $3\times10^{-5}$ counts pix$^{-1}$ s$^{-1}$ and the dQE terms budgeting for photon counting efficiency, cosmic ray efficiency, hot pixel efficiency factors, etc., as defined by Ref. \citenum{morrissey2023}, are estimated to be $\sim0.77$ at end of life\cite{morrissey2023}, consistent with LUVOIR assumptions. Other assumptions remain optimistic. These include clock induced charge, which remains a factor of 10 higher than the LUVOIR assumptions, and most notably, the raw QE near 1000 nm, which is only a few percent for Roman's EMCCD\footnote{\url{https://roman.ipac.caltech.edu/sims/Param_db.html}} but was assumed to be 90\% in the LUVOIR study. In spite of this, there remain paths forward using an EMCCD. The desire for high QE near 1000 nm was motivated by the desire for efficient detection of water vapor, but Ref.~\citenum{stark2023} showed that low QE near 1000 nm can be partially mitigated by searching for water at shorter wavelengths. Alternatively, a dedicated ultra-low-noise NIR detector could be more appropriate for the detection of water vapor. 

Several alternative detector technologies may improve performance beyond the LUVOIR-B assumptions. Here we examine the potential benefits of such detectors. A thorough examination of the impact of different detector technologies, critical to the success of HWO, would require an exhaustive study comparing all detector options, which is beyond the scope of this paper. We therefore choose to adopt parameters consistent with two possible detector options. We start with performance parameters that may be possible with a photon-counting Skipper CCD. Table \ref{table:detector_summary} summarizes the performance parameters we adopted compared to the LUVOIR-B baseline. We adopt a dark current orders of magnitude lower than the LUVOIR assumptions, which has been demonstrated for the Skipper CCD\cite{bebek2015,barak2022}. We note that this reduction in dark current stems from a combination of a high degree of shielding and cosmic ray identification and removal, which may be possible for a traditional EMCCD as well. However, the Skipper CCD also has a clock-induced charge 10$\times$ better than LUVOIR assumptions and has negligible dQE. As a result, the parameters we adopt provide $\sim$30\% higher effective throughput, as well as noise properties that are effectively negligible. 

Skipper CCDs use multiple non-destructive reads to average read noise down to deeply sub-electron levels and thereby count photons \cite{Tiffenberg:2017gsa}. Recent advances in semiconductor fabrication technology have made possible thick, fully depleted, photon counting p-channel Skipper CCDs like those used by Ref.~\citenum{Tiffenberg:2017gsa}. These p-channel Skippers, developed at the Lawrence Berkeley National Laboratory (LBNL), have important advantages for space astrophysics including excellent radiation tolerance and QE greater than 80\% at 940~nm. However, the primary challenge for using LBNL's Skipper CCDs in space is radiation nonetheless. Although p-channel Skipper CCDs do not degrade in space like n-channel CCDs, they require short exposure times to minimize cosmic ray disturbance. This is on account of the $\approx200~\mu$m thick silicon that is used to achieve good near-IR QE. As a practical matter, p-channel Skipper exposure time will need to be on the order of one minute to limit cosmic ray disturbance to about 10\% of pixels, requiring additional amplifier outputs \,---\, the recently developed Multi-Amplifier Sensing CCD\cite{botti2023} is a step toward this.

We define Scenario D as Scenario C with the EMCCD's performance parameters replaced with the Skipper CCD parameters listed in Table \ref{table:detector_summary}. Here we define QE as the traditional quantum efficiency, i.e., the number of photoelectron groups created per photon received. We define dQE as the ``detective" QE, a factor specific to EMCCDs discussed in Ref. \citenum{morrissey2023}.

\begin{figure}[H]
\centering
\includegraphics[width=6in]{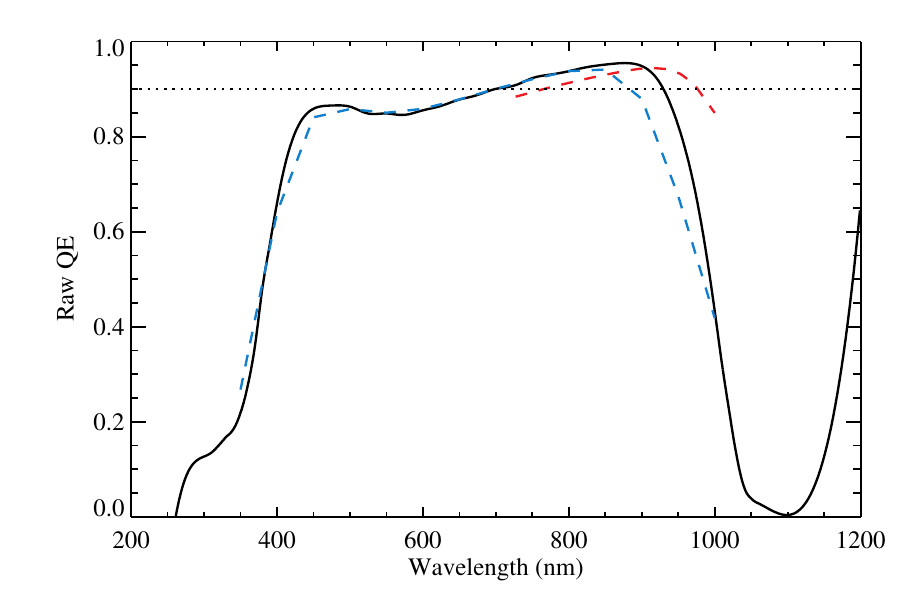}
\caption{Skipper CCD raw QE (solid line) and the raw QE values adopted for the yield code when integrating over a 20\% bandpass for detection (blue dashed) and spectral characterization (red dashed). The raw QE for the LUVOIR-B baseline EMCCD and energy-resolving detector (Scenario E) is shown for comparison (dotted line).\label{fig:QE}}
\end{figure}

Figure \ref{fig:QE} shows our adopted raw QE curve for the Skipper CCD as a black solid line. For yield calculations, we are interested in a bandpass-integrated QE. For exoplanet detections, we assume the detection wavelength is centered in the bandpass as usual and integrate over a bandwidth of 20\%, resulting in the blue dashed line that we adopt for detection raw QE. For spectral characterizations, we work with the longest wavelength of the bandpass to ensure the exoplanet is exterior to the coronagraph IWA over the whole bandpass. Using the single QE value at this wavelength would be doubly conservative, as the QE curve drops rapidly near 1000 nm. Thus, for spectral characterizations, we integrate the QE over a 20\% bandpass with longest wavelength given by the $x$-axis of Figure \ref{fig:QE}, resulting in the red dashed line. We note that this detail is critical: the bandpass-integrated QE over the water vapor absorption feature near 1000 nm is twice that of the QE at 1000 nm. Future studies should investigate the impact of realistic QE curves on water vapor retrievals near 1000 nm. 

\begin{landscape}
\begin{deluxetable}{llccc}
\tablewidth{0pt}
\tablecaption{Detector parameters\label{table:detector_summary}}
\tablehead{
\colhead{Parameter} & \colhead{Units} & \colhead{EMCCD} & \colhead{Skipper} & \colhead{ERD} 
}
\startdata
QE		&							& $0.9$							& See Fig.~\ref{fig:QE} 	& $0.9$ \\
dQE		&							& $0.75$							& $0.99$				& $1.0$ \\ 
DC		& counts pix$^{-1}$ s$^{-1}$ 		& $3\times10^{-5}$ 					& $6.8\times10^{-9}$		& $0$ \\
CIC		& counts pix$^{-1}$ frame$^{-1}$	& $1.3\times10^{-3}$					& $1.5\times10^{-4}$		& $0$ \\
RN		& counts pix$^{-1}$ read$^{-1}$		& $0$							& $0$				& $0$ \\
Scenarios	&							& 0, A, B, C 						& D					& E, F \\
\enddata
\vspace{-0.1in}
\end{deluxetable}
\end{landscape}

Figure \ref{fig:yield_vs_design_change} shows the resulting yield distribution for Scenario D in green. EEC yield increases by 17\%, and $P_{25}$ increases to 0.94 and 0.71 when excluding and including $\sigma_{\eta_{\Earth}}$, respectively. Figure \ref{fig:tdist_vs_design_change} shows that the characterization times decrease as well, by an additional factor of 1.5.

\subsubsection{Scenario E: Adopt an energy-resolving detector\label{section:Cass_2VIS_WFSPSF_TES}}

Next, we show the impact of swapping the Skipper CCD and IFS with a noiseless energy-resolving detector (ERD). An IFS uses a lenslet at each ``pixel" in the image plane to focus light onto a dispersing element, ultimately producing spectra of each image plane pixel in the final detector plane. While this instrument provides spatially resolved spectra over the entire field of view, there are some disadvantages. First, the additional IFS optics have a throughput estimated to be a factor of $0.7$ (c.f., Table \ref{table:missionassumptions}). Second, the PSF's core is spread over potentially hundreds of pixels at the long wavelength end of the channel, effectively amplifying the detector's per-pixel noise properties. 

With an ERD, read noise manifests as part of the energy resolution budget\cite{rauscher2016} and there is no need for an IFS, eliminating throughput-reducing optics. Ref.~\citenum{howe2024} showed that this, combined with negligible dQE, could result in a 30\% increase in EEC yield compared to the LUVOIR-B baseline. Here we examine the incremental improvement of an ERD compared to a Skipper CCD-based IFS, which also has negligible dQE and noise properties, so we will not see the same 30\% increase in EEC yield.  

\begin{figure}[H]
\centering
\includegraphics[width=6in]{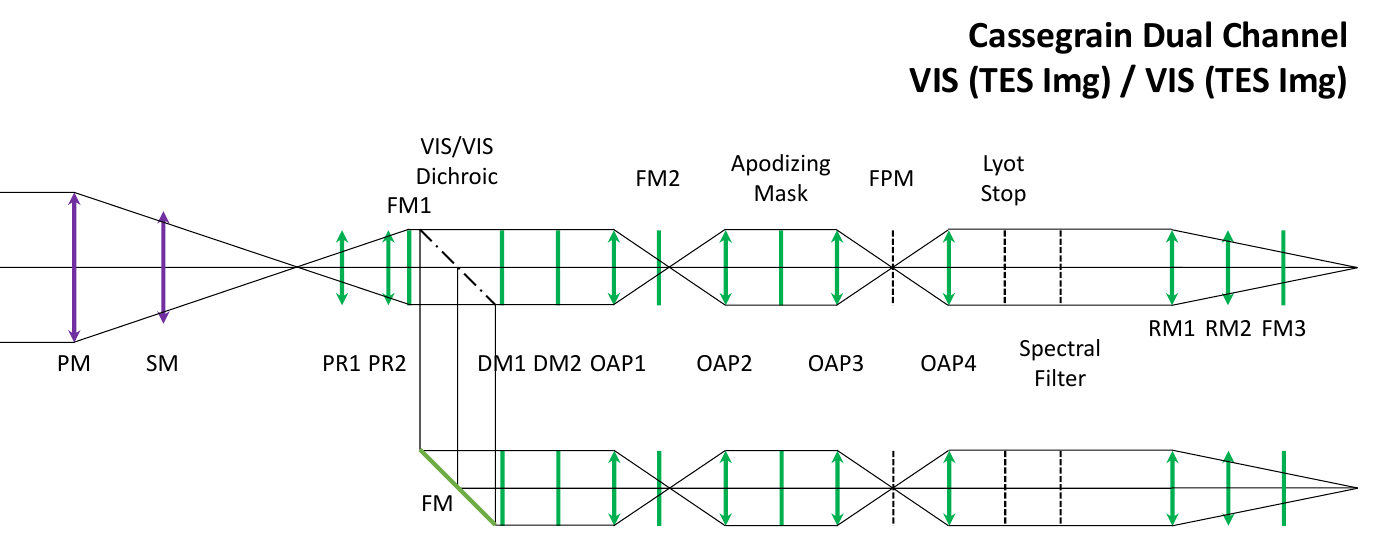}
\caption{Optical layout for a Cassegrain design with two parallel VIS channels with energy-resolving detectors. See Figure \ref{fig:optical_layout-LUVOIR-B} for legend. \label{fig:optical_layout-Cass-2VIS-TES}}
\end{figure}

Figure \ref{fig:optical_layout-Cass-2VIS-TES} illustrates the instrument layout with an ERD. We note that Figure \ref{fig:optical_layout-Cass-2VIS-TES} does not show any optics to thermally isolate the detector from the rest of the instrument, which will be required for an ERD. These optics will reduce the throughput of the system, but a detailed assessment of thermal isolation and the transmissivities of the necessary optics is beyond the scope of this paper. Our estimate of the impact of an ERD should therefore be considered an upper limit.

Figure \ref{fig:throughput-overlaid} shows the assumed optical throughput using an ERD (green). Because there is no separate imaging mode for an ERD, the green line illustrates the throughput for detections and for spectroscopy. An ERD can potentially increase spectroscopic throughput by 40\% compared to an IFS. Table \ref{table:detector_summary} summarizes the adopted performance parameters for our ERD, based off of the expected performance of a transition edge sensor\cite{rauscher2016}. To date, TES arrays have demonstrated $R=90$ at 485 nm\cite{TEStechreport} and MKIDS have demonstrated $R=52$ at 402 nm\cite{devisser2021}. We assume that a future ERD can achieve the $R=140$ requirement that we adopt.

Figure \ref{fig:yield_vs_design_change} shows the resulting yield distribution for Scenario E in blue. The ERD increases EEC yield by 9\% compared to the Skipper CCD scenario, and $P_{25}$ increases to 0.98 and 0.75 when excluding and including $\sigma_{\eta_{\Earth}}$, respectively. The ERD also reduces spectral characterization times compared to the Skipper CCD scenario by another factor of 1.4$\times$, as shown in Figure \ref{fig:tdist_vs_design_change}. Most of these improvements are the result of the increased throughput due to lack of IFS optics, not reduced detector noise, as the adopted noise parameters for the Skipper were already very low. We note that when compared to the LUVOIR-B baseline detector assumptions, the ERD gains are larger, with a 2.2$\times$ reduction in characterization times and a 1.3$\times$ gain in EEC yield.

\subsubsection{Scenario F: Adopt a high-throughput coronagraph\label{section:Cass_2VIS_WFSPSF_TES_coron}}

\begin{figure}[H]
\centering
\includegraphics[width=6in]{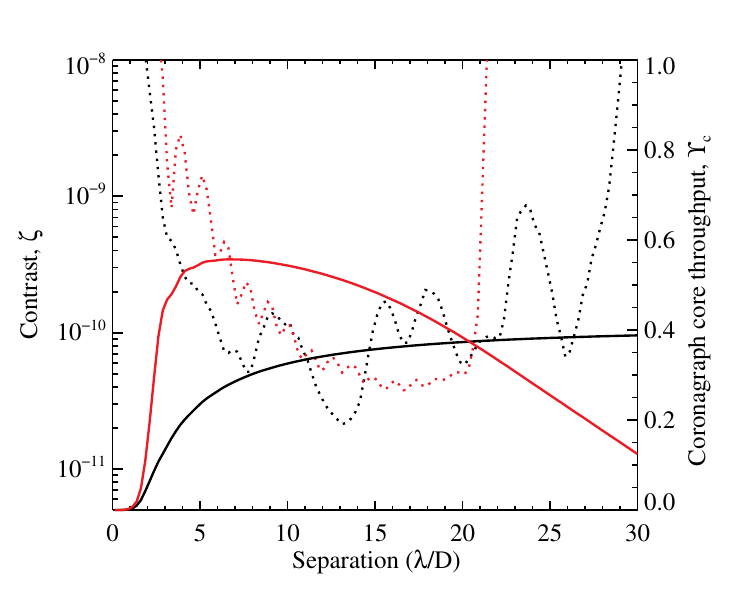}
\caption{Azimuthally averaged contrast for a star with diameter $0.1$ $\lambda/D$ (dotted line) and core throughput (solid line) for the two coronagraphs included in this study. The $x$-axis is in units of $\lambda/D$, where $D$ is the circumscribed diameter of the telescope. The DMVC6 and PIAA-FPM2.5 are shown in black and red, respectively.  \label{fig:coronagraphs}}
\end{figure}

The LUVOIR-B study baselined a DM-assisted charge six vortex coronagraph (DMVC6). Figure \ref{fig:coronagraphs} shows the azimuthally-averaged contrast for a 0.1$\lambda/D$ star as a dotted black line over a 20\% bandwidth for a single polarization. We note that while the DMVC only works for a single polarization, we have implicitly assumed so far that the design allows for parallel polarization channels. The solid black line shows the core throughput of this coronagraph. The IWA for the DMVC6 is $\sim3.5$ $\lambda/D$, where the core throughput reaches half it's maximum value. We remind the reader that $D$ in this study, and that adopted for the $x$-axis of Figure \ref{fig:coronagraphs}, is the circumscribed diameter of the telescope. The DMVC6 is limited to using the inscribed diameter of the telescope, making its IWA in Figure \ref{fig:coronagraphs} larger than that of a circular aperture. Notably, there is useful throughput interior to the IWA, which the yield code takes advantage of for nearby, later type stars (see Figure \ref{fig:target_plots_baseline}). While the core throughput reaches a relatively high maximum value of $\sim0.45$, it rises fairly slowly with working angle, such that it is $\sim$5\% at 2 $\lambda/D$. 

Here we examine one possible alternative coronagraph design that has higher throughput at small working angles: the phase-induced amplitude apodizer (PIAA) coronagraph. We adopt a PIAA design created for the LUVOIR-B aperture which incorporated a focal plane mask with radius $2.5$ $\lambda/D$ and a DM-assisted solution robust to stellar diameters as large as 0.1 $\lambda/D$ (we note that this DM solution may help other coronagraph designs as well). We refer to this design as PIAA-FPM2.5. Figure \ref{fig:coronagraphs} shows the azimuthally averaged performance of the PIAA-FPM2.5 in red. In comparison with the DMVC6, the contrast near $\sim$4 $\lambda/D$ is a factor of $\sim$4 worse and the effective OWA is notably limited by contrast degradations to $\sim$20 $\lambda/D$. Notably, we maintain the same uniform noise floor used throughout this study, which is not proportional to the raw contrast. We choose this single, non-ideal coronagraph design as an example to illustrate a point: in spite of the degraded contrast, if the noise floor remains the same, the increase in core throughput at small working angles will more than compensate on a survey scale, resulting in significantly improved yields. This is because exposure times are reduced for targets in which the leaked starlight doesn't dominate the background count rate (e.g., distant stars).

Unlike the DMVC6, the PIAA-FPM2.5 works for both polarizations simultaneously. This has the potential to reduce the complexity of the instrument by eliminating parallel polarization channels. In principle, the complexity could be maintained and the parallel polarization channels could be replaced by another dichroic split, doubling the total instrument bandwidth. However, here we make the conservative assumption that dual polarization channels are still required to minimize polarization cross-talk and do not adopt any increase in total instrument bandpass. 

The purple curve in Figure \ref{fig:yield_vs_design_change} shows the yield distribution for Scenario F. In spite of the degraded contrast, the higher throughput of the PIAA-FPM2.5 coronagraph design increases EEC yield by 30\%. Combined with all previously discussed changes, $P_{25}$ is now estimated to be 0.99 and 0.85 when excluding and including $\sigma_{\eta_{\Earth}}$, respectively. The PIAA-FPM2.5 reduces spectral characterization times compared to Scenario E by another factor of 1.6$\times$, as shown in Figure \ref{fig:tdist_vs_design_change}.

We note that many other coronagraph designs with smaller IWA and higher core throughput exist in addition to the PIAA-FPM2.5 examined here. Such coronagraphs may also lead to higher yields to varying degrees. A simple example that we do not consider in this paper is combining our baseline charge 6 DMVC with a charge 4 design. A future trade study building off of the Coronagraph Design Survey\cite{belikov2023} and examining a broad range of coronagraphs designed specifically for HWO would be highly valuable.

\subsubsection{Summary of design changes}

We investigated six possible design changes from the LUVOIR-B baseline design. As shown in Fig.~\ref{fig:tdist_vs_design_change}, these improvements significantly reduced exposure times. This in turn allowed the simulated mission to observe a larger portion of the accessible targets at larger distances, as illustrated in Figs.~\ref{fig:target_list} and \ref{fig:targets_deltac}, and achieve much higher yields. 

Notably, all of these design changes produced modest incremental improvements to yield and all are roughly consistent with the scaling relationships between exposure time and yield reported in previous studies\cite{stark2015,stark2019}. Here we showed that while yield is only moderately sensitive to changes in exposure time factors (throughput, bandwidth, etc.), there are many such factors. If multiple factors are improved simultaneously, the impact on exposure time and yield can be large.

Table \ref{table:design_change_summary} lists the incremental and cumulative reductions in exposure times and increases in EEC yield, as well as the value of $P_{25}$ associated with each design change. Incremental changes to yield are broadly consisent with scaling relationships from previous works\cite{stark2014,stark2015,stark2019}. Overall, spectral characterization times can be reduced by more than an order of magnitude while doubling the characterization bandwidth over the visible spectrum. This results in nearly a tripling of the EEC yield. The probability of detecting and characterizing $>$ 25 EECs increases dramatically, with $P_{25}$ increasing from 6\% to 99\% in the case in which $\eta_{\Earth}$ uncertainty is ignored. Given that these six examples are not an exhaustive list, and none of these changes includes increasing the telescope diameter, we conclude that it is possible to establish EEC science margins substantial enough to offset most astrophysical uncertainties for HWO. 

\begin{landscape}
\begin{deluxetable}{clcccccc}
\tablewidth{0pt}
\tabletypesize{\small}
\tablecaption{Summary of design change impacts\label{table:design_change_summary}}
\tablehead{
\colhead{Scenario} & \colhead{Description} & \colhead{Refer to} & \colhead{Reduction in} & \colhead{Increase in} & \colhead{Increase in} & \colhead{$P_{25}$ when} & \colhead{$P_{25}$ when}\\
\colhead{} & \colhead{} & \colhead{Section} & \colhead{char. time to} & \colhead{instantaneous} & \colhead{EEC yield\tablenotemark{\dag}} & \colhead{ignoring $\sigma_{\eta_{\Earth}}$} & \colhead{including $\sigma_{\eta_{\Earth}}$}\\
\colhead{} & \colhead{} & \colhead{} & \colhead{detect H$_2$O\tablenotemark{\dag*}} & \colhead{VIS bandwidth\tablenotemark{\dag}} & \colhead{} & \colhead{} & \colhead{}
}
\startdata
0 & LUVOIR-B baseline (6 m ID) & \ref{section:methods} 								& 1$\!\times$ / 1$\!\times$ & 1$\!\times$ / 1$\!\times$ & 1$\!\times$ / 1$\!\times$ & 0.06 & 0.32\\
A & Minimize Al coatings & \ref{section:Cass_VIS} 							& 1.4$\!\times$ / 1.4$\!\times$ & 1$\!\times$ / 1$\!\times$ & 1.15$\!\times$ / 1.15$\!\times$ & 0.18 & 0.41 \\
B & A + Dual VIS & \ref{section:Cass_2VIS} 								& 1.4$\!\times$ / 1.9$\!\times$ & 2$\!\times$ / 2$\!\times$ & 1.19$\!\times$ / 1.37$\!\times$ & 0.43 & 0.51 \\
C & B + Model-based PSF sub. & \ref{section:Cass_2VIS_WFSPSF} 			& 1.6$\!\times$ / 3.1$\!\times$ & 1$\!\times$ / 2$\!\times$ & 1.23$\!\times$ / 1.69$\!\times$ & 0.79 & 0.63 \\
D & C + Skipper CCD & \ref{section:Cass_2VIS_WFSPSF_Skipper} 			& 1.5$\!\times$ / 4.8$\!\times$ & 1$\!\times$ / 2$\!\times$ & 1.17$\!\times$ / 1.98$\!\times$ & 0.94 & 0.71 \\
E & D + Energy-resolving det. & \ref{section:Cass_2VIS_WFSPSF_TES} 			& 1.4$\!\times$ / 6.8$\!\times$ & 1$\!\times$ / 2$\!\times$ & 1.09$\!\times$ / 2.16$\!\times$ & 0.98 & 0.75 \\
F & E + High-throughput coron. & \ref{section:Cass_2VIS_WFSPSF_TES_coron} 	& 1.6$\!\times$ / 11.1$\!\times$ & 1$\!\times$ / 2$\!\times$ & 1.30$\!\times$ / 2.81$\!\times$ & 0.99 & 0.85 \\
\enddata
\vspace{-0.1in}
\tablenotetext{\dag}{Columns with more than one value separated by a slash indicate the incremental/cumulative change in the quantity for each scenario compared to the previous/baseline scenario.}
\tablenotetext{*}{For the first $<$18 EECs, excluding static overheads}
\end{deluxetable}
\end{landscape}

\section{Budgeting for uncertainties with science margin\label{section:performanceunc}}

\subsection{Budgeting for astrophysical uncertainty\label{section:astrounc}}

\begin{figure}[H]
\centering
\includegraphics[width=6in]{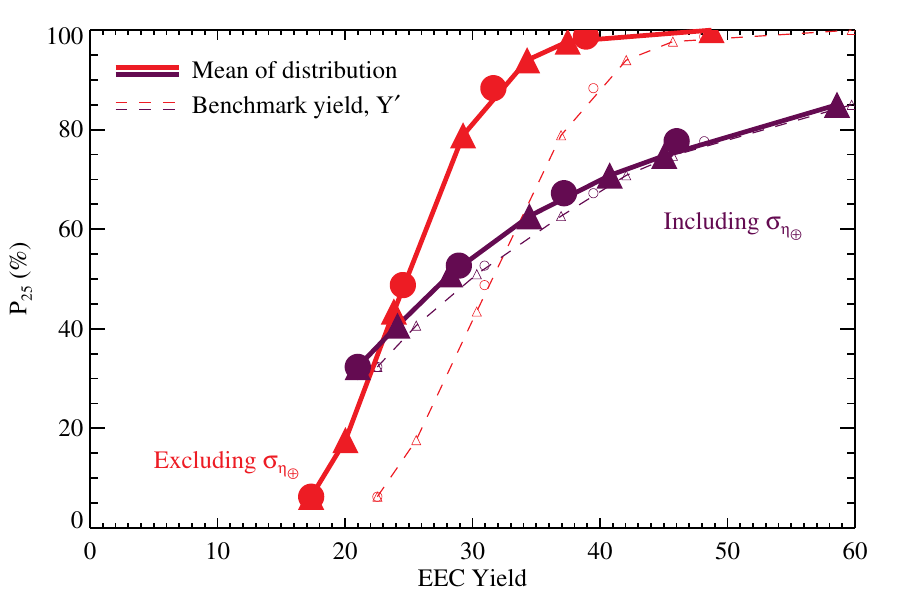}
\caption{Fraction of yield distributions $>$25 EECS, $P_{\rm 25}$, as a function of expected EEC yield excluding (solid red) and including (solid purple) uncertainty in $\eta_{\Earth}$. Filled circles indicate yields obtained by changes to telescope diameter while filled triangles with a connecting line indicate those due to design changes. The agreement between lines and circles suggests $P_{\rm 25}$ is independent of the means used to obtain higher yields. The unfilled symbols and dashed lines show the expectation value of the benchmark yield, $Y'$, which can be used to estimate $P_{25}$ without calculating a yield distribution.\label{fig:conf_vs_yield}}
\end{figure}

Precisely how much science margin is required for HWO to budget for astrophysical uncertainty? This depends in large part on HWO's risk posture, formalized minimum yield goal, whether HWO's formal science goals should account for uncertainty in $\eta_{\Earth}$, the magnitude of that uncertainty, and whether the uncertainty can be reduced with precursor science observations or analyses. These decisions will likely come out of future formalized HWO modeling efforts and are beyond the scope of this paper. However, we can use the results of Sections \ref{sec:bigger_telescope} and \ref{sec:change_design} to provide basic guidance for these future decisions by relating the expectation value of EEC yield to $P_{\rm 25}$. The solid lines in Figure \ref{fig:conf_vs_yield} show $P_{\rm 25}$ as a function of mean EEC yield excluding and including uncertainty in $\eta_{\Earth}$ in red and purple, respectively. The yields obtained by increasing telescope diameter (Section \ref{sec:bigger_telescope}) are shown as filled circles while the yields from improving mission design (Section \ref{sec:change_design}) are shown as filled triangles with a connecting line. The agreement between the lines and circles suggests that $P_{\rm 25}$ is independent of the specific means used to obtain higher yields. Figure \ref{fig:conf_vs_yield} can therefore be used to estimate $P_{\rm 25}$ for a broad range of HWO trade studies by calculating the expectation value of the EEC yield distribution.

We note that calculating the yield distribution including all astrophysical noise sources is critical, as it includes shifts in the expectation value of the yield due to observational biases induced from exoplanet albedo and exozodi uncertainties. However, in practice, calculating a yield distribution is numerically taxing, as it requires hundreds of independent yield calculations to sample the range of possible exozodi and $\eta_{\Earth}$ values; calculating yield distributions including all sources of astrophysical noise could slow trade studies. To aid with this, we ``translate" each of the filled points in Figure \ref{fig:conf_vs_yield} to a much simpler quantity that only requires a single yield calculation, a benchmark yield, $Y'$. We define $Y'$ as the yield assuming 3 zodis of dust around all stars, $\eta_{\Earth}=0.24$, and no albedo or exozodi observational biases included. The empty circles and triangles in Figure \ref{fig:conf_vs_yield}, along with the thin dashed line, show $Y'$ for each of the scenarios shown as filled symbols. By using the dashed curves in Figure \ref{fig:conf_vs_yield}, one can design a mission to achieve a given $P_{25}$ via a single yield calculation, knowing that when astrophysical uncertainties are included the mean yields will shift to the solid curves.

\subsection{Budgeting for performance uncertainty\label{section:performanceunc}}

Science margin can also help reduce risk by budgeting for performance uncertainties. There are unlimited potential causes of on-sky performance degradations, manifesting as, e.g., poor line of sight jitter, coating or detector degradations, unexpected stray light, etc. Here we do not focus on root causes. Instead, we discuss performance degradation in terms of bulk parameters used in our yield analyses, such as raw contrast, throughput, etc. 

As originally shown by Ref.~\citenum{stark2014} and later verified by Ref.~\citenum{stark2019} with updated fidelity, EEC yield decreases relatively gracefully with degradations in most parameters. For the LUVOIR-B baseline parameters, yield is only moderately sensitive to throughput-related terms (scaling roughly to the $\sim$0.37 power\cite{stark2019}); a relatively large factor of two reduction in effective throughput would only reduce yield by $\sim$25\%, though it would increase exposure times by roughly a factor of two. Assuming a noise floor independent of raw contrast, LUVOIR-B was also very insensitive to raw contrast (scaling as raw contrast to the $\sim$-0.07 power\cite{stark2019}); a factor of two degradation in raw contrast would only reduce yield by $\sim$5\%. To first order, assuming the noise floor isn't coupled to raw contrast, degradations in these quantities slow the progress of observing the full target list, but don't limit the boundaries of accessible targets (indicated by the red dashed lines shown in Figure \ref{fig:target_list}). Because this can be partially mitigated by additional survey time, we don't consider these parameters as substantial drivers of performance risk. 

We posit that the astrophysical noise floor, $\Delta{\rm mag_{\rm floor}}$, is a primary driver of performance risk. The noise floor is ultimately determined by one of the largest technological ``tall poles:" the ability to precisely estimate the coronagraphic speckle pattern, which may require picometer stability\cite{luvoirfinalreport}, very precise wavefront sensing if model-based PSF subtraction is used, or a combination of the two. As shown by Fig.~8 in Ref.~\citenum{stark2015}, for $\Delta{\rm mag_{\rm floor}}=26.5$ as assumed in this study, degradations of a factor of two in the flux associated with the noise floor could lead to reductions in EEC yield of $\sim$25\%, on par with throughput factors. However, the impacts of the noise floor differ from throughput factors in two important ways. The first is that the yield's sensitivity to the noise floor appears to be a ``cliff" (c.f. Fig.~8 in Ref.~\citenum{stark2015}), suggesting that additional degradations beyond the initial factor of two become significantly more costly---i.e., it's a slippery slope. 

The second is more fundamental: degradations in the noise floor directly limit the range of accessible targets. The horizontal dashed lines in Fig.~\ref{fig:target_list} mark the luminosity at which a 1.4 $R_{\Earth}$ planet at quadrature at the EEID has a flux equal to the astrophysical noise floor. This is a rough visual guide. In reality, because planets could occupy a range of phase angles, semi-major axes, and radii, the effects of the noise floor can be seen in Fig.~\ref{fig:target_list} as a broad horizontal band that reduces the completeness of targets with luminosities $\gtrsim$2 $L_{\odot}$. Therefore, reductions in the noise floor will move this band to lower luminosities, directly removing targets that cannot be accessed in any other way. Further, reductions in the noise floor would begin to remove the most Sun-like stars, G type stars, from the target list, substantially affecting the mission's ability to survey for Earth-like planets around Sun-like stars.

One approach to mitigating the risk of a degradation in the noise floor is to adopt technologies that relax the system level requirements needed to achieve the desired noise floor. Any technology that relaxes optical stability requirements would aid in this. Of particular note is the model-based PSF subtraction method we considered in Scenario C. Model-based PSF subtraction monitors the WFE at a high cadence so that we can reconstruct the instantaneous science PSF in high fidelity. This would allow us to discard photons collected at times with poor WFE, or at least accurately model how the science PSF varies with WFE to a level better than the Poisson noise. It also would remove the need to maintain an ultrastable WF in the presence of a spacecraft slew or roll, which would be required for the RDI and ADI PSF subtraction methods, respectively. Much work is needed to determine if model-based PSF subtraction is viable for HWO, but the benefit could be significant.

Another approach to mitigating the risk of noise floor degradation is to budget for it with science margin. If degrading the noise floor moves the horizontal dashed lines in Fig.~\ref{fig:target_list} downward, we could maintain our access to a large pool of stars by shifting the curved dashed line downward as well. In other words, the target list would shift toward later type stars. The curved dashed line in Fig.~\ref{fig:target_list} marks where a working angle of 1.5 $\lambda/D$ is at the outer edge of the HZ. This working angle is admittedly fairly extreme already, and it is unlikely that alternative coronagraph designs could reduce it further in units of $\lambda/D$. Additionally, as discussed in Section \ref{section:exozodi_distribution}, there is likely a minimum ``useful" working angle set by spatial resolution requirements interior to which planets blend together too often\cite{saxena2022}---this may be larger than the 1.5$\lambda/D$ shown in Fig.~\ref{fig:target_list}. There are therefore only two ways of shifting the curved dashed line downward: by operating at shorter $\lambda$ or larger $D$. Ref.~\citenum{stark2023} showed that detecting water at shorter $\lambda$ is possible, but not preferable, as it requires significantly longer exposure times. Thus, we conclude that an increase in telescope diameter should be considered as an option to budget for noise floor degradation.

\section{Conclusions}

We identified and estimated the impact of all major sources of astrophysical uncertainty on the exoEarth candidate yield from a blind exoEarth survey by the Habitable Worlds Observatory, where ``yield" was defined as the detection and search for water vapor on all EECs. We find that while $\eta_{\Earth}$ uncertainties dominate the uncertainty in EEC yield, the sampling uncertainties inherent to a blind exoplanet survey are another important source of uncertainty and should be accounted for in mission design. We caution against adopting a science goal of 100 cumulative HZs, which is not equivalent to ignoring only the uncertainty in $\eta_{\Earth}$; such a science goal would effectively ignore \emph{both} dominant sources of astrophysical uncertainty. 

We find that exoplanet albedo uncertainty and exozodi sampling uncertainties shift the expectation value of the yield to lower values. The effect of the uncertainty in the exozodi distribution is less clear. We performed a re-analysis of fits to the LBTI exozodi observations. We find that assuming the exozodi distribution is multi-modal, the uncertainty in the exozodi distribution appears to have a relatively minor impact on yield uncertainty. However, this is only true if the fraction of the exozodi distribution $\lesssim 3$ zodis, which is a better predictor of exoplanet science yield than the median exozodi level, is well-constrained. Details of our ability to precisely subtract exozodi, which we do not address here, may still have a large impact on mission yield as it sets a systematic noise floor for the mission.

By including all astrophysical uncertainties, we estimated the yield distribution for a given mission design scenario and calculated the fraction of the distribution $>$25 EECs, defined as $P_{25}$. Adopting the LUVOIR-B baseline design, we find that an $\sim$8-9 m ID telescope is needed to produce large science margins, resulting in $P_{25}\sim$95\% when ignoring uncertainties in $\eta_{\Earth}$ and $P_{25}\sim$75\% when including $\eta_{\Earth}$ uncertainties. We identified six possible design changes from the LUVOIR-B baseline, each of which focuses on reducing exposure times and provides a modest gain in yield on its own. However, when combined, \emph{these improvements compile to provide significant performance gains}, nearly tripling EEC yield and reducing spectral characterization times by more than an order of magnitude for the highest priority targets. We find that with these changes it is possible for a $\sim$6 m ID telescope to produce substantial science margins, providing $P_{25}>$99\% when ignoring uncertainties in $\eta_{\Earth}$ and $P_{25}\sim$85\% when including $\eta_{\Earth}$ uncertainties. We conclude that a combination of telescope diameter increase and instrument design changes could provide robust exoplanet science margins for HWO. 

We discussed how science margin can help mitigate on-sky performance degradations. Whereas degradation of contrast and throughput lengthen exposure times, we showed that degradation of the noise floor can lead to a large fraction of the target list being permanently unobservable. We identified increasing telescope diameter as a promising path to reducing the risk associated with noise floor degradation.

\label{conclusions}

\section{Data and Code Availability}
NASA regulations govern the release of source code, including what can be released and how it is made available. Readers should contact the corresponding author if they would like copies of the visualization software or data produced for this study.

\acknowledgments

This project was supported by the NASA HQ-directed ExoSpec work package under the Internal Scientist Funding Model (ISFM). N.~L. gratefully acknowledges financial support from an NSF GRFP. N.~W.~T. is supported by an appointment with the NASA Postdoctoral Program at the NASA Goddard Space Flight Center, administered by Oak Ridge Associated Universities under contract with NASA. N.~W.~T. acknowledges support from the GSFC Sellers Exoplanet Environments Collaboration (SEEC), which is supported by NASA's Planetary, Astrophysics and Heliophysics Science Divisions’ Research Program. The Center for Exoplanets and Habitable Worlds and the Penn State Extraterrestrial Intelligence Center are supported by the Pennsylvania State University and the Eberly College of Science. E.~B.~F.'s contributions focused on Bayesian analysis of the distribution of exozodi levels and interpretation of $\eta_{\Earth}$ studies and did not extend to \S4. C.~S. acknowledges the contributions of two anonymous reviewers whose feedback substantially improved this manuscript.

\bibliographystyle{spiejour}   
\bibliography{ms_v2.bbl}

\begin{thebibliography}{10}

\bibitem{astro2020}
E.~{National Academies of Sciences} and Medicine, {\em {Pathways to Discovery
  in Astronomy and Astrophysics for the 2020s}}  (2021).

\bibitem{stark2019}
C.~C. {Stark}, R.~{Belikov}, M.~R. {Bolcar}, E.~{Cady}, B.~P. {Crill},
  S.~{Ertel}, T.~{Groff}, S.~{Hildebrandt}, J.~{Krist}, P.~D. {Lisman},
  J.~{Mazoyer}, B.~{Mennesson}, B.~{Nemati}, L.~{Pueyo}, B.~J. {Rauscher},
  A.~J. {Riggs}, G.~{Ruane}, S.~B. {Shaklan}, D.~{Sirbu}, R.~{Soummer}, K.~S.
  {Laurent}, and N.~{Zimmerman}, ``{ExoEarth yield landscape for future direct
  imaging space telescopes},'' {\em Journal of Astronomical Telescopes,
  Instruments, and Systems} {\bf 5}, 024009  (2019).

\bibitem{morgan2021a}
R.~{Morgan}, D.~{Savransky}, M.~{Turmon}, B.~{Mennesson}, W.~{Dula},
  D.~{Keithly}, E.~E. {Mamajek}, P.~{Newman}, P.~{Plavchan}, T.~D. {Robinson},
  G.~{Roudier}, and C.~{Stark}, ``{Faster Exo-Earth yield for HabEx and LUVOIR
  via extreme precision radial velocity prior knowledge},'' {\em Journal of
  Astronomical Telescopes, Instruments, and Systems} {\bf 7}, 021220  (2021).

\bibitem{savransky2016}
D.~{Savransky} and D.~{Garrett}, ``{WFIRST-AFTA coronagraph science yield
  modeling with EXOSIMS},'' {\em Journal of Astronomical Telescopes,
  Instruments, and Systems} {\bf 2}, 011006  (2016).

\bibitem{stark2014}
C.~C. {Stark}, A.~{Roberge}, A.~{Mandell}, and T.~D. {Robinson}, ``{Maximizing
  the ExoEarth Candidate Yield from a Future Direct Imaging Mission},'' {\em
  \apj} {\bf 795}, 122  (2014).

\bibitem{luvoirfinalreport}
{The LUVOIR Team}, ``{The LUVOIR Mission Concept Study Final Report},'' {\em
  arXiv e-prints} , arXiv:1912.06219  (2019).

\bibitem{habexfinalreport}
B.~S. {Gaudi}, S.~{Seager}, B.~{Mennesson}, A.~{Kiessling}, K.~{Warfield},
  K.~{Cahoy}, J.~T. {Clarke}, S.~{Domagal-Goldman}, L.~{Feinberg}, O.~{Guyon},
  J.~{Kasdin}, D.~{Mawet}, P.~{Plavchan}, T.~{Robinson}, L.~{Rogers},
  P.~{Scowen}, R.~{Somerville}, K.~{Stapelfeldt}, C.~{Stark}, D.~{Stern},
  M.~{Turnbull}, R.~{Amini}, G.~{Kuan}, S.~{Martin}, R.~{Morgan}, D.~{Redding},
  H.~P. {Stahl}, R.~{Webb}, O.~{Alvarez-Salazar}, W.~L. {Arnold}, M.~{Arya},
  B.~{Balasubramanian}, M.~{Baysinger}, R.~{Bell}, C.~{Below}, J.~{Benson},
  L.~{Blais}, J.~{Booth}, R.~{Bourgeois}, C.~{Bradford}, A.~{Brewer},
  T.~{Brooks}, E.~{Cady}, M.~{Caldwell}, R.~{Calvet}, S.~{Carr}, D.~{Chan},
  V.~{Cormarkovic}, K.~{Coste}, C.~{Cox}, R.~{Danner}, J.~{Davis}, L.~{Dewell},
  L.~{Dorsett}, D.~{Dunn}, M.~{East}, M.~{Effinger}, R.~{Eng}, G.~{Freebury},
  J.~{Garcia}, J.~{Gaskin}, S.~{Greene}, J.~{Hennessy}, E.~{Hilgemann},
  B.~{Hood}, W.~{Holota}, S.~{Howe}, P.~{Huang}, T.~{Hull}, R.~{Hunt},
  K.~{Hurd}, S.~{Johnson}, A.~{Kissil}, B.~{Knight}, D.~{Kolenz}, O.~{Kraus},
  J.~{Krist}, M.~{Li}, D.~{Lisman}, M.~{Mandic}, J.~{Mann}, L.~{Marchen},
  C.~{Marrese-Reading}, J.~{McCready}, J.~{McGown}, J.~{Missun},
  A.~{Miyaguchi}, B.~{Moore}, B.~{Nemati}, S.~{Nikzad}, J.~{Nissen},
  M.~{Novicki}, T.~{Perrine}, C.~{Pineda}, O.~{Polanco}, D.~{Putnam},
  A.~{Qureshi}, M.~{Richards}, A.~J. {Eldorado Riggs}, M.~{Rodgers}, M.~{Rud},
  N.~{Saini}, D.~{Scalisi}, D.~{Scharf}, K.~{Schulz}, G.~{Serabyn},
  N.~{Sigrist}, G.~{Sikkia}, A.~{Singleton}, S.~{Shaklan}, S.~{Smith},
  B.~{Southerd}, M.~{Stahl}, J.~{Steeves}, B.~{Sturges}, C.~{Sullivan},
  H.~{Tang}, N.~{Taras}, J.~{Tesch}, M.~{Therrell}, H.~{Tseng}, M.~{Valente},
  D.~{Van Buren}, J.~{Villalvazo}, S.~{Warwick}, D.~{Webb}, T.~{Westerhoff},
  R.~{Wofford}, G.~{Wu}, J.~{Woo}, M.~{Wood}, J.~{Ziemer}, G.~{Arney},
  J.~{Anderson}, J.~{Ma{\'\i}z-Apell{\'a}niz}, J.~{Bartlett}, R.~{Belikov},
  E.~{Bendek}, B.~{Cenko}, E.~{Douglas}, S.~{Dulz}, C.~{Evans}, V.~{Faramaz},
  Y.~K. {Feng}, H.~{Ferguson}, K.~{Follette}, S.~{Ford}, M.~{Garc{\'\i}a},
  M.~{Geha}, D.~{Gelino}, Y.~{G{\"o}tberg}, S.~{Hildebrandt}, R.~{Hu},
  K.~{Jahnke}, G.~{Kennedy}, L.~{Kreidberg}, A.~{Isella}, E.~{Lopez},
  F.~{Marchis}, L.~{Macri}, M.~{Marley}, W.~{Matzko}, J.~{Mazoyer},
  S.~{McCandliss}, T.~{Meshkat}, C.~{Mordasini}, P.~{Morris}, E.~{Nielsen},
  P.~{Newman}, E.~{Petigura}, M.~{Postman}, A.~{Reines}, A.~{Roberge},
  I.~{Roederer}, G.~{Ruane}, E.~{Schwieterman}, D.~{Sirbu}, C.~{Spalding},
  H.~{Teplitz}, J.~{Tumlinson}, N.~{Turner}, J.~{Werk}, A.~{Wofford},
  M.~{Wyatt}, A.~{Young}, and R.~{Zellem}, ``{The Habitable Exoplanet
  Observatory (HabEx) Mission Concept Study Final Report},'' {\em arXiv
  e-prints} , arXiv:2001.06683  (2020).

\bibitem{stark2023}
C.~C. {Stark}, N.~{Latouf}, A.~M. {Mandell}, and A.~{Young}, ``{Optimized
  Bandpasses for the Habitable Worlds Observatory’s ExoEarth Survey}.''
  (accepted).

\bibitem{kopparapu2018}
R.~K. {Kopparapu}, E.~{H{\'e}brard}, R.~{Belikov}, N.~M. {Batalha}, G.~D.
  {Mulders}, C.~{Stark}, D.~{Teal}, S.~{Domagal-Goldman}, and A.~{Mandell},
  ``{Exoplanet Classification and Yield Estimates for Direct Imaging
  Missions},'' {\em \apj} {\bf 856}, 122  (2018).

\bibitem{howe2024}
A.~R. {Howe}, C.~C. {Stark}, and J.~E. {Sadleir}, ``{Scientific impact of a
  noiseless energy-resolving detector for a future exoplanet-imaging
  mission},'' {\em Journal of Astronomical Telescopes, Instruments, and
  Systems} {\bf 10}(2), 025008  (2024).

\bibitem{stark2015}
C.~C. {Stark}, A.~{Roberge}, A.~{Mandell}, M.~{Clampin}, S.~D.
  {Domagal-Goldman}, M.~W. {McElwain}, and K.~R. {Stapelfeldt}, ``{Lower Limits
  on Aperture Size for an ExoEarth Detecting Coronagraphic Mission},'' {\em
  \apj} {\bf 808}, 149  (2015).

\bibitem{hunyadi2007}
S.~L. {Hunyadi}, S.~B. {Shaklan}, and R.~A. {Brown}, ``{The lighter side of
  TPF-C: evaluating the scientific gain from a smaller mission concept},'' in
  {\em Techniques and Instrumentation for Detection of Exoplanets III},  D.~R.
  {Coulter}, Ed., {\em Society of Photo-Optical Instrumentation Engineers
  (SPIE) Conference Series} {\bf 6693}, 66930Q  (2007).

\bibitem{tuchow2024}
N.~W. {Tuchow}, C.~C. {Stark}, and E.~{Mamajek}, ``{HPIC: The Habitable Worlds
  Observatory Preliminary Input Catalog},'' {\em \aj} {\bf 167}, 139  (2024).

\bibitem{stassun2019}
K.~G. {Stassun}, R.~J. {Oelkers}, M.~{Paegert}, G.~{Torres}, J.~{Pepper},
  N.~{De Lee}, K.~{Collins}, D.~W. {Latham}, P.~S. {Muirhead}, J.~{Chittidi},
  B.~{Rojas-Ayala}, S.~W. {Fleming}, M.~E. {Rose}, P.~{Tenenbaum}, E.~B.
  {Ting}, S.~R. {Kane}, T.~{Barclay}, J.~L. {Bean}, C.~E. {Brassuer},
  D.~{Charbonneau}, J.~{Ge}, J.~J. {Lissauer}, A.~W. {Mann}, B.~{McLean},
  S.~{Mullally}, N.~{Narita}, P.~{Plavchan}, G.~R. {Ricker}, D.~{Sasselov},
  S.~{Seager}, S.~{Sharma}, B.~{Shiao}, A.~{Sozzetti}, D.~{Stello},
  R.~{Vanderspek}, G.~{Wallace}, and J.~N. {Winn}, ``{The Revised TESS Input
  Catalog and Candidate Target List},'' {\em \aj} {\bf 158}, 138  (2019).

\bibitem{gaiadr3}
{Gaia Collaboration}, A.~{Vallenari}, A.~G.~A. {Brown}, T.~{Prusti}, J.~H.~J.
  {de Bruijne}, F.~{Arenou}, C.~{Babusiaux}, M.~{Biermann}, O.~L. {Creevey},
  C.~{Ducourant}, D.~W. {Evans}, L.~{Eyer}, R.~{Guerra}, A.~{Hutton},
  C.~{Jordi}, S.~A. {Klioner}, U.~L. {Lammers}, L.~{Lindegren}, X.~{Luri},
  F.~{Mignard}, C.~{Panem}, D.~{Pourbaix}, S.~{Randich}, P.~{Sartoretti},
  C.~{Soubiran}, P.~{Tanga}, N.~A. {Walton}, C.~A.~L. {Bailer-Jones},
  U.~{Bastian}, R.~{Drimmel}, F.~{Jansen}, D.~{Katz}, M.~G. {Lattanzi}, F.~{van
  Leeuwen}, J.~{Bakker}, C.~{Cacciari}, J.~{Casta{\~n}eda}, F.~{De Angeli},
  C.~{Fabricius}, M.~{Fouesneau}, Y.~{Fr{\'e}mat}, L.~{Galluccio},
  A.~{Guerrier}, U.~{Heiter}, E.~{Masana}, R.~{Messineo}, N.~{Mowlavi},
  C.~{Nicolas}, K.~{Nienartowicz}, F.~{Pailler}, P.~{Panuzzo}, F.~{Riclet},
  W.~{Roux}, G.~M. {Seabroke}, R.~{Sordo}, F.~{Th{\'e}venin},
  G.~{Gracia-Abril}, J.~{Portell}, D.~{Teyssier}, M.~{Altmann}, R.~{Andrae},
  M.~{Audard}, I.~{Bellas-Velidis}, K.~{Benson}, J.~{Berthier}, R.~{Blomme},
  P.~W. {Burgess}, D.~{Busonero}, G.~{Busso}, H.~{C{\'a}novas}, B.~{Carry},
  A.~{Cellino}, N.~{Cheek}, G.~{Clementini}, Y.~{Damerdji}, M.~{Davidson},
  P.~{de Teodoro}, M.~{Nu{\~n}ez Campos}, L.~{Delchambre}, A.~{Dell'Oro},
  P.~{Esquej}, J.~{Fern{\'a}ndez-Hern{\'a}ndez}, E.~{Fraile}, D.~{Garabato},
  P.~{Garc{\'\i}a-Lario}, E.~{Gosset}, R.~{Haigron}, J.~L. {Halbwachs}, N.~C.
  {Hambly}, D.~L. {Harrison}, J.~{Hern{\'a}ndez}, D.~{Hestroffer}, S.~T.
  {Hodgkin}, B.~{Holl}, K.~{Jan{\ss}en}, G.~{Jevardat de Fombelle},
  S.~{Jordan}, A.~{Krone-Martins}, A.~C. {Lanzafame}, W.~{L{\"o}ffler},
  O.~{Marchal}, P.~M. {Marrese}, A.~{Moitinho}, K.~{Muinonen}, P.~{Osborne},
  E.~{Pancino}, T.~{Pauwels}, A.~{Recio-Blanco}, C.~{Reyl{\'e}}, M.~{Riello},
  L.~{Rimoldini}, T.~{Roegiers}, J.~{Rybizki}, L.~M. {Sarro}, C.~{Siopis},
  M.~{Smith}, A.~{Sozzetti}, E.~{Utrilla}, M.~{van Leeuwen}, U.~{Abbas},
  P.~{{\'A}brah{\'a}m}, A.~{Abreu Aramburu}, C.~{Aerts}, J.~J. {Aguado},
  M.~{Ajaj}, F.~{Aldea-Montero}, G.~{Altavilla}, M.~A. {{\'A}lvarez},
  J.~{Alves}, F.~{Anders}, R.~I. {Anderson}, E.~{Anglada Varela}, T.~{Antoja},
  D.~{Baines}, S.~G. {Baker}, L.~{Balaguer-N{\'u}{\~n}ez}, E.~{Balbinot},
  Z.~{Balog}, C.~{Barache}, D.~{Barbato}, M.~{Barros}, M.~A. {Barstow},
  S.~{Bartolom{\'e}}, J.~L. {Bassilana}, N.~{Bauchet}, U.~{Becciani},
  M.~{Bellazzini}, A.~{Berihuete}, M.~{Bernet}, S.~{Bertone}, L.~{Bianchi},
  A.~{Binnenfeld}, S.~{Blanco-Cuaresma}, A.~{Blazere}, T.~{Boch}, A.~{Bombrun},
  D.~{Bossini}, S.~{Bouquillon}, A.~{Bragaglia}, L.~{Bramante}, E.~{Breedt},
  A.~{Bressan}, N.~{Brouillet}, E.~{Brugaletta}, B.~{Bucciarelli},
  A.~{Burlacu}, A.~G. {Butkevich}, R.~{Buzzi}, E.~{Caffau}, R.~{Cancelliere},
  T.~{Cantat-Gaudin}, R.~{Carballo}, T.~{Carlucci}, M.~I. {Carnerero}, J.~M.
  {Carrasco}, L.~{Casamiquela}, M.~{Castellani}, A.~{Castro-Ginard},
  L.~{Chaoul}, P.~{Charlot}, L.~{Chemin}, V.~{Chiaramida}, A.~{Chiavassa},
  N.~{Chornay}, G.~{Comoretto}, G.~{Contursi}, W.~J. {Cooper}, T.~{Cornez},
  S.~{Cowell}, F.~{Crifo}, M.~{Cropper}, M.~{Crosta}, C.~{Crowley},
  C.~{Dafonte}, A.~{Dapergolas}, M.~{David}, P.~{David}, P.~{de Laverny},
  F.~{De Luise}, R.~{De March}, J.~{De Ridder}, R.~{de Souza}, A.~{de Torres},
  E.~F. {del Peloso}, E.~{del Pozo}, M.~{Delbo}, A.~{Delgado}, J.~B. {Delisle},
  C.~{Demouchy}, T.~E. {Dharmawardena}, P.~{Di Matteo}, S.~{Diakite},
  C.~{Diener}, E.~{Distefano}, C.~{Dolding}, B.~{Edvardsson}, H.~{Enke},
  C.~{Fabre}, M.~{Fabrizio}, S.~{Faigler}, G.~{Fedorets}, P.~{Fernique},
  A.~{Fienga}, F.~{Figueras}, Y.~{Fournier}, C.~{Fouron}, F.~{Fragkoudi},
  M.~{Gai}, A.~{Garcia-Gutierrez}, M.~{Garcia-Reinaldos},
  M.~{Garc{\'\i}a-Torres}, A.~{Garofalo}, A.~{Gavel}, P.~{Gavras},
  E.~{Gerlach}, R.~{Geyer}, P.~{Giacobbe}, G.~{Gilmore}, S.~{Girona},
  G.~{Giuffrida}, R.~{Gomel}, A.~{Gomez}, J.~{Gonz{\'a}lez-N{\'u}{\~n}ez},
  I.~{Gonz{\'a}lez-Santamar{\'\i}a}, J.~J. {Gonz{\'a}lez-Vidal}, M.~{Granvik},
  P.~{Guillout}, J.~{Guiraud}, R.~{Guti{\'e}rrez-S{\'a}nchez}, L.~P. {Guy},
  D.~{Hatzidimitriou}, M.~{Hauser}, M.~{Haywood}, A.~{Helmer}, A.~{Helmi},
  M.~H. {Sarmiento}, S.~L. {Hidalgo}, T.~{Hilger}, N.~{H{\l}adczuk},
  D.~{Hobbs}, G.~{Holland}, H.~E. {Huckle}, K.~{Jardine}, G.~{Jasniewicz},
  A.~{Jean-Antoine Piccolo}, {\'O}.~{Jim{\'e}nez-Arranz}, A.~{Jorissen},
  J.~{Juaristi Campillo}, F.~{Julbe}, L.~{Karbevska}, P.~{Kervella},
  S.~{Khanna}, M.~{Kontizas}, G.~{Kordopatis}, A.~J. {Korn},
  {\'A}.~{K{\'o}sp{\'a}l}, Z.~{Kostrzewa-Rutkowska}, K.~{Kruszy{\'n}ska},
  M.~{Kun}, P.~{Laizeau}, S.~{Lambert}, A.~F. {Lanza}, Y.~{Lasne}, J.~F. {Le
  Campion}, Y.~{Lebreton}, T.~{Lebzelter}, S.~{Leccia}, N.~{Leclerc},
  I.~{Lecoeur-Taibi}, S.~{Liao}, E.~L. {Licata}, H.~E.~P. {Lindstr{\o}m}, T.~A.
  {Lister}, E.~{Livanou}, A.~{Lobel}, A.~{Lorca}, C.~{Loup}, P.~{Madrero
  Pardo}, A.~{Magdaleno Romeo}, S.~{Managau}, R.~G. {Mann}, M.~{Manteiga},
  J.~M. {Marchant}, M.~{Marconi}, J.~{Marcos}, M.~M.~S. {Marcos Santos},
  D.~{Mar{\'\i}n Pina}, S.~{Marinoni}, F.~{Marocco}, D.~J. {Marshall},
  L.~{Martin Polo}, J.~M. {Mart{\'\i}n-Fleitas}, G.~{Marton}, N.~{Mary},
  A.~{Masip}, D.~{Massari}, A.~{Mastrobuono-Battisti}, T.~{Mazeh}, P.~J.
  {McMillan}, S.~{Messina}, D.~{Michalik}, N.~R. {Millar}, A.~{Mints},
  D.~{Molina}, R.~{Molinaro}, L.~{Moln{\'a}r}, G.~{Monari}, M.~{Mongui{\'o}},
  P.~{Montegriffo}, A.~{Montero}, R.~{Mor}, A.~{Mora}, R.~{Morbidelli},
  T.~{Morel}, D.~{Morris}, T.~{Muraveva}, C.~P. {Murphy}, I.~{Musella},
  Z.~{Nagy}, L.~{Noval}, F.~{Oca{\~n}a}, A.~{Ogden}, C.~{Ordenovic}, J.~O.
  {Osinde}, C.~{Pagani}, I.~{Pagano}, L.~{Palaversa}, P.~A. {Palicio},
  L.~{Pallas-Quintela}, A.~{Panahi}, S.~{Payne-Wardenaar}, X.~{Pe{\~n}alosa
  Esteller}, A.~{Penttil{\"a}}, B.~{Pichon}, A.~M. {Piersimoni}, F.~X.
  {Pineau}, E.~{Plachy}, G.~{Plum}, E.~{Poggio}, A.~{Pr{\v{s}}a}, L.~{Pulone},
  E.~{Racero}, S.~{Ragaini}, M.~{Rainer}, C.~M. {Raiteri}, N.~{Rambaux},
  P.~{Ramos}, M.~{Ramos-Lerate}, P.~{Re Fiorentin}, S.~{Regibo}, P.~J.
  {Richards}, C.~{Rios Diaz}, V.~{Ripepi}, A.~{Riva}, H.~W. {Rix}, G.~{Rixon},
  N.~{Robichon}, A.~C. {Robin}, C.~{Robin}, M.~{Roelens}, H.~R.~O. {Rogues},
  L.~{Rohrbasser}, M.~{Romero-G{\'o}mez}, N.~{Rowell}, F.~{Royer}, D.~{Ruz
  Mieres}, K.~A. {Rybicki}, G.~{Sadowski}, A.~{S{\'a}ez N{\'u}{\~n}ez},
  A.~{Sagrist{\`a} Sell{\'e}s}, J.~{Sahlmann}, E.~{Salguero}, N.~{Samaras},
  V.~{Sanchez Gimenez}, N.~{Sanna}, R.~{Santove{\~n}a}, M.~{Sarasso},
  M.~{Schultheis}, E.~{Sciacca}, M.~{Segol}, J.~C. {Segovia},
  D.~{S{\'e}gransan}, D.~{Semeux}, S.~{Shahaf}, H.~I. {Siddiqui}, A.~{Siebert},
  L.~{Siltala}, A.~{Silvelo}, E.~{Slezak}, I.~{Slezak}, R.~L. {Smart}, O.~N.
  {Snaith}, E.~{Solano}, F.~{Solitro}, D.~{Souami}, J.~{Souchay}, A.~{Spagna},
  L.~{Spina}, F.~{Spoto}, I.~A. {Steele}, H.~{Steidelm{\"u}ller}, C.~A.
  {Stephenson}, M.~{S{\"u}veges}, J.~{Surdej}, L.~{Szabados},
  E.~{Szegedi-Elek}, F.~{Taris}, M.~B. {Taylor}, R.~{Teixeira}, L.~{Tolomei},
  N.~{Tonello}, F.~{Torra}, J.~{Torra}, G.~{Torralba Elipe}, M.~{Trabucchi},
  A.~T. {Tsounis}, C.~{Turon}, A.~{Ulla}, N.~{Unger}, M.~V. {Vaillant}, E.~{van
  Dillen}, W.~{van Reeven}, O.~{Vanel}, A.~{Vecchiato}, Y.~{Viala},
  D.~{Vicente}, S.~{Voutsinas}, M.~{Weiler}, T.~{Wevers}, {\L}.~{Wyrzykowski},
  A.~{Yoldas}, P.~{Yvard}, H.~{Zhao}, J.~{Zorec}, S.~{Zucker}, and
  T.~{Zwitter}, ``{Gaia Data Release 3. Summary of the content and survey
  properties},'' {\em \aap} {\bf 674}, A1  (2023).

\bibitem{gaia2021}
{Gaia Collaboration}, R.~L. {Smart}, L.~M. {Sarro}, J.~{Rybizki},
  C.~{Reyl{\'e}}, A.~C. {Robin}, N.~C. {Hambly}, U.~{Abbas}, M.~A. {Barstow},
  J.~H.~J. {de Bruijne}, B.~{Bucciarelli}, J.~M. {Carrasco}, W.~J. {Cooper},
  S.~T. {Hodgkin}, E.~{Masana}, D.~{Michalik}, J.~{Sahlmann}, A.~{Sozzetti},
  A.~G.~A. {Brown}, A.~{Vallenari}, T.~{Prusti}, C.~{Babusiaux}, M.~{Biermann},
  O.~L. {Creevey}, D.~W. {Evans}, L.~{Eyer}, A.~{Hutton}, F.~{Jansen},
  C.~{Jordi}, S.~A. {Klioner}, U.~{Lammers}, L.~{Lindegren}, X.~{Luri},
  F.~{Mignard}, C.~{Panem}, D.~{Pourbaix}, S.~{Randich}, P.~{Sartoretti},
  C.~{Soubiran}, N.~A. {Walton}, F.~{Arenou}, C.~A.~L. {Bailer-Jones},
  U.~{Bastian}, M.~{Cropper}, R.~{Drimmel}, D.~{Katz}, M.~G. {Lattanzi},
  F.~{van Leeuwen}, J.~{Bakker}, J.~{Casta{\~n}eda}, F.~{De Angeli},
  C.~{Ducourant}, C.~{Fabricius}, M.~{Fouesneau}, Y.~{Fr{\'e}mat}, R.~{Guerra},
  A.~{Guerrier}, J.~{Guiraud}, A.~{Jean-Antoine Piccolo}, R.~{Messineo},
  N.~{Mowlavi}, C.~{Nicolas}, K.~{Nienartowicz}, F.~{Pailler}, P.~{Panuzzo},
  F.~{Riclet}, W.~{Roux}, G.~M. {Seabroke}, R.~{Sordo}, P.~{Tanga},
  F.~{Th{\'e}venin}, G.~{Gracia-Abril}, J.~{Portell}, D.~{Teyssier},
  M.~{Altmann}, R.~{Andrae}, I.~{Bellas-Velidis}, K.~{Benson}, J.~{Berthier},
  R.~{Blomme}, E.~{Brugaletta}, P.~W. {Burgess}, G.~{Busso}, B.~{Carry},
  A.~{Cellino}, N.~{Cheek}, G.~{Clementini}, Y.~{Damerdji}, M.~{Davidson},
  L.~{Delchambre}, A.~{Dell'Oro}, J.~{Fern{\'a}ndez-Hern{\'a}ndez},
  L.~{Galluccio}, P.~{Garc{\'\i}a-Lario}, M.~{Garcia-Reinaldos},
  J.~{Gonz{\'a}lez-N{\'u}{\~n}ez}, E.~{Gosset}, R.~{Haigron}, J.~L.
  {Halbwachs}, D.~L. {Harrison}, D.~{Hatzidimitriou}, U.~{Heiter},
  J.~{Hern{\'a}ndez}, D.~{Hestroffer}, B.~{Holl}, K.~{Jan{\ss}en}, G.~{Jevardat
  de Fombelle}, S.~{Jordan}, A.~{Krone-Martins}, A.~C. {Lanzafame},
  W.~{L{\"o}ffler}, A.~{Lorca}, M.~{Manteiga}, O.~{Marchal}, P.~M. {Marrese},
  A.~{Moitinho}, A.~{Mora}, K.~{Muinonen}, P.~{Osborne}, E.~{Pancino},
  T.~{Pauwels}, A.~{Recio-Blanco}, P.~J. {Richards}, M.~{Riello},
  L.~{Rimoldini}, T.~{Roegiers}, C.~{Siopis}, M.~{Smith}, A.~{Ulla},
  E.~{Utrilla}, M.~{van Leeuwen}, W.~{van Reeven}, A.~{Abreu Aramburu},
  S.~{Accart}, C.~{Aerts}, J.~J. {Aguado}, M.~{Ajaj}, G.~{Altavilla}, M.~A.
  {{\'A}lvarez}, J.~{{\'A}lvarez Cid-Fuentes}, J.~{Alves}, R.~I. {Anderson},
  E.~{Anglada Varela}, T.~{Antoja}, M.~{Audard}, D.~{Baines}, S.~G. {Baker},
  L.~{Balaguer-N{\'u}{\~n}ez}, E.~{Balbinot}, Z.~{Balog}, C.~{Barache},
  D.~{Barbato}, M.~{Barros}, S.~{Bartolom{\'e}}, J.~L. {Bassilana},
  N.~{Bauchet}, A.~{Baudesson-Stella}, U.~{Becciani}, M.~{Bellazzini},
  M.~{Bernet}, S.~{Bertone}, L.~{Bianchi}, S.~{Blanco-Cuaresma}, T.~{Boch},
  A.~{Bombrun}, D.~{Bossini}, S.~{Bouquillon}, A.~{Bragaglia}, L.~{Bramante},
  E.~{Breedt}, A.~{Bressan}, N.~{Brouillet}, A.~{Burlacu}, D.~{Busonero}, A.~G.
  {Butkevich}, R.~{Buzzi}, E.~{Caffau}, R.~{Cancelliere}, H.~{C{\'a}novas},
  T.~{Cantat-Gaudin}, R.~{Carballo}, T.~{Carlucci}, M.~I. {Carnerero},
  L.~{Casamiquela}, M.~{Castellani}, A.~{Castro-Ginard}, P.~{Castro Sampol},
  L.~{Chaoul}, P.~{Charlot}, L.~{Chemin}, A.~{Chiavassa}, M.~R.~L. {Cioni},
  G.~{Comoretto}, T.~{Cornez}, S.~{Cowell}, F.~{Crifo}, M.~{Crosta},
  C.~{Crowley}, C.~{Dafonte}, A.~{Dapergolas}, M.~{David}, P.~{David}, P.~{de
  Laverny}, F.~{De Luise}, R.~{De March}, J.~{De Ridder}, R.~{de Souza}, P.~{de
  Teodoro}, A.~{de Torres}, E.~F. {del Peloso}, E.~{del Pozo}, A.~{Delgado},
  H.~E. {Delgado}, J.~B. {Delisle}, P.~{Di Matteo}, S.~{Diakite}, C.~{Diener},
  E.~{Distefano}, C.~{Dolding}, D.~{Eappachen}, B.~{Edvardsson}, H.~{Enke},
  P.~{Esquej}, C.~{Fabre}, M.~{Fabrizio}, S.~{Faigler}, G.~{Fedorets},
  P.~{Fernique}, A.~{Fienga}, F.~{Figueras}, C.~{Fouron}, F.~{Fragkoudi},
  E.~{Fraile}, F.~{Franke}, M.~{Gai}, D.~{Garabato}, A.~{Garcia-Gutierrez},
  M.~{Garc{\'\i}a-Torres}, A.~{Garofalo}, P.~{Gavras}, E.~{Gerlach},
  R.~{Geyer}, P.~{Giacobbe}, G.~{Gilmore}, S.~{Girona}, G.~{Giuffrida},
  R.~{Gomel}, A.~{Gomez}, I.~{Gonzalez-Santamaria}, J.~J. {Gonz{\'a}lez-Vidal},
  M.~{Granvik}, R.~{Guti{\'e}rrez-S{\'a}nchez}, L.~P. {Guy}, M.~{Hauser},
  M.~{Haywood}, A.~{Helmi}, S.~L. {Hidalgo}, T.~{Hilger}, N.~{H{\l}adczuk},
  D.~{Hobbs}, G.~{Holland}, H.~E. {Huckle}, G.~{Jasniewicz}, P.~G. {Jonker},
  J.~{Juaristi Campillo}, F.~{Julbe}, L.~{Karbevska}, P.~{Kervella},
  S.~{Khanna}, A.~{Kochoska}, M.~{Kontizas}, G.~{Kordopatis}, A.~J. {Korn},
  Z.~{Kostrzewa-Rutkowska}, K.~{Kruszy{\'n}ska}, S.~{Lambert}, A.~F. {Lanza},
  Y.~{Lasne}, J.~F. {Le Campion}, Y.~{Le Fustec}, Y.~{Lebreton},
  T.~{Lebzelter}, S.~{Leccia}, N.~{Leclerc}, I.~{Lecoeur-Taibi}, S.~{Liao},
  E.~{Licata}, H.~E.~P. {Lindstr{\o}m}, T.~A. {Lister}, E.~{Livanou},
  A.~{Lobel}, P.~{Madrero Pardo}, S.~{Managau}, R.~G. {Mann}, J.~M. {Marchant},
  M.~{Marconi}, M.~M.~S. {Marcos Santos}, S.~{Marinoni}, F.~{Marocco}, D.~J.
  {Marshall}, L.~{Martin Polo}, J.~M. {Mart{\'\i}n-Fleitas}, A.~{Masip},
  D.~{Massari}, A.~{Mastrobuono-Battisti}, T.~{Mazeh}, P.~J. {McMillan},
  S.~{Messina}, N.~R. {Millar}, A.~{Mints}, D.~{Molina}, R.~{Molinaro},
  L.~{Moln{\'a}r}, P.~{Montegriffo}, R.~{Mor}, R.~{Morbidelli}, T.~{Morel},
  D.~{Morris}, A.~F. {Mulone}, D.~{Munoz}, T.~{Muraveva}, C.~P. {Murphy},
  I.~{Musella}, L.~{Noval}, C.~{Ord{\'e}novic}, G.~{Orr{\`u}}, J.~{Osinde},
  C.~{Pagani}, I.~{Pagano}, L.~{Palaversa}, P.~A. {Palicio}, A.~{Panahi},
  M.~{Pawlak}, X.~{Pe{\~n}alosa Esteller}, A.~{Penttil{\"a}}, A.~M.
  {Piersimoni}, F.~X. {Pineau}, E.~{Plachy}, G.~{Plum}, E.~{Poggio},
  E.~{Poretti}, E.~{Poujoulet}, A.~{Pr{\v{s}}a}, L.~{Pulone}, E.~{Racero},
  S.~{Ragaini}, M.~{Rainer}, C.~M. {Raiteri}, N.~{Rambaux}, P.~{Ramos},
  M.~{Ramos-Lerate}, P.~{Re Fiorentin}, S.~{Regibo}, V.~{Ripepi}, A.~{Riva},
  G.~{Rixon}, N.~{Robichon}, C.~{Robin}, M.~{Roelens}, L.~{Rohrbasser},
  M.~{Romero-G{\'o}mez}, N.~{Rowell}, F.~{Royer}, K.~A. {Rybicki},
  G.~{Sadowski}, A.~{Sagrist{\`a} Sell{\'e}s}, J.~{Salgado}, E.~{Salguero},
  N.~{Samaras}, V.~{Sanchez Gimenez}, N.~{Sanna}, R.~{Santove{\~n}a},
  M.~{Sarasso}, M.~{Schultheis}, E.~{Sciacca}, M.~{Segol}, J.~C. {Segovia},
  D.~{S{\'e}gransan}, D.~{Semeux}, S.~{Shahaf}, H.~I. {Siddiqui}, A.~{Siebert},
  L.~{Siltala}, E.~{Slezak}, E.~{Solano}, F.~{Solitro}, D.~{Souami},
  J.~{Souchay}, A.~{Spagna}, F.~{Spoto}, I.~A. {Steele},
  H.~{Steidelm{\"u}ller}, C.~A. {Stephenson}, M.~{S{\"u}veges}, L.~{Szabados},
  E.~{Szegedi-Elek}, F.~{Taris}, G.~{Tauran}, M.~B. {Taylor}, R.~{Teixeira},
  W.~{Thuillot}, N.~{Tonello}, F.~{Torra}, J.~{Torra}, C.~{Turon}, N.~{Unger},
  M.~{Vaillant}, E.~{van Dillen}, O.~{Vanel}, A.~{Vecchiato}, Y.~{Viala},
  D.~{Vicente}, S.~{Voutsinas}, M.~{Weiler}, T.~{Wevers}, {\L}.~{Wyrzykowski},
  A.~{Yoldas}, P.~{Yvard}, H.~{Zhao}, J.~{Zorec}, S.~{Zucker}, C.~{Zurbach},
  and T.~{Zwitter}, ``{Gaia Early Data Release 3. The Gaia Catalogue of Nearby
  Stars},'' {\em \aap} {\bf 649}, A6  (2021).

\bibitem{sirbu2018}
D.~{Sirbu}, R.~{Belikov}, E.~{Bendeck}, C.~{Henze}, A.~J. {Eldorado-Riggs}, and
  S.~{Shaklan}, ``{Multi-star wavefront control for the wide-field infrared
  survey telescope},'' in {\em Space Telescopes and Instrumentation 2018:
  Optical, Infrared, and Millimeter Wave},  M.~Lystrup, H.~A. MacEwen, G.~G.
  Fazio, N.~Batalha, N.~Siegler, and E.~C. Tong, Eds.,  {\bf 10698}, 106982F,
  International Society for Optics and Photonics, SPIE  (2018).

\bibitem{mamajek2024}
E.~{Mamajek} and K.~{Stapelfeldt}, ``{NASA Exoplanet Exploration Program (ExEP)
  Mission Star List for the Habitable Worlds Observatory (2023)},'' {\em arXiv
  e-prints} , arXiv:2402.12414  (2024).

\bibitem{kopparapu2013}
R.~K. {Kopparapu}, R.~{Ramirez}, J.~F. {Kasting}, V.~{Eymet}, T.~D. {Robinson},
  S.~{Mahadevan}, R.~C. {Terrien}, S.~{Domagal-Goldman}, V.~{Meadows}, and
  R.~{Deshpande}, ``{Habitable Zones around Main-sequence Stars: New
  Estimates},'' {\em \apj} {\bf 765}, 131  (2013).

\bibitem{kopparapu2014}
R.~K. {Kopparapu}, R.~M. {Ramirez}, J.~{SchottelKotte}, J.~F. {Kasting},
  S.~{Domagal-Goldman}, and V.~{Eymet}, ``{Habitable Zones around Main-sequence
  Stars: Dependence on Planetary Mass},'' {\em \apjl} {\bf 787}, L29  (2014).

\bibitem{sag13_report}
R.~{Belikov}, C.~C. {Stark}, N.~{Batalha}, C.~{Burke}, and {SAG13 Members},
  ``{ExoPAG SAG13: Exoplanet Occurrence Rates and Distributions, Closeout
  Report},''  (2017).

\bibitem{bryson2021}
S.~{Bryson}, M.~{Kunimoto}, R.~K. {Kopparapu}, J.~L. {Coughlin}, W.~J.
  {Borucki}, D.~{Koch}, V.~S. {Aguirre}, C.~{Allen}, G.~{Barentsen}, N.~M.
  {Batalha}, T.~{Berger}, A.~{Boss}, L.~A. {Buchhave}, C.~J. {Burke}, D.~A.
  {Caldwell}, J.~R. {Campbell}, J.~{Catanzarite}, H.~{Chandrasekaran}, W.~J.
  {Chaplin}, J.~L. {Christiansen}, J.~{Christensen-Dalsgaard}, D.~R. {Ciardi},
  B.~D. {Clarke}, W.~D. {Cochran}, J.~L. {Dotson}, L.~R. {Doyle}, E.~S.
  {Duarte}, E.~W. {Dunham}, A.~K. {Dupree}, M.~{Endl}, J.~L. {Fanson}, E.~B.
  {Ford}, M.~{Fujieh}, I.~{Gautier}, Thomas~N., J.~C. {Geary}, R.~L.
  {Gilliland}, F.~R. {Girouard}, A.~{Gould}, M.~R. {Haas}, C.~E. {Henze}, M.~J.
  {Holman}, A.~W. {Howard}, S.~B. {Howell}, D.~{Huber}, R.~C. {Hunter}, J.~M.
  {Jenkins}, H.~{Kjeldsen}, J.~{Kolodziejczak}, K.~{Larson}, D.~W. {Latham},
  J.~{Li}, S.~{Mathur}, S.~{Meibom}, C.~{Middour}, R.~L. {Morris}, T.~D.
  {Morton}, F.~{Mullally}, S.~E. {Mullally}, D.~{Pletcher}, A.~{Prsa}, S.~N.
  {Quinn}, E.~V. {Quintana}, D.~{Ragozzine}, S.~V. {Ramirez}, D.~T.
  {Sanderfer}, D.~{Sasselov}, S.~E. {Seader}, M.~{Shabram}, A.~{Shporer}, J.~C.
  {Smith}, J.~H. {Steffen}, M.~{Still}, G.~{Torres}, J.~{Troeltzsch}, J.~D.
  {Twicken}, A.~K. {Uddin}, J.~E. {Van Cleve}, J.~{Voss}, L.~M. {Weiss}, W.~F.
  {Welsh}, B.~{Wohler}, and K.~A. {Zamudio}, ``{The Occurrence of Rocky
  Habitable-zone Planets around Solar-like Stars from Kepler Data},'' {\em \aj}
  {\bf 161}, 36  (2021).

\bibitem{ertel2020}
S.~{Ertel}, D.~{Defr{\`e}re}, P.~{Hinz}, B.~{Mennesson}, G.~M. {Kennedy}, W.~C.
  {Danchi}, C.~{Gelino}, J.~M. {Hill}, W.~F. {Hoffmann}, J.~{Mazoyer},
  G.~{Rieke}, A.~{Shannon}, K.~{Stapelfeldt}, E.~{Spalding}, J.~M. {Stone},
  A.~{Vaz}, A.~J. {Weinberger}, P.~{Willems}, O.~{Absil}, P.~{Arbo}, V.~P.
  {Bailey}, C.~{Beichman}, G.~{Bryden}, E.~C. {Downey}, O.~{Durney},
  S.~{Esposito}, A.~{Gaspar}, P.~{Grenz}, C.~A. {Haniff}, J.~M. {Leisenring},
  L.~{Marion}, T.~J. {McMahon}, R.~{Millan-Gabet}, M.~{Montoya}, K.~M.
  {Morzinski}, S.~{Perera}, E.~{Pinna}, J.~U. {Pott}, J.~{Power}, A.~{Puglisi},
  A.~{Roberge}, E.~{Serabyn}, A.~J. {Skemer}, K.~Y.~L. {Su},
  V.~{Vaitheeswaran}, and M.~C. {Wyatt}, ``{The HOSTS Survey for Exozodiacal
  Dust: Observational Results from the Complete Survey},'' {\em \aj} {\bf 159},
  177  (2020).

\bibitem{latouf2023}
N.~{Latouf}, A.~M. {Mandell}, G.~L. {Villanueva}, M.~D. {Moore},
  N.~{Susemiehl}, V.~{Kofman}, and M.~D. {Himes}, ``{Bayesian Analysis for
  Remote Biosignature Identification on exoEarths (BARBIE). I. Using Grid-based
  Nested Sampling in Coronagraphy Observation Simulations for H$_{2}$O},'' {\em
  \aj} {\bf 166}, 129  (2023).

\bibitem{latouf2024}
N.~{Latouf}, A.~M. {Mandell}, G.~L. {Villanueva}, M.~D. {Himes}, M.~D. {Moore},
  N.~{Susemiehl}, J.~{Crouse}, S.~{Domagal-Goldman}, G.~{Arney}, V.~{Kofman},
  and A.~V. {Young}, ``{Bayesian Analysis for Remote Biosignature
  Identification on exoEarths (BARBIE). II. Using Grid-based Nested Sampling in
  Coronagraphy Observation Simulations for O$_{2}$ and O$_{3}$},'' {\em \aj}
  {\bf 167}, 27  (2024).

\bibitem{bijan2020}
B.~{Nemati}, H.~P. {Stahl}, M.~T. {Stahl}, G.~J. {Ruane}, and L.~J. {Sheldon},
  ``{Method for deriving optical telescope performance specifications for
  Earth-detecting coronagraphs},'' {\em Journal of Astronomical Telescopes,
  Instruments, and Systems} {\bf 6}, 039002  (2020).

\bibitem{bruna2023}
M.~{Bruna}, N.~B. {Cowan}, J.~{Sheffler}, H.~M. {Haggard}, A.~{Bourdon}, and
  M.~{M{\^a}lin}, ``{Combining photometry and astrometry to improve orbit
  retrieval of directly imaged exoplanets},'' {\em MNRAS} {\bf 519}, 460--470
  (2023).

\bibitem{stark2016}
C.~C. {Stark}, E.~J. {Cady}, M.~{Clampin}, S.~{Domagal-Goldman}, D.~{Lisman},
  A.~M. {Mandell}, M.~W. {McElwain}, A.~{Roberge}, T.~D. {Robinson},
  D.~{Savransky}, S.~B. {Shaklan}, and K.~R. {Stapelfeldt}, ``{A direct
  comparison of exoEarth yields for starshades and coronagraphs},'' in {\em
  Space Telescopes and Instrumentation 2016: Optical, Infrared, and Millimeter
  Wave},  H.~A. {MacEwen}, G.~G. {Fazio}, M.~{Lystrup}, N.~{Batalha},
  N.~{Siegler}, and E.~C. {Tong}, Eds., {\em Society of Photo-Optical
  Instrumentation Engineers (SPIE) Conference Series} {\bf 9904}, 99041U
  (2016).

\bibitem{crass2021}
J.~{Crass}, B.~S. {Gaudi}, S.~{Leifer}, C.~{Beichman}, C.~{Bender},
  G.~{Blackwood}, J.~A. {Burt}, J.~L. {Callas}, H.~M. {Cegla}, S.~A. {Diddams},
  X.~{Dumusque}, J.~D. {Eastman}, E.~B. {Ford}, B.~{Fulton}, R.~{Gibson},
  S.~{Halverson}, R.~D. {Haywood}, F.~{Hearty}, A.~W. {Howard}, D.~W. {Latham},
  J.~{L{\"o}hner-B{\"o}ttcher}, E.~E. {Mamajek}, A.~{Mortier}, P.~{Newman},
  P.~{Plavchan}, A.~{Quirrenbach}, A.~{Reiners}, P.~{Robertson}, A.~{Roy},
  C.~{Schwab}, A.~{Seifahrt}, A.~{Szentgyorgyi}, R.~{Terrien}, J.~K. {Teske},
  S.~{Thompson}, and G.~{Vasisht}, ``{Extreme Precision Radial Velocity Working
  Group Final Report},'' {\em arXiv e-prints} , arXiv:2107.14291  (2021).

\bibitem{robinson2011}
T.~D. {Robinson}, V.~S. {Meadows}, D.~{Crisp}, D.~{Deming}, M.~F. {A'Hearn},
  D.~{Charbonneau}, T.~A. {Livengood}, S.~{Seager}, R.~K. {Barry}, T.~{Hearty},
  T.~{Hewagama}, C.~M. {Lisse}, L.~A. {McFadden}, and D.~D. {Wellnitz},
  ``{Earth as an Extrasolar Planet: Earth Model Validation Using EPOXI Earth
  Observations},'' {\em Astrobiology} {\bf 11}, 393--408  (2011).

\bibitem{tinetti2005}
G.~{Tinetti}, V.~S. {Meadows}, D.~{Crisp}, W.~{Fong }, T.~{Velusamy}, and
  H.~{Snively}, ``{Disk-Averaged Synthetic Spectra of Mars},'' {\em
  Astrobiology} {\bf 5}, 461--482  (2005).

\bibitem{tinetti2006}
G.~{Tinetti}, V.~S. {Meadows}, D.~{Crisp}, W.~{Fong}, E.~{Fishbein},
  M.~{Turnbull}, and J.-P. {Bibring}, ``{Detectability of Planetary
  Characteristics in Disk-Averaged Spectra. I: The Earth Model},'' {\em
  Astrobiology} {\bf 6}, 34--47  (2006).

\bibitem{cox1954}
C.~{Cox} and W.~{Munk}, ``{Measurement of the roughness of the sea surface from
  photographs of the sun's glitter},'' {\em Journal of the Optical Society of
  America (1917-1983)} {\bf 44}, 838  (1954).

\bibitem{mennesson2014}
B.~{Mennesson}, R.~{Millan-Gabet}, E.~{Serabyn}, M.~M. {Colavita}, O.~{Absil},
  G.~{Bryden}, M.~{Wyatt}, W.~{Danchi}, D.~{Defr{\`e}re}, O.~{Dor{\'e}},
  P.~{Hinz}, M.~{Kuchner}, S.~{Ragland}, N.~{Scott}, K.~{Stapelfeldt},
  W.~{Traub}, and J.~{Woillez}, ``{Constraining the Exozodiacal Luminosity
  Function of Main-sequence Stars: Complete Results from the Keck Nuller
  Mid-infrared Surveys},'' {\em \apj} {\bf 797}, 119  (2014).

\bibitem{kammerer2022}
J.~{Kammerer}, C.~C. {Stark}, K.~J. {Ludwick}, R.~{Juanola-Parramon}, and
  B.~{Nemati}, ``{Simulating Reflected Light Coronagraphy of Earth-like
  Exoplanets with a Large IR/O/UV Space Telescope: Impact and Calibration of
  Smooth Exozodiacal Dust},'' {\em \aj} {\bf 164}, 235  (2022).

\bibitem{defrere2012}
D.~{Defr{\`e}re}, C.~{Stark}, K.~{Cahoy}, and I.~{Beerer}, ``{Direct imaging of
  exoEarths embedded in clumpy debris disks},'' in {\em Space Telescopes and
  Instrumentation 2012: Optical, Infrared, and Millimeter Wave},  M.~C.
  {Clampin}, G.~G. {Fazio}, H.~A. {MacEwen}, and J.~{Oschmann}, Jacobus~M.,
  Eds., {\em Society of Photo-Optical Instrumentation Engineers (SPIE)
  Conference Series} {\bf 8442}, 84420M  (2012).

\bibitem{currie2023}
M.~H. {Currie}, C.~C. {Stark}, J.~{Kammerer}, R.~{Juanola-Parramon}, and V.~S.
  {Meadows}, ``{Mitigating Worst-case Exozodiacal Dust Structure in
  High-contrast Images of Earth-like Exoplanets},'' {\em \aj} {\bf 166}, 197
  (2023).

\bibitem{absil2013}
O.~{Absil}, D.~{Defr{\`e}re}, V.~{Coud{\'e} du Foresto}, E.~{Di Folco},
  A.~{M{\'e}rand}, J.~C. {Augereau}, S.~{Ertel}, C.~{Hanot}, P.~{Kervella},
  B.~{Mollier}, N.~{Scott}, X.~{Che}, J.~D. {Monnier}, N.~{Thureau}, P.~G.
  {Tuthill}, T.~A. {ten Brummelaar}, H.~A. {McAlister}, J.~{Sturmann},
  L.~{Sturmann}, and N.~{Turner}, ``{A near-infrared interferometric survey of
  debris-disc stars. III. First statistics based on 42 stars observed with
  CHARA/FLUOR},'' {\em \aap} {\bf 555}, A104  (2013).

\bibitem{ertel2014}
S.~{Ertel}, O.~{Absil}, D.~{Defr{\`e}re}, J.~B. {Le Bouquin}, J.~C. {Augereau},
  L.~{Marion}, N.~{Blind}, A.~{Bonsor}, G.~{Bryden}, J.~{Lebreton}, and
  J.~{Milli}, ``{A near-infrared interferometric survey of debris-disk stars.
  IV. An unbiased sample of 92 southern stars observed in H band with
  VLTI/PIONIER},'' {\em \aap} {\bf 570}, A128  (2014).

\bibitem{pseudozodi}
C.~C. {Stark}, M.~J. {Kuchner}, and A.~{Lincowski}, ``{The Pseudo-Zodi Problem
  for Edge-On Planetary Systems},'' {\em \apj} {\bf 801}, 128  (2015).

\bibitem{thompson2018}
S.~E. {Thompson}, J.~L. {Coughlin}, K.~{Hoffman}, F.~{Mullally}, J.~L.
  {Christiansen}, C.~J. {Burke}, S.~{Bryson}, N.~{Batalha}, M.~R. {Haas},
  J.~{Catanzarite}, J.~F. {Rowe}, G.~{Barentsen}, D.~A. {Caldwell}, B.~D.
  {Clarke}, J.~M. {Jenkins}, J.~{Li}, D.~W. {Latham}, J.~J. {Lissauer},
  S.~{Mathur}, R.~L. {Morris}, S.~E. {Seader}, J.~C. {Smith}, T.~C. {Klaus},
  J.~D. {Twicken}, J.~E. {Van Cleve}, B.~{Wohler}, R.~{Akeson}, D.~R. {Ciardi},
  W.~D. {Cochran}, C.~E. {Henze}, S.~B. {Howell}, D.~{Huber}, A.~{Pr{\v{s}}a},
  S.~V. {Ram{\'\i}rez}, T.~D. {Morton}, T.~{Barclay}, J.~R. {Campbell}, W.~J.
  {Chaplin}, D.~{Charbonneau}, J.~{Christensen-Dalsgaard}, J.~L. {Dotson},
  L.~{Doyle}, E.~W. {Dunham}, A.~K. {Dupree}, E.~B. {Ford}, J.~C. {Geary},
  F.~R. {Girouard}, H.~{Isaacson}, H.~{Kjeldsen}, E.~V. {Quintana},
  D.~{Ragozzine}, M.~{Shabram}, A.~{Shporer}, V.~{Silva Aguirre}, J.~H.
  {Steffen}, M.~{Still}, P.~{Tenenbaum}, W.~F. {Welsh}, A.~{Wolfgang}, K.~A.
  {Zamudio}, D.~G. {Koch}, and W.~J. {Borucki}, ``{Planetary Candidates
  Observed by Kepler. VIII. A Fully Automated Catalog with Measured
  Completeness and Reliability Based on Data Release 25},'' {\em \apjs} {\bf
  235}, 38  (2018).

\bibitem{berger2020a}
T.~A. {Berger}, D.~{Huber}, E.~{Gaidos}, J.~L. {van Saders}, and L.~M. {Weiss},
  ``{The Gaia-Kepler Stellar Properties Catalog. II. Planet Radius Demographics
  as a Function of Stellar Mass and Age},'' {\em \aj} {\bf 160}, 108  (2020).

\bibitem{berger2020b}
T.~A. {Berger}, D.~{Huber}, J.~L. {van Saders}, E.~{Gaidos}, J.~{Tayar}, and
  A.~L. {Kraus}, ``{The Gaia-Kepler Stellar Properties Catalog. I. Homogeneous
  Fundamental Properties for 186,301 Kepler Stars},'' {\em \aj} {\bf 159}, 280
  (2020).

\bibitem{saxena2022}
P.~{Saxena}, ``{Photobombing Earth 2.0: Diffraction-limit-related Contamination
  and Uncertainty in Habitable Planet Spectra},'' {\em \apjl} {\bf 934}, L32
  (2022).

\bibitem{krissansentotton2016}
J.~{Krissansen-Totton}, E.~W. {Schwieterman}, B.~{Charnay}, G.~{Arney}, T.~D.
  {Robinson}, V.~{Meadows}, and D.~C. {Catling}, ``{Is the Pale Blue Dot
  Unique? Optimized Photometric Bands for Identifying Earth-like Exoplanets},''
  {\em \apj} {\bf 817}, 31  (2016).

\bibitem{stlaurent2018}
K.~{St. Laurent}, K.~{Fogarty}, N.~T. {Zimmerman}, M.~{N'Diaye}, C.~C. {Stark},
  J.~{Mazoyer}, A.~{Sivaramakrishnan}, L.~{Pueyo}, S.~{Shaklan},
  R.~{Vanderbei}, and R.~{Soummer}, ``{Apodized pupil Lyot coronagraphs designs
  for future segmented space telescopes},'' in {\em Space Telescopes and
  Instrumentation 2018: Optical, Infrared, and Millimeter Wave},  M.~{Lystrup},
  H.~A. {MacEwen}, G.~G. {Fazio}, N.~{Batalha}, N.~{Siegler}, and E.~C. {Tong},
  Eds., {\em Society of Photo-Optical Instrumentation Engineers (SPIE)
  Conference Series} {\bf 10698}, 106982W  (2018).

\bibitem{brown2005}
R.~A. {Brown}, ``{Single-Visit Photometric and Obscurational Completeness},''
  {\em \apj} {\bf 624}, 1010--1024  (2005).

\bibitem{guyon2021}
O.~{Guyon}, B.~{Norris}, M.-A. {Martinod}, K.~{Ahn}, P.~{Tuthill}, J.~{Males},
  A.~{Wong}, N.~{Skaf}, T.~{Currie}, K.~{Miller}, S.~{Bos}, J.~{Lozi},
  V.~{Deo}, S.~{Vievard}, R.~{Belikov}, K.~{van Gorkom}, S.~{Haffert},
  B.~{Mazin}, M.~{Bottom}, R.~{Frazin}, A.~{Rodack}, T.~{Groff},
  N.~{Jovanovic}, and F.~{Martinache}, ``{High contrast imaging at the photon
  noise limit with self-calibrating WFS/C systems},'' in {\em Techniques and
  Instrumentation for Detection of Exoplanets X},  S.~B. {Shaklan} and G.~J.
  {Ruane}, Eds., {\em Society of Photo-Optical Instrumentation Engineers (SPIE)
  Conference Series} {\bf 11823}, 1182318  (2021).

\bibitem{soummer2011}
R.~{Soummer}, J.~B. {Hagan}, L.~{Pueyo}, A.~{Thormann}, A.~{Rajan}, and
  C.~{Marois}, ``{Orbital Motion of HR 8799 b, c, d Using Hubble Space
  Telescope Data from 1998: Constraints on Inclination, Eccentricity, and
  Stability},'' {\em \apj} {\bf 741}, 55  (2011).

\bibitem{fergus2014}
R.~{Fergus}, D.~W. {Hogg}, R.~{Oppenheimer}, D.~{Brenner}, and L.~{Pueyo},
  ``{S4: A Spatial-spectral model for Speckle Suppression},'' {\em \apj} {\bf
  794}, 161  (2014).

\bibitem{morrissey2023}
P.~{Morrissey}, L.~K. {Harding}, N.~L. {Bush}, M.~{Bottom}, B.~{Nemati},
  A.~{Daniel}, B.~{Jun}, L.~{Maria Martinez-Sierra}, N.~{Desai}, D.~{Barry},
  R.-T. {Davis}, R.~T. {Demers}, D.~J. {Hall}, A.~{Holland}, P.~{Turner}, and
  B.~{Shortt}, ``{Flight photon counting electron multiplying charge coupled
  device development for the Roman Space Telescope coronagraph instrument},''
  {\em Journal of Astronomical Telescopes, Instruments, and Systems} {\bf
  9}(1), 016003  (2023).

\bibitem{bebek2015}
C.~Bebek, J.~Emes, D.~Groom, S.~Haque, S.~Holland, A.~Karcher, W.~Kolbe,
  J.~Lee, N.~Palaio, and G.~Wang, ``Ccd development for the dark energy
  spectroscopic instrument,'' {\em Journal of Instrumentation} {\bf 10}, C05026
   (2015).

\bibitem{barak2022}
L.~{Barak}, I.~M. {Bloch}, A.~{Botti}, M.~{Cababie}, G.~{Cancelo},
  L.~{Chaplinsky}, F.~{Chierchie}, M.~{Crisler}, A.~{Drlica-Wagner},
  R.~{Essig}, J.~{Estrada}, E.~{Etzion}, G.~{Fernandez Moroni}, D.~{Gift},
  S.~E. {Holland}, S.~{Munagavalasa}, A.~{Orly}, D.~{Rodrigues}, A.~{Singal},
  M.~S. {Haro}, L.~{Stefanazzi}, J.~{Tiffenberg}, S.~{Uemura}, T.~{Volansky},
  T.-T. {Yu}, and {Sensei Collaboration}, ``{SENSEI: Characterization of
  Single-Electron Events Using a Skipper Charge-Coupled Device},'' {\em
  Physical Review Applied} {\bf 17}, 014022  (2022).

\bibitem{Tiffenberg:2017gsa}
J.~Tiffenberg, M.~Sofo-Haro, A.~Drlica-Wagner, R.~Essig, Y.~Guardincerri,
  S.~Holland, T.~Volansky, and T.-T. Yu, ``{Single-Electron and Single-Photon
  Sensitivity with a Silicon Skipper CCD},'' {\em Phys. Rev. Lett.} {\bf 119},
  131802--131806  (2017).

\bibitem{botti2023}
A.~M. Botti, B.~A. Cervantes-Vergara, C.~R. Chavez, F.~Chierchie,
  A.~Drlica-Wagner, J.~Estrada, G.~F. Moroni, S.~E. Holland, B.~J.~I. Gimenez,
  A.~J. Lapi, E.~M. Villalpando, M.~S. Haro, J.~Tiffenberg, and S.~Uemura,
  ``{Fast Single-Quantum Measurement with a Multi-Amplifier Sensing
  Charge-Coupled Device},'' 1--8  (2023).

\bibitem{rauscher2016}
B.~J. {Rauscher}, E.~R. {Canavan}, S.~H. {Moseley}, J.~E. {Sadleir}, and
  T.~{Stevenson}, ``{Detectors and cooling technology for direct spectroscopic
  biosignature characterization},'' {\em Journal of Astronomical Telescopes,
  Instruments, and Systems} {\bf 2}, 041212  (2016).

\bibitem{TEStechreport}
``{Ultra-high Efficiency Noiseless Quantum Sensors for HWO and QIS}.'' Webpage,
  \url{https://techport.nasa.gov/view/146757}.
\newblock (Accessed: 17 July 2024).

\bibitem{devisser2021}
P.~J. {de Visser}, S.~A.~H. {de Rooij}, V.~{Murugesan}, D.~J. {Thoen}, and
  J.~J.~A. {Baselmans}, ``{Phonon-Trapping-Enhanced Energy Resolution in
  Superconducting Single-Photon Detectors},'' {\em Physical Review Applied}
  {\bf 16}, 034051  (2021).

\bibitem{belikov2023}
R.~{Belikov}, C.~{Stark}, N.~{Siegler}, E.~{Por}, B.~{Mennesson}, P.~{Chen},
  K.~{Fogarty}, O.~{Guyon}, R.~{Juanola-Parramon}, J.~{Krist}, D.~{Mawet},
  C.~{Mejia Prada}, J.~{Kasdin}, L.~{Pueyo}, S.~{Redmond}, G.~{Ruane},
  D.~{Sirbu}, K.~{Stapelfeldt}, J.~{Trauger}, and N.~{Zimmerman},
  ``{Coronagraph design survey for future exoplanet direct imaging space
  missions: interim update},'' in {\em Techniques and Instrumentation for
  Detection of Exoplanets XI},  G.~J. Ruane, Ed.,  {\bf 12680}, 126802G,
  International Society for Optics and Photonics, SPIE  (2023).

\end{thebibliography}

\listoffigures
\listoftables

\end{spacing}
\end{document}